\documentclass[a4paper,11pt]{article}
\usepackage{jheppub} 
\usepackage{lineno}
\usepackage{graphicx,float,wrapfig,subfigure}
\usepackage{amsfonts,amsmath,amssymb,amstext}
\usepackage{latexsym}
\usepackage{bm}
\usepackage{color}
\usepackage[normalem]{ulem}
\usepackage{MnSymbol}
\usepackage[colorlinks=true,linktocpage=true,linkcolor=blue,citecolor=blue]{hyperref}
\usepackage{slashed}
\usepackage{orcidlink}    
\usepackage{comment}
\usepackage{booktabs}
\usepackage{siunitx}
\allowdisplaybreaks[1]

\newcommand{\sign}{ \mbox{sign}}
\renewcommand{\Im}{\mbox{Im}}

\definecolor{red}{rgb}{0.7,0,0}
\definecolor{green}{rgb}{0,0.5,0}

\newcommand{\Tr}{\operatorname{Tr}}

\arxivnumber{2609.00157} 

\title{Magnetic field-induced enhancement and quenching of Urca emission in quark matter} 

\author[a]{William Gyory\,\orcidlink{0000-0002-6936-6911}}
\author[a,b,1]{Igor A. Shovkovy\,\orcidlink{0000-0002-5230-6891}\note{Corresponding author.}}
\affiliation[a]{College of Integrative Sciences and Arts, Arizona State University, Mesa, Arizona 85212, USA}
\affiliation[b]{Department of Physics, Arizona State University, Tempe, Arizona 85287, USA}

\emailAdd{wgyory@asu.edu}
\emailAdd{igor.shovkovy@asu.edu}

\abstract{Using first-principles field-theoretic methods, we investigate neutrino emission from strongly magnetized dense quark matter under conditions relevant to compact stars. We account for Landau-level quantization of both electron and quark states and show that it strongly modifies the kinematics of Urca processes. In particular, quark quantization restricts the available phase space in which both quark and electron energies can simultaneously lie near their respective Fermi surfaces. At moderately strong magnetic fields, before pronounced quark quantization sets in, the emission rate tends to increase on average with increasing field strength. In the regime of very strong fields, however, the increasingly restricted phase space first gives rise to Shubnikov--de Haas-type oscillations and then to resonance-like spikes near a discrete sequence of Urca-resonant magnetic field values, separated by regions of strong suppression. Finally, the emission rate becomes nearly completely quenched once $|eB| \gtrsim 6\mu_{e}\mu_{u}$, corresponding to approximately $B\gtrsim 1.5\times 10^{19}~\mbox{G}$ for the representative set of model parameters considered. We also find significant anisotropy in the longitudinal momentum emission near the Urca-resonant magnetic field values.} 


\begin{document}
\maketitle
\flushbottom

\section{Introduction}
\label{sec:Introduction}

There is growing evidence that the cores of the most massive compact stars may contain deconfined quark matter~\cite{Annala:2019puf,Fujimoto:2022ohj,Annala:2023cwx}. The presence of such matter can substantially modify the macroscopic and transport properties of these objects relative to purely hadronic neutron stars \cite{Schmitt:2017efp,Baym:2019iky}. Among the affected physical phenomena, cooling is particularly important \cite{Yakovlev:2000jp,Potekhin:2015qsa}, since neutrino emission from deconfined quark matter can provide an efficient channel for energy loss \cite{Iwamoto:1980eb,Iwamoto:1982zz}.

\begin{figure}
\centering
  \subfigure[]{\includegraphics[width=0.525\textwidth]{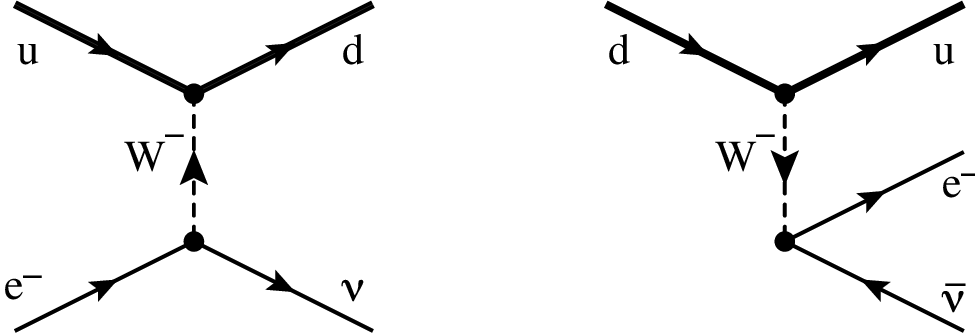}}
  \hspace{0.1\textwidth}
  \subfigure[]{\includegraphics[width=0.35\textwidth]{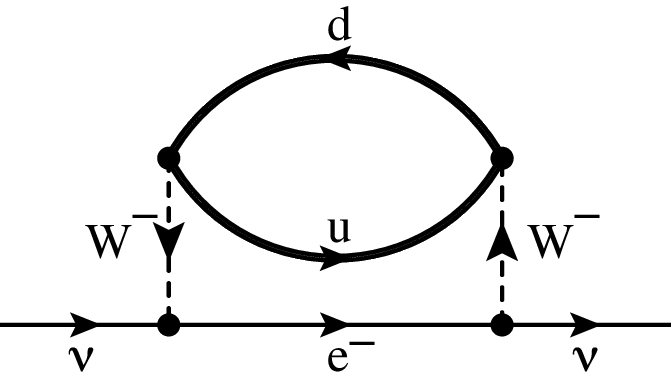}}
\caption{(a) Direct Urca processes that dominate (anti)neutrino emission from dense quark matter.
(b) Neutrino self-energy diagram used to calculate the corresponding (anti)neutrino emission rate.}
\label{fig.NuSelfEne}
\end{figure}

Neutrino emission from dense quark matter is dominated by direct Urca processes involving weak interactions among quarks and leptons~\cite{Iwamoto:1980eb,Iwamoto:1982zz,Schafer:2004jp}. The relevant reactions are electron capture by an up quark, $u+e^- \rightarrow d+\nu_{e}$,
and $\beta$-decay of a down quark, $ d \rightarrow u+e^-+\bar{\nu}_{e}$. These processes are illustrated in figure~\ref{fig.NuSelfEne}(a). Since neutrinos and antineutrinos have extremely long mean free paths in sufficiently cold stellar matter, they can escape without rescattering and efficiently carry away energy. Direct Urca emission therefore provides the principal cooling mechanism for dense quark matter in compact-star interiors. In the absence of magnetic fields, the density and temperature dependence of the corresponding emissivities has been well understood since the pioneering works of Iwamoto~\cite{Iwamoto:1980eb,Iwamoto:1982zz}. 

The situation becomes considerably richer in the presence of a strong magnetic field. While sizable magnetic fields are ubiquitous in compact stars, their effects are expected to be most pronounced in magnetars, where surface fields can reach $10^{15}~\mbox{G}$ and interior fields may be substantially stronger~\cite{Lai_1991,Duncan:1992hi,Turolla:2015mwa,Kaspi:2017fwg}. In the case of nuclear matter, the effects of strong magnetic fields on neutrino emission and neutron-star cooling have been investigated in refs.~\cite{Baiko:1998jq,Leinson:1998yr,Duan:2005fc,Riquelme:2005ac,Potekhin:2017ufy,Dehman:2022rpa}. Recent studies have revisited the Urca rates from nuclear matter in strong magnetic fields, demonstrating that magnetic effects can be especially important near or below the  direct-Urca threshold~\cite{Tambe:2024usx,Kumamoto:2024jiq}. In addition, the possible role of magnetic fields in generating anisotropies in neutrino transport and pulsar kicks has also been discussed in refs.~\cite{Lai:1998sz,Arras:1998cv,Sagert:2007as,Ayala:2018kie,Ayala:2024wgb,Ghosh:2025sjn}.

In this work, we study the Urca rate in strongly magnetized dense quark matter using first-principles field-theoretic methods. By quantizing the spectra of charged particles into Landau levels (LLs), the magnetic field modifies the available phase space for Urca reactions. Unlike in nuclear matter, three of the four fermions participating in direct Urca processes in quark matter are electrically charged. In ref.~\cite{Ghosh:2025sjn}, this problem was investigated by treating the LL quantization of electrons exactly, while neglecting the analogous quantization of quark states. This approximation is justified for magnetic fields below roughly $10^{17}~\mbox{G}$. Indeed, the spacing between adjacent quark LLs near the Fermi surface scales as $|e_{f} B|/\mu_{f}$, where $f=u,d$ denotes the quark flavor. Since the typical quark chemical potentials, $\mu_{f}\simeq 300~\mbox{MeV}$, are much larger than the electron chemical potential, $\mu_{e}\simeq 50~\mbox{MeV}$, the quark LL spacing remains comparatively small. Thus, for fields below about $10^{17}~\mbox{G}$, the quark spectrum is effectively quasicontinuous, whereas the electron spectrum can already be strongly quantized. 

At magnetic fields on the order of $10^{18}~\mbox{G}$ or larger, LL quantization of quarks can no longer be neglected. For reference, table~\ref{tab:landau-levels} lists the approximate maximum indices $n_{\rm max}$ of occupied LLs for electrons and quarks at several representative magnetic field strengths. The resulting impact on neutrino emission is not obvious a priori. On the one hand, quantization imposes kinematic restrictions on the Urca processes, potentially suppressing the emissivity. On the other hand, by modifying the momentum structure of the participating quarks, Landau quantization effectively relaxes the near-collinearity constraints that otherwise limit the available phase space \cite{Ghosh:2025sjn}. In this respect, its effect can be qualitatively analogous to the role of Fermi-liquid corrections in strongly interacting quark matter, which also open up the Urca phase space and enhance the rate~\cite{Iwamoto:1980eb,Iwamoto:1982zz}. Determining which tendency dominates requires a systematic calculation.

In this study, we compute both the energy emission rate and the net longitudinal momentum emission rate of neutrinos from magnetized dense quark matter, fully incorporating the LL quantization of electrons and quarks. This approach allows us to assess how ultra-strong magnetic fields modify the phase space, kinematics, and anisotropy of direct Urca emission. The resulting framework provides a quantitative basis for studying neutrino cooling and momentum transport in dense quark matter under magnetic field strengths beyond the regime considered in ref.~\cite{Ghosh:2025sjn}.

\begin{table}[t]
\centering
\label{tab:landau-levels}
\renewcommand{\arraystretch}{1.2}
\setlength{\tabcolsep}{12pt}
\begin{tabular}{lcccc}
\hline
Species & $n_{\rm max}$ & $B=10^{17}\,\mathrm{G}$ & $B=10^{18}\,\mathrm{G}$ & $B=10^{19}\,\mathrm{G}$ \\
\hline
electrons  &  $\mu_{e}^2/(2|eB|)$ & $\approx 2$ & $\approx 0$ & $\approx 0$\\
down quarks & $ 3\mu_{d}^2/(2|eB|)$ & $\approx 311$ & $\approx 31$ & $\approx 3$\\
up quarks & $3\mu_{u}^2/(4|eB|)$ & $\approx 114$ &   $\approx 11$ & $\approx 1$\\
\hline
\end{tabular}
\caption{Approximate maximum indices $n_{\rm max}$ of occupied LLs for electrons and quarks at selected magnetic field strengths. We use $\mu_{e}=50~\text{MeV}$, $\mu_{d}=350~\text{MeV}$, and $\mu_{u}=300~\text{MeV}$.}
\end{table}

This paper is organized as follows. In section~\ref{sec:formalism}, we develop the general formalism, based on the Kadanoff--Baym transport equation, for calculating neutrino emission from unpaired dense quark matter in a background magnetic field. In section~\ref{sec:Numerics}, we present our main numerical results and investigate the dependence of the emission rates on the magnetic field strength and temperature. We identify and explain Shubnikov--de Haas-type oscillations associated with the thresholds for populating successive $u$-quark LLs. We also examine how the calculated rates approach the weak-field regime, in which LL quantization of the quark states becomes negligible. Section~\ref{sec:Summary} summarizes the main findings of the study. Several appendices provide additional technical details of our calculations.

\section{General Formalism}
\label{sec:formalism}

Following an approach similar to that of refs.~\cite{Sedrakian:1999jh,Schmitt:2005wg,Jaikumar:2005hy,Ghosh:2025sjn,Shovkovy:2026aci}, here we employ the Kadanoff--Baym transport equation for calculating the (anti-)neutrino emission rate. Unlike the conventional method, which relies on amplitudes constructed from cumbersome Landau-level wave functions in a background magnetic field, the Kadanoff--Baym formalism is formulated in terms of Green functions, making the computation of neutrino emission rates more efficient.

We assume unpaired dense quark matter composed of the two lightest quark flavors, up and down. We also consider the regime with no neutrino trapping in quark matter, implying that the temperature is sufficiently low to render the neutrino mean free path comparable to or larger than the star size, and hence the neutrino chemical potential vanishes ($\mu_\nu =0$). We will account for the fact that, in electrically neutral matter, a small but nonzero density of electrons is present. In $\beta$-equilibrium, the chemical potentials of the up and down quarks ($\mu_{u}$ and $\mu_{d}$) are related to the electron chemical potential ($\mu_{e}$) by $\mu_{d} = \mu_{u} + \mu_{e}$.

\subsection{Kadanoff--Baym transport equation for neutrinos}

We assume that dense quark matter in the stellar core is approximately spatially homogeneous within regions that are small compared to the stellar radius but large compared to the microscopic length scales relevant to neutrino emission. The out-of-equilibrium neutrino Green functions $G^{\lessgtr}_\nu(t,P_\nu)$ then depend only on time and can be expressed in terms of the (left-handed) neutrino distribution function $f_\nu(t,\bm{p}_\nu)$ as follows \cite{Sedrakian:1999jh,Schmitt:2005wg}:
\begin{align}
i G^{<}_\nu(t,P_\nu) &= -\frac{\pi}{p_\nu}\frac{1-\gamma_5}{2} (\gamma^\lambda P_{\nu,\lambda}+\mu_\nu \gamma_{0}) \Big\{f_\nu(t,\bm{p}_\nu)\delta(p_{\nu,0}+\mu_\nu-p_\nu) \nonumber \\ 
&-\left[1-f_{\bar{\nu}}(t,-\bm{p}_\nu)\right]\delta(p_{\nu,0}+\mu_\nu+p_\nu)\Big\} ,  \label{G-less} \\
i G^{>}_\nu(t,P_\nu) &= \frac{\pi}{p_\nu} \frac{1-\gamma_5}{2} (\gamma^\lambda P_{\nu,\lambda}+\mu_\nu \gamma_{0}) \Big\{\left[1-f_\nu(t,\bm{p}_\nu)\right] \delta(p_{\nu,0}+\mu_\nu-p_\nu)\nonumber \\ 
&-f_{\bar{\nu}}(t,-\bm{p}_\nu)\delta(p_{\nu,0}+\mu_\nu+p_\nu)\Big\} .  \label{G-greater}
\end{align}
Here $P_\nu =(p_{\nu,0},\bm{p}_\nu)$ is by definition the neutrino four-momentum and $p_\nu =|\bm{p}_\nu|$ is the magnitude of its three-momentum. Since we consider quark matter without neutrino trapping, we set $\mu_\nu =0$ in this study. To avoid potential confusion, throughout this study we use the subscript $\nu$ exclusively to denote a neutrino and never as a Lorentz index.

The Green functions satisfy the Kadanoff--Baym transport equation,
\begin{equation}
i \partial_t \mbox{Tr}[\gamma^0 G_\nu^{<} (t,P_\nu)] = -\mbox{Tr}[G_\nu^{>} (t,P_\nu)\Sigma^{<}_\nu(t,P_\nu)-\Sigma^{>}_\nu(t,P_\nu)G_\nu^{<} (t,P_\nu)],
\label{KB-kinetic-eq}
\end{equation}
where $\Sigma_\nu^{\lessgtr}$ are the lesser and greater neutrino self-energies. The Urca processes in figure~\ref{fig.NuSelfEne}(a) are determined by the following charged-current weak-interaction Lagrangian \cite{Weinberg:1996kr}:
\begin{equation}
\mathcal{L}=\frac{G_F \cos \theta_C}{\sqrt{2}} \bar{u} \gamma^\mu(1-\gamma_5) d \, \bar{e} \gamma_\mu (1-\gamma_5)\nu_{e},
\label{interaction-Lagrangian}
\end{equation}
where $G_F\approx 1.166\times 10^{-11}~\mbox{MeV}^{-2}$ is the Fermi coupling constant and $\theta_C$ is the Cabibbo angle (note that $\cos^2\theta_C \approx 0.948$). At leading one-loop order, the corresponding contribution to the neutrino self-energy is given by
\begin{equation}
\Sigma_\nu^{\lessgtr}(P_\nu) = i  \frac{G_F^2\cos^2\theta_C}{2} \int \frac{d^4 Q}{(2\pi)^4} \gamma^\delta (1-\gamma^5)\bar{S}_{e}^{\lessgtr}(P_\nu + Q)\gamma^\sigma(1-\gamma^5)\bar{\Pi}^{\gtrless }_{\delta\sigma}(Q), 
\label{Sigma-gtr}
\end{equation}
where $\bar{S}_{e}^\lessgtr$ and $\bar{\Pi}_{\delta\sigma}^\lessgtr$ are the electron propagator and W-boson self-energy, respectively. The corresponding Feynman diagram is shown in figure~\ref{fig.NuSelfEne}(b).

In the presence of a background magnetic field, the coordinate-space representations of Green functions for charged particles take the forms $S_{e}^{\lessgtr}(u,u^{\prime})=e^{i\Phi(u,u^{\prime})}\bar{S}_{e}^{\lessgtr}(u-u^{\prime})$ and $\Pi^{\lessgtr}_{\delta\sigma}(u^{\prime},u)=e^{-i\Phi(u,u^{\prime})}\bar{\Pi}^{\lessgtr}_{\delta\sigma}(u^{\prime}-u)$, respectively, where $\Phi(u,u^{\prime})$ is the Schwinger phase and $u=(t,\bm{r})$ represents the spacetime coordinates. While the Schwinger phases formally break the translational invariance of the Green functions, they cancel out in the product of the two Green functions. Taking this cancellation into account and applying the Fourier transforms of the translationally invariant parts of these quantities,
\begin{align}
\bar{S}_{e}^{\lessgtr}(P_{e}) &= \int  d^4 u \, e^{i P_{e}^\lambda u_\lambda }\bar{S}_{e}^{\lessgtr}(u) ,
\label{S-Fourier} \\
\bar{\Pi}^{\gtrless}_{\delta\sigma}(Q)&= \int  d^4 u \, e^{i Q^\lambda u_\lambda } \bar{\Pi}^{\gtrless}_{\delta\sigma}(u),
\label{Pi-Fourier}
\end{align}
one arrives at Eq.~(\ref{Sigma-gtr}), which formally resembles the zero-field case.

Using the spectral function representation, the electron Green functions can be written as follows:
\begin{align}
i\bar{S}_{e}^{>}(P_{e})&=[1-n_F(p_{e,0})]A_{e}(p_{e,0}+\mu_{e},\bm{p}_{e}), \\
i\bar{S}_{e}^{<}(P_{e})&= -n_F(p_{e,0})A_{e}(p_{e,0}+\mu_{e},\bm{p}_{e}),
\end{align}
where $n_{F}(p_{0}) = 1/\left[\exp(p_{0}/T)+1\right]$ is the Fermi--Dirac distribution and $A_{e}(p_{e,0}+\mu_{e},\bm{p}_{e})$ is the electron spectral function. In the LL representation, the explicit expression for the spectral function reads \cite{Miransky:2015ava}:
\begin{align}
    A_{e}(p_{e,0}+\mu_{e},\bm{p}_{e}) 
    &= 2\pi e^{-p_{e,\perp}^2\ell^2}  \sum_{\lambda_{e}=\pm 1}\sum_{n=0}^\infty
    \frac{(-1)^n}{E_{e,n}}
    \Big\{
        \left[
            E_{e,n}\gamma^{0} 
            -\lambda_{e} p_{e,z}\gamma^3+ \lambda_{e} m_{e} 
        \right]
        \Big[
            {\cal P}_{+}L_{n}
            \left(2 p_{e,\perp}^2\ell^2\right)\nonumber \\ 
&-{\cal P}_{-}L_{n-1}
            \left(2p_{e,\perp}^2\ell^2\right)
        \Big]+2\lambda_{e}(\bm{p}_{e,\perp} \cdot \bm\gamma_{\perp}) L_{n-1}^1
        \left(2 p_{e,\perp}^2\ell^{2} \right)
    \Big\}\delta(p_{e,0}+\mu_{e}-\lambda_{e} E_{e,n}) ,
\label{electron-spectral-density}
\end{align}
where $ E_{e,n}=\sqrt{2n|eB|+p_{e,z}^2+ m_{e}^{2} }$ are the electron LL energies, $\ell =1/\sqrt{|eB|}$ is the magnetic length for the electron, ${\cal P}_{\pm}=(1\pm i s_{\perp} \gamma^1\gamma^2)/2$ are the spin projectors, $s_{\perp} = \mbox{sign}(eB)$, and $L_{n}^{\alpha}\left(z\right)$ are the generalized Laguerre polynomials \cite{Gradshteyn:1943cpj} with the convention that $L_{-1}^{\alpha}(z)=0$. The sum over $\lambda_{e}$ accounts for the contributions from particles ($\lambda_{e}=+1$) and antiparticles ($\lambda_{e}=-1$). In dense matter with $T\ll \mu_{e}$, the antiparticle contribution is negligible and will eventually be omitted in the derivations of the emission rate below.

Similarly, the $W$-boson self-energies can be expressed as follows \cite{Bellac:2011kqa}:
\begin{align}
 i \bar{\Pi}_{\delta\sigma}^>(Q)&= 2[1+n_B(q_{0})] \Im \left[\bar{\Pi}^R_{\delta\sigma}(Q)\right],\\
 i\bar{\Pi}_{\delta\sigma}^<(Q)&= 2n_B(q_{0})\Im \left[\bar{\Pi}^R_{\delta\sigma}(Q)\right],
\end{align}
where $n_B(q_{0}) =1/\left[\exp(q_{0}/T)-1\right]$ is the Bose--Einstein distribution function and $\bar{\Pi}^R_{\delta\sigma}(Q)$ denotes the Fourier transform of the translationally invariant part of the $W$-boson retarded Green function. As seen from figure~\ref{fig.NuSelfEne}(b), this self-energy is determined by the one-loop quark diagram. An explicit expression for $\mbox{Im}\left[\bar{\Pi}_{\delta\sigma}^R(Q)\right]$ is derived in appendix~\ref{sec:Im-Pi} and presented in the following subsection.

For electrons, LL quantization becomes important already at magnetic fields $B\gtrsim 10^{16}~\mbox{G}$ for realistic values of the electron chemical potential. In fact, for $\mu_{e}=50~\mbox{MeV}$, only the lowest Landau level (LLL) is occupied when $|eB|\gtrsim \frac{1}{2}\mu_{e}^2$, i.e., $B\gtrsim 2.11\times 10^{17}~\mbox{G}$. When converting between magnetic field units, it is convenient to use the following approximate relation:
\begin{equation}
B\approx 1.69\times 10^{14}\frac{|eB|}{\mbox{MeV}^2}~\mbox{G} .
\label{eq:mag-field-conv}
\end{equation}
Since quarks also carry electric charge, the background magnetic field should be included in their propagators when calculating the retarded $W$-boson self-energy. A simple estimate of the energy spacing between neighboring LLs near the Fermi surface gives $\Delta \epsilon_{f} \simeq |e_{f}B|/\mu_{f}$, where $f=u,d$ labels the quark flavor. It is then clear that LL quantization becomes important when this spacing is comparable to or exceeds the thermal broadening scale, i.e., $\Delta \epsilon_{f} \gtrsim 2\pi T$. For typical quark chemical potentials, $\mu_{u},\mu_{d}\sim 300~\mbox{MeV}$, this estimate suggests that quark LL quantization becomes relevant at magnetic fields of order $10^{17}~\mbox{G}$ for $T=0.1~\mbox{MeV}$ and of order $10^{18}~\mbox{G}$ for $T=1~\mbox{MeV}$. 

As shown in table~\ref{tab:landau-levels}, the number of occupied quark LLs decreases substantially for magnetic fields $B\gtrsim 10^{18}~\mbox{G}$. In this regime, LL quantization of quark states becomes essential for an accurate calculation of the Urca emission rate. As the magnetic field increases further, up quarks become confined to the LLL at $B\gtrsim 1.14\times 10^{19}~\mbox{G}$, while down quarks enter the LLL regime at $B\gtrsim 3.11\times 10^{19}~\mbox{G}$. 

Previous studies \cite{Ghosh:2025sjn,Shovkovy:2026aci} focused on magnetic fields $B \lesssim 5 \times 10^{17}~\mbox{G}$, where the quark spectrum was treated as effectively continuous, while electron LL quantization was already included explicitly. In the present work, we focus instead on the complementary regime $B \gtrsim 5 \times 10^{17}~\mbox{G}$, where quark LL quantization can no longer be neglected and electrons are confined to the LLL.

\subsection{Neutrino-number production rates}
\label{sec:nu-number-production}

Substituting the spectral representations of the neutrino and electron Green functions, as well as the $W$-boson self-energy, into the kinetic equation (\ref{KB-kinetic-eq}), one derives the expression for neutrino-number production rate \cite{Ghosh:2025sjn,Shovkovy:2026aci}:
\begin{align}
    \frac{\partial f_\nu(t,\bm{p}_\nu)}{\partial t} &= -\frac{G_F^2\cos^2\theta_C}{2}
    \hspace{-4pt}
    \sum_{\lambda_{e}=\pm}
    \sum_{n_{e}=0}^\infty (-1)^{n_{e}} 
    \hspace{-4pt}
    \int \hspace{-4pt} \frac{d^3\bm{p}_{e} e^{-p_{e,\perp}^2\ell^2}}{(2\pi)^3 E_\nu E_{e,n_{e}}}
    n_F(E_{e,n_{e}}-\mu_{e})  n_B(E_\nu+\mu_{e}-E_{e,n_{e}}) \nonumber \\ 
    &\times {\cal L}_{n_{e},\lambda_{e}}^{\delta\sigma}(\bm{p}_{e},\bm{p}_\nu)  \Im \left[ \bar{\Pi}^R_{\delta\sigma}(Q) \right] ,
    \label{rate-01}
\end{align}
where $Q\equiv(E_{e,n_{e}} -\mu_{e}-E_\nu,\bm{p}_{e}-\bm{p}_\nu)$, and we now use $E_\nu = |\bm{p}_\nu|$ to denote the on-shell neutrino energy. The distribution function for antineutrinos satisfies a similar equation. Note that the integer $n_{e}$ denotes the electron LL index. Similarly, we use $n_{u}$ and $n_{d}$ in the following as indices for the $u$- and $d$-quark LLs, respectively. 

The LL-dependent lepton tensor is defined by
\begin{align}
   & {\cal L}^{\delta\sigma}_{n_{e},\lambda_{e}}(\bm{p}_{e},\bm{p}_\nu)
    =
    \Tr
    \Big[
        \gamma^\delta (1-\gamma^5)
        \Big\{
            \left[E_{e,n_{e}}\gamma^0 -\lambda_{e} p_{e,z}\gamma^3+\lambda_{e} m_{e}\right] 
            \Big[
                {\cal P}_{+} L_{n_{e}}\left(2 p_{e,\perp}^{2} \ell^2\right)\nonumber \\ 
    &\quad -{\cal P}_{-}L_{n_{e}-1}\left(2 p_{e,\perp}^2\ell^2\right)
            \Big] 
            +2\lambda_{e}        
            (\bm{p}_{e,\perp} \cdot\bm\gamma_{\perp}) L_{n_{e}-1}^1 \left(2 p_{e,\perp}^2\ell^2\right)
        \Big\}
        \gamma^\sigma(1-\gamma^5)(\gamma_{0} E_\nu-\bm\gamma\cdot \bm{p}_\nu)
    \Big],
    \label{Lmunu}
\end{align}
and the imaginary part of the retarded $W$-boson self-energy, or the quark tensor, can be expressed as
\begin{align}
    \operatorname{Im}
    \left[
        \bar{\Pi}_R^{\delta\sigma}(Q)
    \right] 
    &= 
    \frac{N_c}2 \int 
    \frac{d^4 K}{(2\pi)^4} 
    \left[n_F(k_{0}) - n_F(p_{0})
    \right] 
    \mbox{Tr}
    \Big[
        \gamma^\delta(1-\gamma_5) A_{u} (k_{0} + \mu_{u},\bm k)\nonumber \\ 
    &\times \gamma^\sigma(1-\gamma_5) A_{d} (p_{0} + \mu_{d},\bm p) 
    \Big],
\label{Im-Pi}
\end{align}
as shown in appendix~\ref{sec:Im-Pi}. Here, $K = (k_{0},\bm k)$ and $P = (p_{0},\bm p)$ denote the four-momenta associated with the $u$- and $d$-quarks, respectively, and the momentum $P$ is not independent, but is related to $K$ and $Q$ through $P = K + Q$. Note that the transverse components of $K$ and $P$ do not represent the physical momentum components of the charged quarks. Rather, they are Fourier variables associated with the translationally invariant parts of the corresponding propagators. 

The quark spectral functions have a structure analogous to Eq.~(\ref{electron-spectral-density}), with the electron charge and mass replaced by the appropriate quark charges and masses: 
\begin{align}
    A_{f} (k_{0,f} + \mu_{f},\bm k_{f}) 
    &= 2\pi e^{-k_{f,\perp}^{2} \ell_{f}^2} 
    \sum_{\lambda_{f}=\pm} 
    \sum_{n_{f}=0}^\infty
    \frac{(-1)^{n_{f}}}{E_{f,n_{f}}}
    \Big\{
        \left[
        E_{f,n_{f}}\gamma^0-\lambda_{f} k_{f,z}\gamma^3 + \lambda_{f} m_{f}
        \right]\nonumber \\ 
    &\times 
        \left[
            {\cal P}_{f}^+ L_{n_{f}}
            \left(2k_{f,\perp}^{2} \ell_{f}^2
            \right)
            -{\cal P}_{f}^- L_{n_{f}-1}
            \left(2k_{f,\perp}^{2} \ell_{f}^2
            \right)
        \right] +  2\lambda_{f}
        (\bm k_{f,\perp} \cdot \bm \gamma_{\perp}) 
        L_{n_{f}-1}^1
        \left(2 k_{f,\perp}^{2} \ell_{f}^2
        \right)
    \Big\}     \nonumber \\ 
    &\times 
    \delta \left(k_{f,0} + \mu_{f} - \lambda_{f} E_{f,n_{f}} \right) .
\label{quark-spectral-density}
\end{align}
Here 
$ E_{f,n_{f}}=\sqrt{2n_{f}|e_{f} B|+k_{f,z}^2+ m_{f}^{2} }$ are the quark LL energies, $\ell_{f} = 1/\sqrt{|e_{f}B|}$ is the flavor-dependent magnetic length, $\mathcal P_{f}^\pm = \frac{1}{2}[1 \pm \sign(e_{f}B) i \gamma^1\gamma^2]$ are the flavor-dependent spin projectors, and $(k_{f,0},\bm k_{f})$ denotes $(k_{0},\bm k)$ or $(p_{0},\bm p)$ according to whether $f = u$ or $d$, respectively. 

Rather than evaluating the lepton and quark tensors separately, it is convenient to calculate their Lorentz contraction directly, ${\cal L}_{n_{e},\lambda_{e}}^{\delta\sigma}(\bm{p}_{e},\bm{p}_\nu)  \Im \left[ \bar{\Pi}^R_{\delta\sigma}(Q) \right]$, which enters the emission rate in Eq.~(\ref{rate-01}). An explicit derivation is provided in appendix~\ref{sec:L-Im-Pi}. The resulting expression is
\begin{align}
    {\cal L}^{\delta\sigma}_{n_{e},\lambda_{e}}(\bm{p}_{e},\bm{p}_\nu)
    \mbox{Im}\left[ \bar{\Pi}_R^{\delta\sigma}(Q)\right] 
    &= 64\pi N_c 
    \sum_{\lambda_{u}=\pm}
    \sum_{\lambda_{d}=\pm}
    \sum_{n_{u}=0}^{\infty}  
    \sum_{n_{d}=0}^{\infty}  
    \int \frac{d^3 \bm{k}}{(2\pi)^3}    e^{-k_{\perp}^2\ell_{u}^2-p_{\perp}^2\ell_{d}^2}
    \frac{(-1)^{n_{u}+n_{d}}}
    {E_{u,n_{u}} E_{d,n_{d}}}\nonumber \\ 
    &\times 
    {\cal X}(\bm{p}_{e},\bm{k}) {\cal Y}(\bm{p}_\nu,\bm{p}) \left[
    n_F(\lambda_{u}E_{u,n_{u}}-\mu_{u}) - n_F(\lambda_{d}E_{d,n_{d}}-\mu_{d})
    \right]
    \nonumber \\ 
    &\times
 \delta\left(\lambda_{d}E_{d,n_{d}}-\lambda_{u}E_{u,n_{u}} - \lambda_{e}E_{e,n_{e}} + E_\nu \right) ,
\label{Tr-Tr}
\end{align}
where 
\begin{align}
{\cal X}(\bm{p}_{e},\bm{k})  &=    \Big[
        (E_{e,n_{e}}+\lambda_{e} p_{e,z})
        (E_{u,n_{u}}-\lambda_{u} k_{z})
        L_{n_{e}}(2p_{e,\perp}^2\ell^2)
        L_{n_{u}}(2k_{\perp}^2\ell_{u}^2) \nonumber \\ 
        &+ (E_{e,n_{e}}-\lambda_{e} p_{e,z})
        (E_{u,n_{u}}+\lambda_{u} k_{z})
        L_{n_{e}-1}(2p_{e,\perp}^2\ell^2)
        L_{n_{u}-1}(2k_{\perp}^2\ell_{u}^2)
        \nonumber \\ 
        & -  8\lambda_{e} \lambda_{u} (\bm{p}_{e,\perp}\cdot\bm k_{\perp})
        L_{n_{e}-1}^1(2p_{e,\perp}^2\ell^2)
        L_{n_{u}-1}^1(2k_{\perp}^2\ell_{u}^2)
    \Big] ,
    \label{X-expr}
\end{align}
and
\begin{align}
{\cal Y}(\bm{p}_\nu,\bm{p}) &=
    \Big[
        (E_\nu-p_{\nu,z})
        (E_{d,n_{d}}+\lambda_{d} p_{z})
        L_{n_{d}}(2p_{\perp}^2\ell_{d}^2)
        -(E_\nu+p_{\nu,z})
        (E_{d,n_{d}}-\lambda_{d} p_{z})
        L_{n_{d}-1}(2p_{\perp}^2\ell_{d}^2) \nonumber \\ 
         &+  4 \lambda_{d} (\bm{p}_{\nu,\perp}\cdot\bm{p}_{\perp})
        L_{n_{d}-1}^1(2p_{\perp}^2\ell_{d}^2)
    \Big].
    \label{Y-expr}
\end{align}
By comparing the Lorentz contraction of the lepton and quark tensors with its counterpart in the zero-field limit, we identify the combination ${\cal X}(\bm{p}_{e},\bm{k}) {\cal Y}(\bm{p}_\nu,\bm{p})$ as the kinematic part of the squared matrix element for the Urca weak process. It is structurally analogous to the zero-field expression $|{\cal M}|^{2} \propto (P_{u} \cdot P_{e}) (P_{d} \cdot P_\nu)$, where $P_i$, with $i=d,u,e,\nu$, denote the corresponding four-momenta \cite{Iwamoto:1982zz}. As in the zero-field case, the squared matrix element factorizes into two functions: one depending on the quantum numbers of the $u$-quark and electron, and the other on those of the $d$-quark and neutrino momenta. The magnetic field selects a preferred direction, which we have taken as the $z$-direction, and so the corresponding expressions in Eqs.~(\ref{X-expr})--(\ref{Y-expr}) exhibit an anisotropy between the longitudinal and transverse momenta. It should also be emphasized that the combination ${\cal X}(\bm{p}_{e},\bm{k}) {\cal Y}(\bm{p}_\nu,\bm{p})$ by itself should not be interpreted as the physical squared matrix element before the integrations over the transverse momenta are performed. Indeed, these transverse momentum components are auxiliary Fourier variables rather than physical quantum numbers characterizing the quark states.

The $\delta$ function in Eq.~(\ref{Tr-Tr}) enforces energy conservation. Its roots with respect to the longitudinal quark momentum can be determined analytically, allowing the $k_{z}$ integration to be carried out in closed form. The remaining integrations over the transverse components of the $u$-quark momentum can likewise be evaluated analytically. Furthermore, after substituting the resulting expression into the rate in Eq.~(\ref{rate-01}), the integrations over the transverse components of the electron momentum, $\bm{p}_{e,\perp}$, can also be performed analytically. Details of these calculations are presented in appendix~\ref{sec:L-Im-Pi}, where the final expression for the neutrino-number production rate is given in Eq.~(\ref{dfdt-app3}). An equivalent but slightly more explicit form of the result is 
\begin{align}
\frac{\partial f_\nu(t,\bm{p}_\nu)}{\partial t} &= \frac{N_cG_F^2\cos^2\theta_C}{72\pi^3 \ell^4  E_\nu }  
\sum_{n_{e},n_{u},n_{d}=0}^{\infty} \sum_{s=\pm}  
\int \frac{d p_{e,z}\, \Theta_{n_{u},n_{d}}^{(s)}\left(\bar{q}_{0},q_{z}\right) }
{E_{e,n_{e}} \sqrt{\left(q_{+}^2-\bar{q}_{0}^2+q_{z}^2\right)
    \left(q_{-}^2-\bar{q}_{0}^2+q_{z}^2\right)}} 
\nonumber \\ 
& \times \frac{e^{-\frac{1}{2}\frac{E_\nu}{T}} }
    {\cosh\left(\frac{E_{e,n_{e}}-\mu_{e}}{2T}\right)
    \cosh\left(\frac{E_{u,n_{u}}^{(s)}-\mu_{u}}{2T}\right)
    \cosh\left(\frac{E_{d,n_{d}}^{(s)}-\mu_{d}}{2T}\right)} \nonumber \\ 
&\times  \Bigg[
(E_{e,n_{e}}+p_{e,z})(E_{u,n_{u}}^{(s)}-k_{z}^{(s)})
(E_{d,n_{d}}^{(s)}+p_{z}^{(s)})(E_{\nu}-p_{\nu,z}) 
\mathcal{K}^{0,0}_{n_{e},n_{u},n_{d}} (\xi)
\nonumber \\ 
& +(E_{e,n_{e}}+p_{e,z})(E_{u,n_{u}}^{(s)}-k_{z}^{(s)})
(E_{d,n_{d}}^{(s)}-p_{z}^{(s)})(E_{\nu}+p_{\nu,z}) 
\mathcal{K}^{0,0}_{n_{e},n_{u},n_{d}-1} (\xi)
\nonumber \\ 
& + (E_{e,n_{e}}-p_{e,z})(E_{u,n_{u}}^{(s)}+k_{z}^{(s)})
(E_{d,n_{d}}^{(s)}+p_{z}^{(s)})(E_{\nu}-p_{\nu,z}) 
\mathcal{K}^{0,0}_{n_{e}-1,n_{u}-1,n_{d}} (\xi)
\nonumber \\ 
& + (E_{e,n_{e}}-p_{e,z})(E_{u,n_{u}}^{(s)}+k_{z}^{(s)})
(E_{d,n_{d}}^{(s)}-p_{z}^{(s)})(E_{\nu}+p_{\nu,z}) 
\mathcal{K}^{0,0}_{n_{e}-1,n_{u}-1,n_{d}-1} (\xi)
\nonumber \\ 
&
+ \frac{2}{3} p_{\nu,\perp}^{2} (E_{e,n_{e}}+p_{e,z})(E_{u,n_{u}}^{(s)}-k_{z}^{(s)})
\mathcal{K}^{0,1}_{n_{e},n_{u},n_{d}-1} (\xi)
\nonumber \\ 
&+ \frac{2}{3} p_{\nu,\perp}^{2} (E_{e,n_{e}}-p_{e,z})(E_{u,n_{u}}^{(s)}+k_{z}^{(s)})
\mathcal{K}^{0,1}_{n_{e}-1,n_{u}-1,n_{d}-1} (\xi)
\nonumber \\ 
& -\frac{8}{3\ell^2}(E_{d,n_{d}}^{(s)}+p_{z}^{(s)})(E_{\nu}-p_{\nu,z}) 
\mathcal{K}^{1,0}_{n_{e}-1,n_{u}-1,n_{d}} (\xi)
\nonumber \\ 
&+\frac{8}{3\ell^2}(E_{d,n_{d}}^{(s)}-p_{z}^{(s)})(E_{\nu}+p_{\nu,z}) 
\mathcal{K}^{1,0}_{n_{e}-1,n_{u}-1,n_{d}-1} (\xi)
-\frac{16 p_{\nu,\perp}^2}{9\ell^2} \mathcal{K}^{1,1}_{n_{e}-1,n_{u}-1,n_{d}-1} (\xi)
\Bigg]  ,
\label{der-rate-main}
\end{align}
where $\xi=\frac{1}{2}p_{\nu,\perp}^2\ell^2$, $\bar{q}_{0}=E_{e,n_{e}}-E_\nu$, $q_{z}=p_{e,z}-p_{\nu,z}$, and $q_\pm = \sqrt{2n_{d}|e_{d}B|+m_{d}^2}  \pm \sqrt{2n_{u}|e_{u}B|+m_{u}^2}$. The explicit expressions for the $u$- and $d$-quark energies satisfying the energy conservation relation, $E_{u,n_{u}}^{(s)}$ and $E_{d,n_{d}}^{(s)}$ with $s=\pm1$, are given in Eqs.~(\ref{Eu-pm}) and (\ref{Ed-pm}), respectively. The corresponding expressions for the longitudinal momenta, $k_{z}^{(s)}$ and $p_{z}^{(s)}$, are given in Eqs.~(\ref{kz-pm}) and (\ref{pz-pm}), respectively. Note that the antiparticle contributions ($\lambda_{e}=-1$, $\lambda_{d}=-1$, and $\lambda_{u}=-1$) have been neglected, as they are strongly suppressed by the corresponding Fermi-Dirac distribution functions. This approximation is well justified in dense quark matter when the temperature is much smaller than the relevant chemical potentials.

The integrand in the above expression for the rate contains the following product of step functions, 
\begin{equation}
\Theta_{n_{u},n_{d}}^{(s)}(\bar{q}_{0},q_{z}) = \theta\left[
        \left(q_{+}^2-\bar{q}_{0}^2+q_{z}^2\right)
        \left(q_{-}^2-\bar{q}_{0}^2+q_{z}^2\right)\right]
        \theta\left[E_{u,n_{u}}^{(s)}\right]
        \theta\left[E_{d,n_{d}}^{(s)}\right] ,
\label{Theta-function}
\end{equation}
which restricts the allowed solutions to positive quark energies and ensures that the combination of the zeroth and longitudinal components of the virtual $W$-boson four-momentum, $\bar{q}_{0}^2-q_{z}^2$, lies within the kinematically allowed phase space determined by the electron and neutrino on-shell conditions.

The explicit expressions for the functions $\mathcal{K}^{\alpha,\beta}_{n_{e},n_{u},n_{d}}(\xi) $, with $\alpha,\beta=0,1$, are given by
{\small
\begin{equation}
\mathcal{K}^{\alpha,\beta}_{n_{e},n_{u},n_{d}} (\xi)
=  e^{-\xi }
\sum_{i=0}^{n_{u}}
\sum_{j=0}^{n_{d}} 
\sum_{k=0}^{n_{e}}
(-1)^{i+j+k}
\frac{2^{i}}{3^{i+j}}  
\binom{n_{u}+\alpha}{i+\alpha} 
\binom{n_{d}+\beta}{j+\beta}
\binom{n_{e}+\alpha}{k+\alpha} 
\frac{(i+j+k+\alpha)!}{i! j! k!}
L_{i+j+k+\alpha}^{\beta}\left(\xi\right).
\label{K-functions}
\end{equation}}
The neutrino-number production rate in Eq.~(\ref{der-rate-main}) is one of the main results of this work. It provides an expression in which the LL quantization of both electrons and quarks is treated exactly within the approximations adopted in this study.

Formally, the expression for the rate in Eq.~(\ref{der-rate-main}) is valid over a broad range of magnetic field strengths, potentially extending to arbitrarily weak fields. In its present form, however, it does not include Fermi-liquid corrections to the quark dispersion relations. This is not a critical limitation, as those corrections can be readily included in principle; see section~\ref{sec:Fermi-liquid}. More importantly, the practical use of Eq.~(\ref{der-rate-main}) in the weak-field regime becomes computationally prohibitive, since an increasingly large number of quark LLs must be included. In this regime, it is therefore more efficient to employ an approximation in which the LL quantization of quarks is neglected \cite{Ghosh:2025sjn,Shovkovy:2026aci}.

In the present work, we focus instead on the regime of sufficiently strong magnetic fields, $B \gtrsim 10^{18}$ G, where electrons occupy only the LLL ($n_{e}=0$) and the LL quantization of quarks can no longer be neglected. The expression for $\mathcal{K}_{n_{e},n_{u},n_{d}}^{\alpha,\beta}(\xi)$ in Eq.~(\ref{K-functions}) is understood to vanish when any subscript is negative, and it follows immediately that several terms in Eq.~(\ref{der-rate-main}) vanish when $n_{e} = 0$. In this regime, the on-mass-shell condition for electrons reduces to $E_{e,0}=|p_{e,z}|$, where, for simplicity, we neglect the electron mass, as it is small compared to the corresponding chemical potential. Substituting $E_{e,0}=|p_{e,z}|$ into the rate and examining the nonvanishing terms for $n_{e}=0$, we find that only electrons with $p_{e,z}>0$ contribute. This is consistent with the fact that only left-handed electrons participate in the weak Urca processes. Indeed, since the electron spin in the LLL is oriented antiparallel to the magnetic field, an ultra-relativistic left-handed electron must have its momentum directed along the magnetic field.

By noting that typical energies of emitted neutrinos are of the order of temperature, which is much smaller than the electron and quark chemical potentials, the exact expression for the rate can be further simplified. In particular, considering that $T^2\ll |eB|$, we can approximate $L_{i+j+k}(\xi) \approx L_{i+j+k}(0) =1$ and $L^1_{i+j+k}(\xi) \approx L^1_{i+j+k}(0) =i+j+k+1$. Additionally, noting that $T^{2} \ll \mu_{e} T \ll |eB|$ and $m_{u/d}^2\ll |eB|$, and keeping only the leading-order terms in the integrand outside the Fermi distribution functions, we derive 
\begin{align}
    \frac{\partial f_\nu(t,\bm{p}_\nu)}{\partial t}
    &\approx
    \frac{N_cG_F^2\cos^2\theta_C T}{36\pi^3\ell^4 E_\nu}
    e^{-\frac{E_\nu}{2T}} 
    \hspace{-3pt}
    \sum_{n_{u}=0}^\infty
    \sum_{n_{d}=2n_{u}+1}^\infty
    \frac{2^{n_{d}}}{3^{n_{u}+n_{d}}}
    \binom{n_{u}+n_{d}-1}{n_{u}} \mathcal{F}_{n_{u},n_{d}}(p_{\nu,\perp},p_{\nu,z}) \nonumber \\ 
    & \times
    \left(
        \frac{n_{u}+n_{d}}{n_{d}-2n_{u}}
        (E_\nu-p_{\nu,z})
        +
        \frac{n_{d}-2n_{u}}{4\mu_{e}^2\ell^2}
        (E_\nu+p_{\nu,z})
    \right) , 
    \label{dfdt-approx}
\end{align}
where we have introduced the dimensionless function:
\begin{equation}
\mathcal{F}_{n_{u},n_{d}}(p_{\nu,\perp},p_{\nu,z})
= \int_{0}^{\infty} \frac{d p_{e,z} }{ T \cosh\left(\frac{p_{e,z}-\mu_{e}}{2T}\right) \cosh\left(\frac{E_{u,n_{u}}^{*}-\mu_{u}}{2T}\right) \cosh\left(\frac{E_{d,n_{d}}^{*}-\mu_{d}}{2T}\right) } ,
\label{function-F}
\end{equation}
together with the notation
\begin{align}
    E_{u,n_{u}}^*
    &= -\frac{\bar{q}_{0}}2
    \left(
        1-\frac{q_{+}q_{-}}{\bar{q}_{0}^2-q_{z}^2}
    \right)
    -\frac{1}{2}\frac{q_{z}}{\bar{q}_{0}^2-q_{z}^2}
        \sqrt{
        (q_{+}^{2} - \bar{q}_{0}^{2} + q_{z}^2)
        (q_{-}^{2} - \bar{q}_{0}^{2} + q_{z}^2)
        } ,
    \label{Eu-star}
    \\
    E_{d,n_{d}}^*
    &= \frac{\bar{q}_{0}}2
    \left(
        1+\frac{q_{+}q_{-}}{\bar{q}_{0}^2-q_{z}^2}
    \right)
    -\frac{1}{2}\frac{q_{z}}{\bar{q}_{0}^2-q_{z}^2}
        \sqrt{
        (q_{+}^{2} - \bar{q}_{0}^{2} + q_{z}^2)
        (q_{-}^{2} - \bar{q}_{0}^{2} + q_{z}^2)
        }.
    \label{Ed-star}
\end{align}
For details of the derivation, see appendix~\ref{sec:L-Im-Pi}. Note that the sum over quark LLs in Eq.~(\ref{dfdt-approx}) explicitly excludes the cases where $n_{d} < 2n_{u}$ and $n_{d} = 2n_{u}$. The transitions with $n_{d} < 2n_{u}$ are essentially kinematically forbidden; see section~\ref{subsec:discrete-urca-B} and appendix~\ref{subsec:calc-dfdt}, in particular Eqs.~(\ref{kz-star-app1})--(\ref{Ed-star-app1}) and the surrounding discussion. The transitions with $n_{d} = 2n_{u}$, on the other hand, are kinematically allowed over a very narrow region of phase space. In appendix~\ref{sec:Anom-transitions}, we show that their contribution to the total neutrino emission rate is relatively small (at most a few percent within the range of parameter choices under consideration), so we omit this contribution from the main results presented in this paper.

\subsection{Neutrino energy and momentum emission rates}
\label{sec:energy-momentum-rates}

Having obtained the neutrino-number production rate in the preceding subsection, we can now derive the corresponding neutrino energy and net longitudinal-momentum emission rates as follows:
\begin{align}
\dot{\cal E}_\nu 
    &= 2 \int \frac{d^3\bm{p}_\nu}{(2\pi)^3} E_\nu 
    \frac{\partial f_\nu(t,\bm{p}_\nu)}{\partial t},
    \\
    \dot{\cal P}_{\nu,z} 
    &= 2 \int \frac{d^3\bm{p}_\nu}{(2\pi)^3} p_{\nu,z} \frac{\partial f_\nu(t,\bm{p}_\nu)}{\partial t}.
\end{align}
Note that we have included a factor of $2$ to account for the combined rate from both Urca processes, $u+ e^- \rightarrow d+\nu_{e}$ and $d\rightarrow u +e^- + \bar\nu_{e}$, one of which produces neutrinos and the other antineutrinos. 

From Eq.~(\ref{dfdt-approx}), which is valid in the low-temperature regime, $T^2\ll \mu_{e} T\ll |eB|$, and for sufficiently strong magnetic fields such that electrons occupy only the LLL, we derive the following expression for the neutrino energy emission rate:
\begin{align}
\dot{\cal E}_\nu 
    & \approx \frac{N_cG_F^2\cos^2\theta_C T}{72\pi^5\ell^4}
    \int_{-\infty}^{\infty} p_{\nu,z}
    \int_{0}^{\infty} p_{\nu,\perp} dp_{\nu,\perp} 
     e^{-\frac{E_\nu}{2T}} 
    \hspace{-3pt}
    \sum_{n_{u}=0}^\infty
    \sum_{n_{d}=2n_{u}+1}^\infty
    \frac{2^{n_{d}}}{3^{n_{u}+n_{d}}}
    \binom{n_{u}+n_{d}-1}{n_{u}}
    \nonumber \\ 
    &\times
    \mathcal{F}_{n_{u},n_{d}}(p_{\nu,\perp},p_{\nu,z})
    \left(
        \frac{n_{u}+n_{d}}{n_{d}-2n_{u}}
        (E_\nu-p_{\nu,z})
        +
        \frac{n_{d}-2n_{u}}{4\mu_{e}^2\ell^2}
        (E_\nu+p_{\nu,z})
    \right) .
    \label{rate-approx}
\end{align}
The corresponding expression for the net longitudinal momentum emission rate, $\dot{\mathcal{P}}_{\nu,z}$, is similar but contains an additional factor of $p_{\nu,z}/E_\nu$ in the integrand. Since the characteristic energies of the emitted neutrinos are of order $T$, the above rate scales with temperature as $T^5$.

\subsection{Discrete Urca-resonant magnetic field values}
\label{subsec:discrete-urca-B}

In the calculation of the Urca rate defined by Eq.~(\ref{rate-approx}), one of the strongest kinematic constraints arises from the particle distribution functions. Their effect is encoded in the function $\mathcal{F}_{n_{u},n_{d}}(p_{\nu,\perp},p_{\nu,z})$, which appears in the integrand and is defined in Eq.~(\ref{dfdt-approx}). At low temperatures, this function exponentially suppresses neutrino emission unless all particles participating in the Urca process lie close to their respective Fermi surfaces. In other words, while the typical neutrino energy must be of order $T$, the electron and quarks must have energies close to their respective Fermi energies.

However, since the $u$- and $d$-quark energies given by Eqs.~(\ref{Eu-star}) and (\ref{Ed-star}) also depend on $p_{e,z}$, the corresponding near-Fermi-surface constraints cannot, in general, be satisfied for arbitrary model parameters. Using energy conservation together with $\beta$-equilibrium, we have $E_{d}^* - \mu_{d} =  (E_{u}^* - \mu_{u})+(p_{e,z} - \mu_{e}) - E_\nu$, which shows that the Fermi-surface constraint for the $d$-quark is satisfied whenever the corresponding constraints are satisfied for the other three particles. In the low-temperature limit, $T\ll \mu_{e}$, the near-Fermi-surface approximation corresponds to setting $E_\nu = 0$, $p_{e,z} = \mu_{e}$, and $E_{u}^* = \mu_{u}$ in Eq.~(\ref{Eu-star}), which yields the following relation:
\begin{align}
    \mu_{u} 
    &= \frac{q_{+}q_{-}}{4\mu_{e}}
    + \frac{(q_{+} - q_{-})^2}{4q_{+}q_{-}}\mu_{e}
    \nonumber \\ 
    &\approx  \frac{n_{d}-2n_{u}}{6\mu_{e}}|eB|
    +\frac{2n_{u}}
    {n_{d}-2n_{u}} \mu_{e} .
    \label{eq-relation-Eu-muu}
\end{align}
Here the final approximate expression is obtained by taking the massless limit, $m_{u},m_{d} \to 0$. This relation can be satisfied only if $n_{d}>2n_{u}$, as expected from the requirement that the $d$-quark LL lie above the corresponding $u$-quark LL. More importantly, for fixed chemical potentials and a given LL transition specified by $(n_{d},n_{u})$, Eq.~(\ref{eq-relation-Eu-muu}) can be satisfied only at specific values of the magnetic field. Thus, an appreciable Urca emission rate is possible only when the field strength is finely tuned to the following value:
{\small
\begin{align}
    eB_{n_{d},n_{u}}
    &= 3\mu_{e}
        \frac{(n_{d}\mu_{u}-2n_{u}\mu_{d})
        +\sqrt{
            (n_{d}\mu_{u}-2n_{u}\mu_{d})^2
            -(n_{d}-2n_{u})(n_{d}m_{u}^2-2n_{u}m_{d}^2)
        }}{(n_{d}-2n_{u})^2}
        -\frac32\frac{m_{d}^2-m_{u}^2}{n_{d}-2n_{u}} \nonumber  \\
    &\approx
    6\mu_{e}
        \frac{n_{d}\mu_{u}-2n_{u}\mu_{d}
        }{(n_{d}-2n_{u})^2} \theta\left(n_{d}\mu_{u}-2n_{u}\mu_{d}\right),
    \qquad (n_{e} = 0),
\label{resonant-B}
\end{align}}
where the final approximate form is obtained in the limit $m_{u},m_{d} \to 0$.
We refer to such discrete magnetic field values for which the Landau-quantized electron and quark states satisfy the kinematic matching conditions of the Urca process as {\em Urca-resonant} fields. The term ``resonant" is used here in a kinematic sense and does not imply the presence of any intermediate propagating states, as in conventional particle-physics resonances. Instead, these resonances are analogous to optical resonances, in which the energy carried away by the emitted particles matches the separation between the quantized initial and final states.

\begin{figure}[b]
\centering
\includegraphics[width=0.7\textwidth]{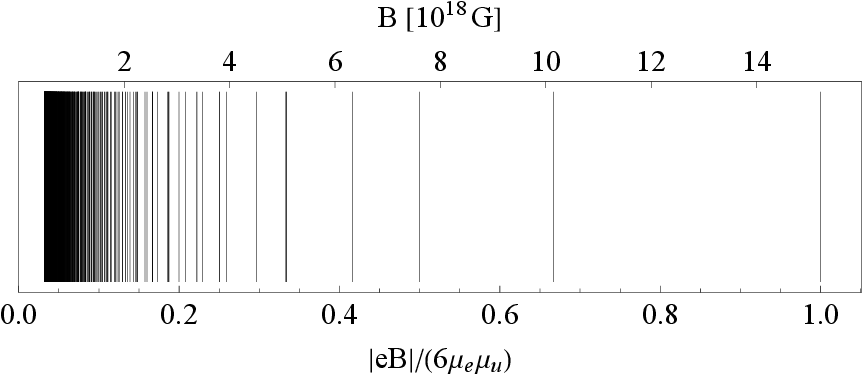}
\caption{Discrete Urca-resonant magnetic field values, defined in Eq.~(\ref{resonant-B}), for which the kinematics is fine-tuned so that both quark flavors and electrons remain near their respective Fermi surfaces.}
\label{fig:special-B}
\end{figure}

The largest such magnetic field is $eB_{1,0} \approx 6\mu_{e}\mu_{u}$, which corresponds to $1.52 \times 10^{19}$ G for our choice of representative values of $\mu_{e}$ and $\mu_{u}$. We find that there are 124 transitions such that $B_{n_{d},n_{u}} > 10^{18}$ G, and 2490 transitions in the range $2.11 \times 10^{17}\mbox{ G} < B_{n_{d},n_{u}} < 10^{18}$ G, where the lower magnetic field value corresponds to the point where electron higher LLs start to be populated, i.e., $|eB| = \mu_{e}^2/2$. The twelve largest Urca-resonant magnetic fields are listed in table~\ref{tab:special-B}, while the 500 largest values are shown in figure~\ref{fig:special-B}. As $B$ decreases, these resonant fields become increasingly dense. Consequently, at a fixed temperature $T$, the corresponding emission peaks are expected to become unresolved below a certain threshold magnetic field strength.

\begin{table}
\centering 
\setlength{\tabcolsep}{0.5pt} 
{\small
\begin{tabular}{c *{12}{S[table-format=2.3]}} 
\toprule & 
\multicolumn{12}{c}{Discrete Urca-resonant magnetic field values} \\ 
\cmidrule(lr){2-13} 
{$n_{d}$} & 1 & 3 & 2 & 4 & 5 & 6 & 3 & 5 & 7 & 8 & 4 & 6 \\ 
{$n_{u}$} & 0 & 1 & 0 & 1 & 2 & 2 & 0 & 1 & 2 & 3 & 0 & 1 \\ 
{$eB_{n_{d},n_{u}}/(100\,\mathrm{MeV})^2$} & 8.997 &  5.998 &  4.499 &  3.749 &  3.002 &  2.999 &  2.999 &  2.666 &  2.333 &  2.250 &  2.249 &  2.062 \\
{$B_{n_{d},n_{u}}/(10^{18}\,\mathrm{G})$} & 15.209 & 10.140 & 7.605 & 6.337 & 5.074 & 5.070 & 5.070 & 4.506 & 3.943 & 3.803 & 3.802 & 3.485 \\ 
\bottomrule 
\end{tabular}}
\caption{The twelve largest Urca-resonant magnetic field values $B_{n_{d},n_{u}}$, defined in Eq.~(\ref{resonant-B}). We use the representative model parameters $(\mu_{e},\mu_{d},\mu_{u})=(50,350,300)\,\mathrm{MeV}$ and $(m_{e},m_{d},m_{u})=(0,5,3)\,\mathrm{MeV}$. The values $B_{6,2}$ and $B_{3,0}$ are close but not identical: $B_{6,2}=5.06994\times10^{18}\,\mathrm{G}$ and $B_{3,0}=5.06976\times10^{18}\,\mathrm{G}$.} 
\label{tab:special-B} 
\end{table}

As we argue in this study, one of the most striking consequences of quark LL quantization is the pronounced magnetic field dependence of the Urca neutrino emission rate, characterized by sharply enhanced peaks near the Urca-resonant field values. As we will show, finite temperatures smear the exact matching condition, broadening these resonances into peaks with finite characteristic widths. In principle, nonzero quasiparticle widths should also contribute to this smearing, although such effects are not included in the present study. In addition, the constraints imposed by exact LL quantization become progressively more relaxed as the magnetic field decreases, since transitions to neighboring LLs, e.g., those obtained by replacing $n_{d} \to n_{d} \pm \delta n$ with $\delta n = 1,2,\ldots$, can also occur without conflicting too strongly with the requirement that the particle energies remain close to their respective Fermi energies.

\begin{figure}
\centering
  \subfigure[]{\includegraphics[width=0.49\textwidth]{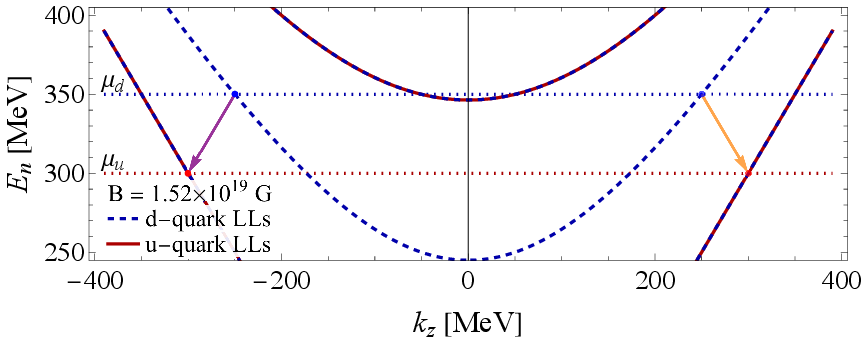}}
\hspace{0.001\textwidth}
  \subfigure[]{\includegraphics[width=0.49\textwidth]{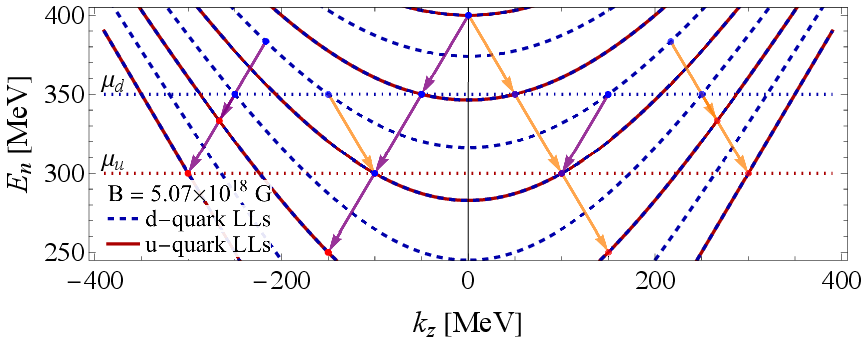}}\\
  \subfigure[]{\includegraphics[width=0.49\textwidth]{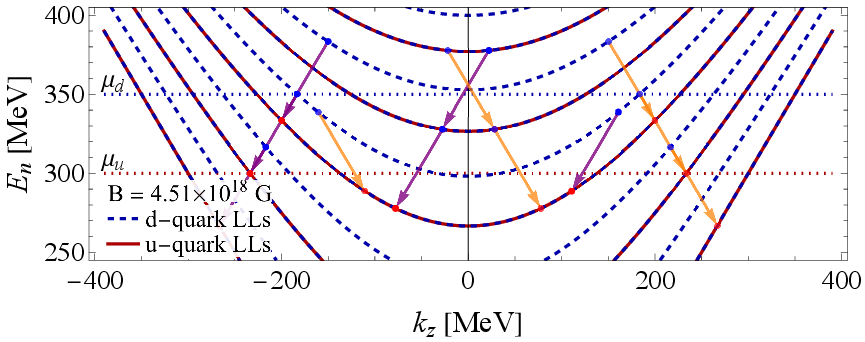}}\hspace{0.001\textwidth}
  \subfigure[]{\includegraphics[width=0.49\textwidth]{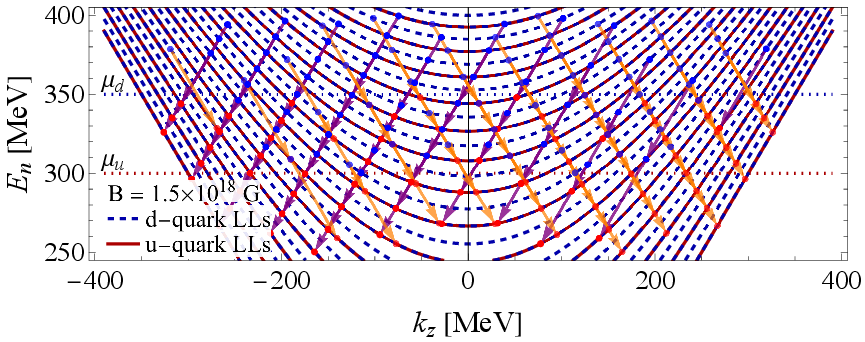}} 
\caption{Illustration of LL transitions between $d$- and $u$-quark states for several representative magnetic field strengths:
(a) $B=B_{1,0}\approx 1.52\times 10^{19}~\mbox{G}$, (b) $B=B_{3,0}\approx B_{5,2}\approx B_{6,2}\approx 5.07\times 10^{18}~\mbox{G}$, 
(c) $B=B_{5,1}\approx 4.51\times 10^{18}~\mbox{G}$, and (d) $B=B_{15,3}\approx B_{18,5}\approx 1.5\times 10^{18}~\mbox{G}$. 
Purple and orange arrows denote allowed and forbidden weak processes, respectively, involving left-handed and right-handed electrons in the LLL.
Each panel displays all transitions satisfying energy conservation under the approximation $\bm p_\nu = 0$, and such that the initial and final states both lie within $50~\mbox{MeV}$ of the corresponding quark Fermi surfaces.}
\label{fig.LLs-transitions}
\end{figure}

To illustrate the constrained kinematics of the Urca processes involving LL-quantized quark states, we show the $d$- and $u$-quark spectra, together with the LL transitions allowed by energy conservation, in figure~\ref{fig.LLs-transitions}. For the purposes of this illustration, the neutrino energy is assumed to be negligible compared with those of the electron and quarks. Since the electrons occupy the LLL and their mass is negligible, for all transitions shown, the energy and longitudinal momentum transfers have the same magnitude, both equal to the electron chemical potential, $\mu_{e}$. This is consistent with the approximation used in deriving the Urca-resonant fields in Eq.~(\ref{resonant-B}). In addition to the above constraints, each panel displays only those transitions whose initial and final states both lie within $50~\mbox{MeV}$ of the corresponding quark Fermi surfaces.

It should be noted that energy conservation alone is not sufficient to distinguish between transitions that are allowed or forbidden by weak interactions involving only left-handed particles. As mentioned earlier, for approximately massless electrons in the LLL, the spin must be anti-aligned with the magnetic field, which requires the longitudinal momentum to be positive. Consequently, only half of the transitions shown in figure~\ref{fig.LLs-transitions} are allowed, while the other half are forbidden. The allowed and forbidden transitions are indicated by purple and orange arrows, respectively. 

The four panels in figure~\ref{fig.LLs-transitions} correspond to different fixed values of the magnetic field. For example, panel (a) shows the case of the highest Urca-resonant field, $B_{1,0}\approx 1.52\times 10^{19}~\mbox{G}$, for which only a single quantum transition is kinematically fine-tuned so that both quark energies and the electron energy lie at their respective Fermi surfaces. As the magnetic field is decreased or increased away from $B_{1,0}$, the separation between the LLs changes. Energy conservation then requires a corresponding change in the electron or quark energies to compensate for this mismatch. However, this drives the corresponding particle away from its Fermi energy and consequently suppresses the emission rate.

Essentially the same behavior occurs for other magnetic field values. In some cases, different Urca-resonant fields can accidentally become nearly degenerate. This occurs, for example, for $B_{5,2}$, $B_{6,2}$, and $B_{3,0}$. The corresponding quark spectra for $B=B_{3,0}\approx B_{5,2}\approx B_{6,2}\approx 5.07\times 10^{18}~\mbox{G}$ are shown in panel (b) of figure~\ref{fig.LLs-transitions}. In this case, there are three distinct LL transitions for which the quark energies are simultaneously fine-tuned to lie close to their respective Fermi surfaces. The main qualitative difference at lower magnetic fields is the rapid increase in the number of distinct transitions involving quark states with energies near their respective Fermi surfaces, e.g., see panel (d) in figure~\ref{fig.LLs-transitions}.

In general, transitions involving states closest to the Fermi surfaces are expected to provide the dominant contributions. At small but nonzero temperatures, however, LL transitions involving quark states within finite energy intervals around the Fermi surfaces can also make nonnegligible contributions to the rate. In addition, when the thermal energy becomes comparable to or larger than the quark LL spacing, $|e_{f}B|/\mu_{f}$, transitions involving neighboring LLs begin to contribute as well. For example, at $T=1~\mbox{MeV}$, the $d$-quark LL quantization becomes effectively unresolved when $|eB|\lesssim 6\pi T\mu_{d}$, corresponding to a magnetic field of approximately $1.1\times 10^{18}~\mbox{G}$. Thus, while only a few transitions typically contribute near the strongest Urca-resonant fields, the number of relevant LL transitions proliferates rapidly as the magnetic field decreases toward $10^{18}~\mbox{G}$, particularly at finite temperature.

At sufficiently low magnetic field strengths, electrons begin to populate higher LLs with $n_{e}\geq 1$. Although the kinematic details are modified, the essential mechanism remains unchanged: the quark LL quantization continues to select discrete magnetic field values for which the quark energies are finely tuned to their respective Fermi surfaces. The corresponding approximate expression for the Urca-resonant magnetic fields is
\begin{equation}
eB_{n_{d},n_{u},n_{e}} \approx \frac{6\left(n_{d} \mu_{e} \mu_{u} - 2 n_{u} \mu_{e} \mu_{d} - 3 n_{e} \mu_{u} \mu_{d}\right)}{(n_{d}-2n_{u}+3n_{e})^2-12 n_{e} n_{d}}\left(1+O\left(\frac{m_{f}^2}{\mu_{f}^2}\right)\right),
\label{resonant-B-nd-nu-ne}
\end{equation}
when the right-hand side is positive, and zero otherwise. For simplicity, in deriving this approximate expression, we have neglected the quark masses. Noting that the higher electron LLs ($n_{e}\geq 1$) begin to be populated only at sufficiently low magnetic fields, i.e., $B< \mu_{e}^2/(2|e|)\approx 2.11\times 10^{17}~\mbox{G}$, we can argue that the Urca-resonant magnetic fields in Eq.~(\ref{resonant-B-nd-nu-ne}) can become relevant only at sufficiently low temperatures, $T\lesssim 
|eB|/(6\pi\mu_{d}) < \mu_{e}^2/(12\pi\mu_{d})$, where $\mu_{e}^2/(12\pi\mu_{d}) \approx 0.14~\mbox{MeV}$. At higher temperatures, the $d$-quark LL quantization is no longer resolved and therefore does not play a significant role, causing these Urca-resonant magnetic fields to lose their relevance.

\subsection{Fermi-liquid corrections}
\label{sec:Fermi-liquid}

In the zero-field case, the quark momenta $\bm{k}$ and $\bm{p}$ and the electron momentum $\bm{p}_{e}$ are approximately collinear if strong interaction of quark matter is neglected. The resulting kinematic constraint parametrically suppresses the Urca rate \cite{Burrows:1980ec}. This constraint is relaxed once Fermi-liquid corrections to the quark dispersion relations are included \cite{Iwamoto:1980eb,Iwamoto:1982zz}. Such corrections reduce the quark Fermi momenta and velocities, leading to the modified energy relation: $E_{p,f}=\mu_{f}+v_F(p-p_F)$, where $f=u,d$ labels the quark flavors, $p_F=v_F\mu_{f}$ is the Fermi momentum, $v_F= 1-\kappa$ is the Fermi velocity, and $\kappa = 2\alpha_s/(3\pi)$  \cite{Baym:1975va,Schafer:2004jp}. 

In deriving the Urca emission rate for strongly magnetized quark matter in this study, we neglected Fermi-liquid corrections. To our knowledge, such effects have not previously been investigated for quark matter in the regime of strong (quantizing) magnetic fields. Nevertheless, an analogy with the zero-field case allows one to make an educated conjecture regarding the qualitative form of Fermi-liquid corrections for LL-quantized quarks. In particular, the dispersion relation of interacting quarks near the Fermi surface may be expected to take the generic form:
\begin{equation}
E_{n,p_{z}}^{\pm} \simeq \mu_{f} + v_{z,F}^{(n,\pm)} \left(p_{z} -p_{z,F}^{(n,\pm)} \right) ,
\end{equation}
where $v_{z,F}^{(n,\pm)}$ and $p_{z,F}^{(n,\pm)}$ are the LL-dependent longitudinal components of the Fermi velocity and Fermi momentum, respectively. 
For higher LLs, the two signs denote the two possible spin states, which need not remain degenerate in interacting quark matter.

In general, $v_{z,F}^{(n,\pm)}$ should be smaller than the speed of light and may be expected to decrease with the LL index $n$. Indeed, the minimum energy of a higher LL increases with $n$ and progressively approaches the Fermi-level energy, where $v_{z,F}^{(n,\pm)}\approx 0$. As for $p_{z,F}^{(n,\pm)}$, similarly to the zero-field case, it is expected to be smaller than its value in noninteracting matter. These two generic features appear sufficient to infer the qualitative influence of Fermi-liquid corrections on the Urca rate.

The reduction of the longitudinal quark momenta due to Fermi-liquid corrections tends to increase the mismatch in the momentum conservation condition, since the electron momenta remain largely unaffected. Indeed, analogous corrections for electrons are expected to be negligible because electrons, unlike quarks, do not participate in strong interactions. Compensating for the increased momentum mismatch requires a larger effective separation between the quark LLs, which is naturally achieved by increasing the magnetic field strength. These arguments suggest that the principal effect of quark Fermi-liquid corrections is to shift the Urca-resonant magnetic field values toward slightly larger fields. Therefore, the resonance-like enhancement of Urca emission should also persist in interacting quark matter, although the magnetic field values at which the resonances occur will be shifted.

An additional qualitatively new effect may arise from the interaction-induced splitting of the two spin states within each higher quark LL. Although this splitting is not expected to be large, it could, in principle, split the Urca-resonant magnetic field values into pairs of closely spaced resonances. While conceptually interesting, both effects are unlikely to have a major impact on the qualitative magnetic field dependence of the Urca neutrino emission rate, including its tendency to increase at moderately strong fields and its quenching in the ultra-strong-field limit. Nor are they expected to lead to dramatically different phenomenological implications for hybrid and quark stars.

\section{Numerical results}
\label{sec:Numerics}

Using the analytical results presented in the preceding section, we now study the Urca emission rates numerically. We focus on quark matter under conditions relevant to the cores of compact stars, where the density may reach several times the nuclear saturation density. In this regime, the quark chemical potentials are typically of order $300~\mbox{MeV}$ or larger, whereas the electron chemical potential is of order $50~\mbox{MeV}$. For the representative model analysis presented below, we take $\mu_{u}=300~\mbox{MeV}$ and $\mu_{e}=50~\mbox{MeV}$. Imposing the condition of $\beta$-equilibrium then gives $\mu_{d} = \mu_{u} + \mu_{e} = 350~\mbox{MeV}$. The corresponding model baryon chemical potential is
$\mu_B = \mu_{u} + 2\mu_{d} = 1000~\mbox{MeV}$.

We concentrate primarily on the low-temperature regime, $T\lesssim 1~\mbox{MeV}$, in which neutrino trapping can be neglected. To illustrate the temperature dependence of the emission rate, we present numerical results for several fixed temperatures between $T=0.1~\mbox{MeV}$ and $T=1~\mbox{MeV}$. This range is sufficiently broad to capture the qualitative features of Urca emission relevant to compact star cooling. The numerical analysis can be straightforwardly extended to either lower or higher temperatures, if necessary.

In contrast to the earlier studies of refs.~\cite{Ghosh:2025sjn,Shovkovy:2026aci}, which were restricted to moderately strong magnetic fields for which the LL quantization of quarks could be neglected, here we extend the analysis to fully incorporate quark LL quantization effects. In particular, our numerical analysis focuses on magnetic fields above approximately $8\times 10^{17}~\mbox{G}$, corresponding to $\sqrt{|eB|}\gtrsim 69~\mbox{MeV}$. In this regime, LL quantization becomes essential and substantially alters the kinematics of the Urca processes. 

Although the analytical expressions derived in this study are valid for arbitrary magnetic field strengths, the computational cost grows rapidly as the field decreases, owing to the increasing number of LL transitions that must be included. Calculations remain readily tractable for fields of order $10^{18}~\mbox{G}$, whereas extending the analysis to weaker fields requires substantially greater computational resources. As shown below, however, the effects of quark LL quantization become progressively less pronounced in this regime, and the results, after averaging over the oscillatory structure, approach those obtained in refs.~\cite{Ghosh:2025sjn,Shovkovy:2026aci}, as expected.

Because quarks are strongly interacting at densities relevant to compact stars, interaction effects should generally be taken into account, at least at the level of Fermi-liquid corrections \cite{Baym:1975va}. In the absence of a magnetic field, such corrections play a crucial role by relaxing the otherwise highly restrictive collinear kinematics of the Urca processes \cite{Iwamoto:1980eb,Iwamoto:1982zz}. They also affect Urca emission in magnetized quark matter \cite{Ghosh:2025sjn,Shovkovy:2026aci}, although their importance is reduced because the electron transverse momentum is not conserved in the presence of a magnetic field. As discussed in section~\ref{sec:Fermi-liquid}, Fermi-liquid corrections are not expected to play a major role in the regime of very strong magnetic fields, where quark LL quantization controls both the available phase space and the relevant kinematic matching conditions. This expectation is consistent with the general findings of ref.~\cite{Ghosh:2025sjn}, which showed that, even without quark LL quantization, the relative importance of Fermi-liquid corrections decreases as the magnetic field strength increases.

\subsection{Numerical algorithm}
\label{sec:numerical-algorithm}

In principle, the emission rate includes contributions from an infinite number of LL transitions, as is apparent, e.g., from the double sum appearing in Eq.~(\ref{rate-approx}). As far as practical numerical calculations are concerned, it is clear from general considerations that only a finite number of possible transitions contribute substantially to the rate. These are the transitions where the participating $u$- and $d$-quarks have energies not too far from their respective Fermi surfaces. Mathematically, this restriction is enforced by the distribution functions appearing in the expression for the rate. From a physical viewpoint, the quark states far below their Fermi surfaces are Pauli blocked from participating in weak processes, while the states far above Fermi surfaces are not sufficiently populated. 

In practice, when performing the calculation numerically for a fixed value of the magnetic field, we begin by identifying the most relevant LL transitions. This is done by requiring that the $u$- and $d$-quark energies, defined by Eqs.~(\ref{Eu-star}) and (\ref{Ed-star}), lie sufficiently close to their respective Fermi surfaces. By assuming a wide band of energies around the Fermi surfaces and neglecting the neutrino contributions to Eqs.~(\ref{Eu-star}) and (\ref{Ed-star}), this procedure yields an initial set of $(n_{d},n_{u})$ pairs corresponding to the relevant LL transitions and generally works very well for ultra-strong magnetic fields and low temperatures. As the field strength decreases or the temperature increases, however, additional LL transitions may become important. In particular, since the spacing between the $d$-quark LLs is smaller than that between the $u$-quark LLs, transitions to neighboring $d$-quark LLs, obtained by replacing $n_{d}$ with $n_{d} \pm \delta n$, where $\delta n = 1,2,\ldots$, can contribute substantially.

For the highest temperature considered in this study, $T=1~\mbox{MeV}$, we begin by restricting the energy band around the Fermi surface to approximately $\pm 85~\mbox{MeV}$. At lower temperatures, we narrow this band slightly, scaling its width proportionally to $\sqrt{T}$. While this range may be overly broad in the regime of ultra-strong magnetic fields above $5\times 10^{18}~\mbox{G}$, it becomes increasingly important to include additional LL transitions as the field strength decreases. In particular, we find that the initial set of $(n_{d},n_{u})$ transitions must be supplemented by additional sets with $n_{d}$ replaced by $n_{d} \pm 1$. With a further decrease in the field strength, transitions involving $n_{d} \pm 2$, and eventually numerous other LL transitions, also begin to contribute substantially. To keep the numerical calculations manageable, we restrict our analysis to magnetic fields above approximately $8\times 10^{17}~\mbox{G}$. Despite this restriction, even when hundreds of LL transitions are included in the rate, we observe a gradual deterioration in numerical precision toward the lower end of the magnetic field range.

\subsection{Regime of very strong magnetic fields}
\label{sec:strong-B-regime}

\begin{figure}
\centering
 \subfigure[]{\includegraphics[width=0.85\textwidth]{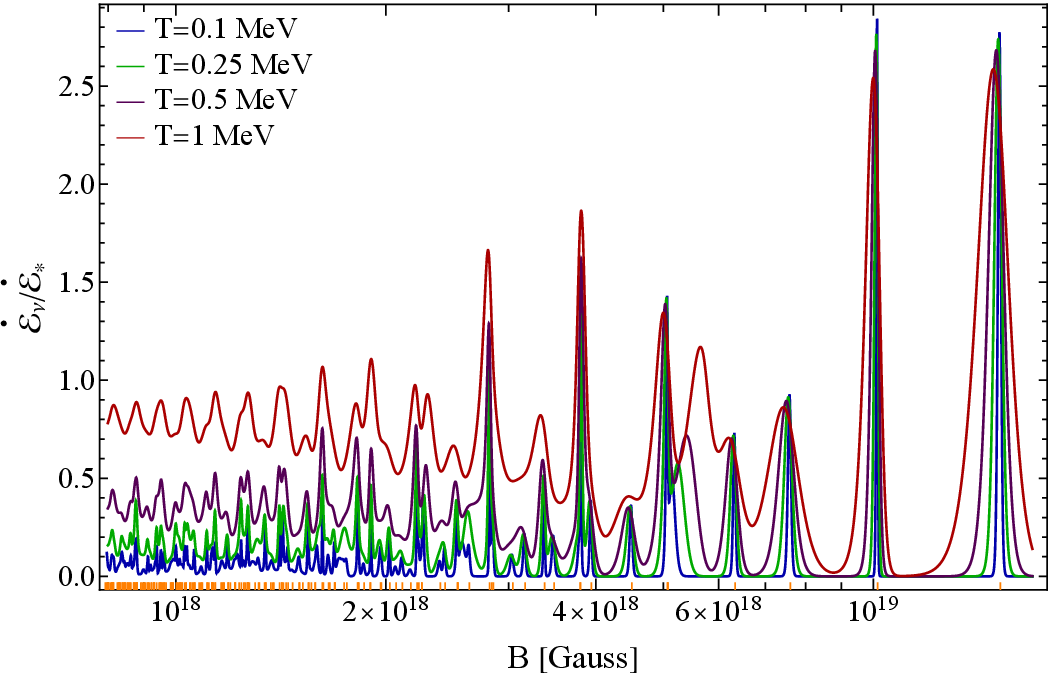}}
\\
  \subfigure[]{\includegraphics[width=0.85\textwidth]{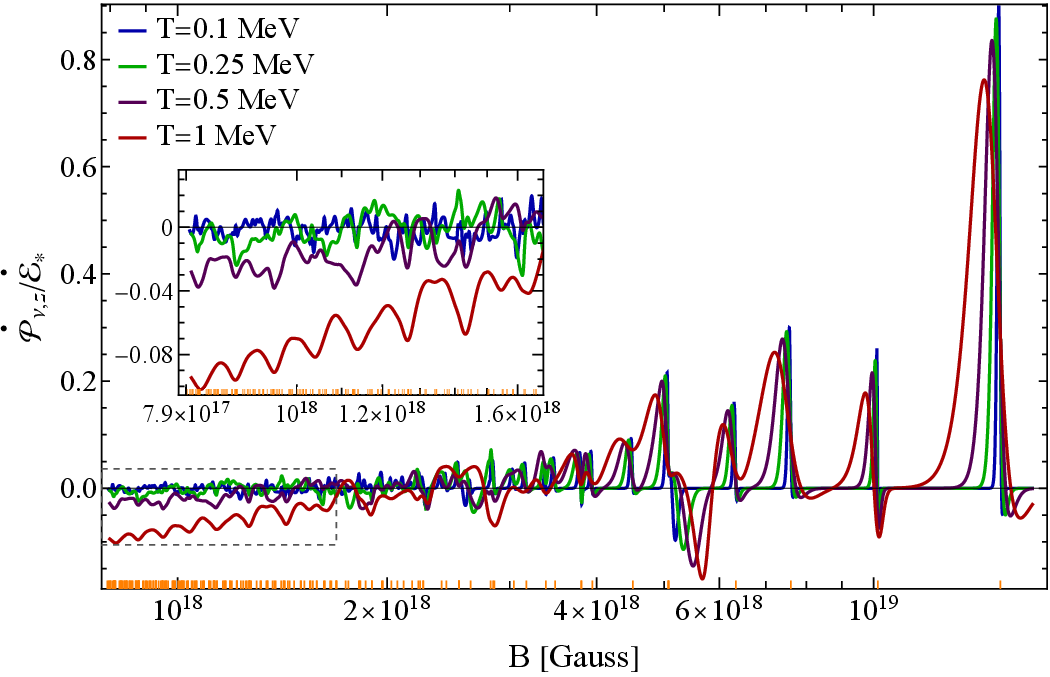}}
  \caption{The neutrino energy emission rate (a) and longitudinal momentum emission rate (b) as functions of the magnetic field strength. Results are shown for four fixed temperatures ranging from $T=0.1~\mbox{MeV}$ to $T=1~\mbox{MeV}$. All rates are normalized by the reference emission rate $\dot{\cal E}_{*}$ defined in Eq.~(\ref{E-dot-ref}). The inset in panel (b) provides an enlarged view of the low-field region.}
\label{fig.mag-Urca-rates}
\end{figure}

Our numerical results for the neutrino energy and longitudinal-momentum emission rates are presented in figure~\ref{fig.mag-Urca-rates}. We show the rates for four fixed temperatures: $T=0.1~\mbox{MeV}$ (blue), $T=0.25~\mbox{MeV}$ (green), $T=0.5~\mbox{MeV}$ (purple), and $T=1~\mbox{MeV}$ (red). The results are normalized by the characteristic emission-rate scale
\begin{equation}
\dot{\cal E}_{*} = \frac{N_cG_F^2\cos^2\theta_C}{\pi^5} \mu_{u} \mu_{d} |eB| T^5 ,
\label{E-dot-ref}    
\end{equation}
which captures the leading parametric dependence on the relevant model parameters in the strong-field regime, where quark LL quantization is important. 

Notably, the characteristic rate $\dot{\cal E}_{*}$ scales as $T^5$, in contrast to the $T^6$ dependence of the Iwamoto rate \cite{Iwamoto:1980eb,Iwamoto:1982zz}. This scaling is supported by our numerical results at the upper end of the magnetic field range in figure~\ref{fig.mag-Urca-rates}(a): the corresponding peak values of the energy emission rates, normalized by $\dot{\cal E}_{*}$, are approximately temperature independent at the largest Urca-resonant magnetic fields. As the magnetic field decreases and LL quantization of the quarks becomes less important, however, the emission rates gradually approach the conventional $T^6$ scaling. This crossover is evident at the lower end of the magnetic field range in figure~\ref{fig.mag-Urca-rates}(a), where the normalized rates at different temperatures begin to separate.

The Urca-resonant magnetic field values, $B_{n_{d},n_{u}}$, defined by Eq.~(\ref{resonant-B}), are marked by additional orange ticks along the abscissas in figure~\ref{fig.mag-Urca-rates}. As shown in panel (a), the Urca emission rates exhibit resonance-like peaks near the discrete field values indicated by these ticks. As the magnetic field strength decreases, the density of Urca-resonant values increases, causing neighboring peaks to overlap and eventually merge into smooth curves. As expected, thermal effects broaden the individual peaks and further enhance their overlap. At the lowest temperature, $T=0.1~\mbox{MeV}$, nearly all peaks remain resolved down to magnetic field strengths of approximately $10^{18}~\mbox{G}$. Some groups of nearly degenerate Urca-resonant fields, however, cannot be completely resolved even at this temperature. One example is the group near $5.07\times 10^{18}~\mbox{G}$ associated with $B_{5,2}$, $B_{6,2}$, and $B_{3,0}$; see table~\ref{tab:special-B}. These resonances appear to produce two peaks, with the higher-field peak shifting away from the other as the temperature increases. By contrast, at the highest temperature, $T=1~\mbox{MeV}$, most neighboring peaks merge, with only the few largest ones remaining clearly resolved. Contrary to naive expectations, the partial contributions of the resonant peaks vary substantially. A detailed analysis shows that this variation arises from the sensitivity of the corresponding transition amplitudes to the specific properties of the initial and final quark states, including their LL indices and longitudinal momenta.

As follows from the analysis in the preceding section, no additional peaks appear in the Urca emission rates for magnetic fields above $1.52\times 10^{19}~\mbox{G}$, which corresponds to the largest Urca-resonant magnetic field, $eB_{1,0}\approx 6\mu_{e}\mu_{u}$. At stronger fields, the rate becomes exponentially suppressed because the energy conservation condition cannot be satisfied by quark and electron states near their respective Fermi surfaces when LL quantization is taken into account. In other words, the Urca emission rate is quenched at sufficiently large magnetic fields.

\subsection{Emergent Shubnikov--de Haas-type oscillations}
\label{sec:SdH-oscillations}

\begin{figure}
\centering
  \includegraphics[width=0.85\textwidth]{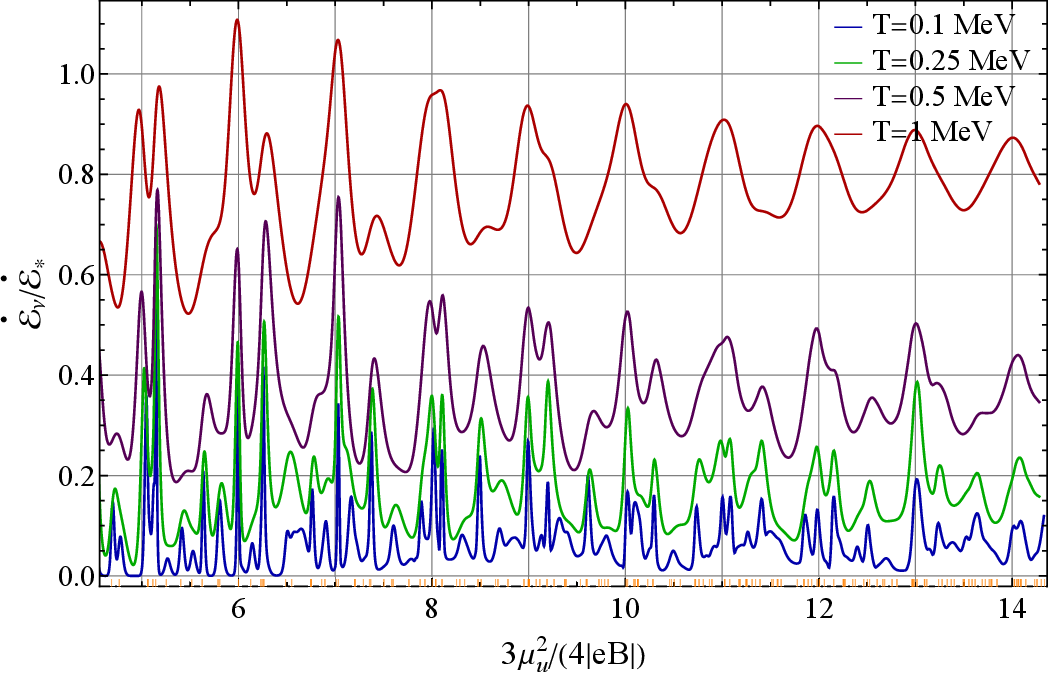}
  \caption{Neutrino energy emission rate as a function of the inverse magnetic field, exhibiting emergent SdH-type oscillations associated with the LL quantization of $u$-quarks. The inverse magnetic field is normalized so that the predicted oscillation peaks occur at integer values.}
\label{fig.rates-SdH}
\end{figure}

It may appear surprising that at the lower end of the magnetic field range shown in figure~\ref{fig.mag-Urca-rates}, the highest-temperature rate ($T=1~\mbox{MeV}$) develops a quasiperiodic dependence on the field strength. This behavior does not appear to correlate with the Urca-resonant field values or their density. Instead, it closely resembles Shubnikov--de Haas (SdH) oscillations of resistivity in metals \cite{Schubnikov:1930aa}, which arise from variations in the density of states at the Fermi surface as successive LLs become populated. We argue that the observed structure has a similar origin. Indeed, the locations of the first few peaks in the $T=1~\mbox{MeV}$ rate are found to lie close to the magnetic field values
\begin{equation}
    |eB|=\frac{3\mu_{u}^2}{4n_{u}},\qquad \text{(SdH peaks)}.
\end{equation}
with the $u$-quark LL index ranging approximately from $n_{u}=6$ to $n_{u}=14$. These values correspond to the thresholds at which the respective $u$-quark LLs begin to be populated. To illustrate this correspondence more clearly, figure~\ref{fig.rates-SdH} shows the energy emission rate as a function of $3\mu_{u}^2/(4|eB|)$. As anticipated, the peaks occur close to integer values of this variable. The figure focuses on the lower-field portion of the range considered previously in figure~\ref{fig.mag-Urca-rates}, namely $B\lesssim 2.5\times 10^{18}~\mbox{G}$.

One may then ask why the $u$-quark LL thresholds give rise to resolvable SdH-type peaks, whereas the corresponding $d$-quark thresholds generally remain unresolved. In fact, as seen from figure~\ref{fig.rates-SdH}, some of the latter may also become visible as secondary peaks, especially at lower temperatures. In general, however, the combined effect of LL quantization of both $u$- and $d$-quarks tends to complicate the emission-rate dependence on the field and obscure the contributions associated with individual LL thresholds. This is further compounded by the fact that the corresponding would-be periods in the inverse magnetic field, $4/(3\mu_{u}^2)$ and $2/(3\mu_{d}^2)$, are generally incommensurate for $\mu_{d}>\mu_{u}$.

An additional and essential distinction between the two flavors arises from different thermal smearing of their LLs. In the regime of moderately strong magnetic fields, the energy spacing between adjacent $u$-quark LLs, $2|eB|/(3\mu_{u})$, is slightly more than twice that between adjacent $d$-quark LLs, $|eB|/(3\mu_{d})$. When the $d$-quark LL spacing falls below approximately $2\pi T$, thermal broadening largely washes out not only the associated threshold structure but also the broader signatures of $d$-quark LL quantization. By contrast, the larger spacing between adjacent $u$-quark LLs may still exceed $2\pi T$, allowing the corresponding thresholds to remain well resolved. As successive $u$-quark LLs cross the Fermi level, the resulting modulation of the density of states produces oscillations in the emission rate. As the temperature is lowered, thermal smearing becomes less effective, allowing analogous quasiperiodic SdH-type oscillations to remain visible down to progressively weaker magnetic fields. At sufficiently weak magnetic fields, however, the oscillations are gradually attenuated as thermal broadening washes out the $u$-quark LL structure. This occurs when $|eB|\lesssim 3\pi T \mu_{u}$. For $T=1~\mbox{MeV}$, for example, this condition  corresponds approximately to $B\lesssim 4.8\times10^{17}~\mbox{G}$.

By combining the approximate conditions for the onset of the $u$-quark-driven SdH-type oscillations and their eventual suppression by thermal smearing, we estimate that these oscillations should be most clearly visible within the magnetic field range
\begin{equation}
    3\pi T\mu_{u} \lesssim |eB| \lesssim 6\pi T\mu_{d} .
    \label{eq-SdH-range}
\end{equation}
Although this interval is relatively narrow, it may still encompass a substantial number of resolvable oscillation periods. For the highest temperature considered, $T=1~\mbox{MeV}$, we estimate that more than a dozen oscillations may be resolved. As the temperature decreases, this number may increase to several dozen, although the oscillations then occur at progressively lower magnetic fields. Therefore, the estimate in Eq.~(\ref{eq-SdH-range}) should be treated with caution at low temperatures. Indeed, for the two lowest temperatures considered, $T=0.1~\mbox{MeV}$ and $T=0.25~\mbox{MeV}$, the magnetic field interval defined by Eq.~(\ref{eq-SdH-range}) formally extends into a sufficiently weak-field regime in which electrons begin to populate LLs above the LLL. The system then crosses over to a qualitatively different regime, where the oscillatory structure of the emission rate is expected to be dominated by the more pronounced effects of electron LL quantization, as investigated in refs.~\cite{Ghosh:2025sjn,Shovkovy:2026aci}.

\subsection{Regime of moderate magnetic fields}
\label{sec:moderate-B-regime}

\begin{figure}
\centering
\subfigure[]{\includegraphics[width=0.85\textwidth]{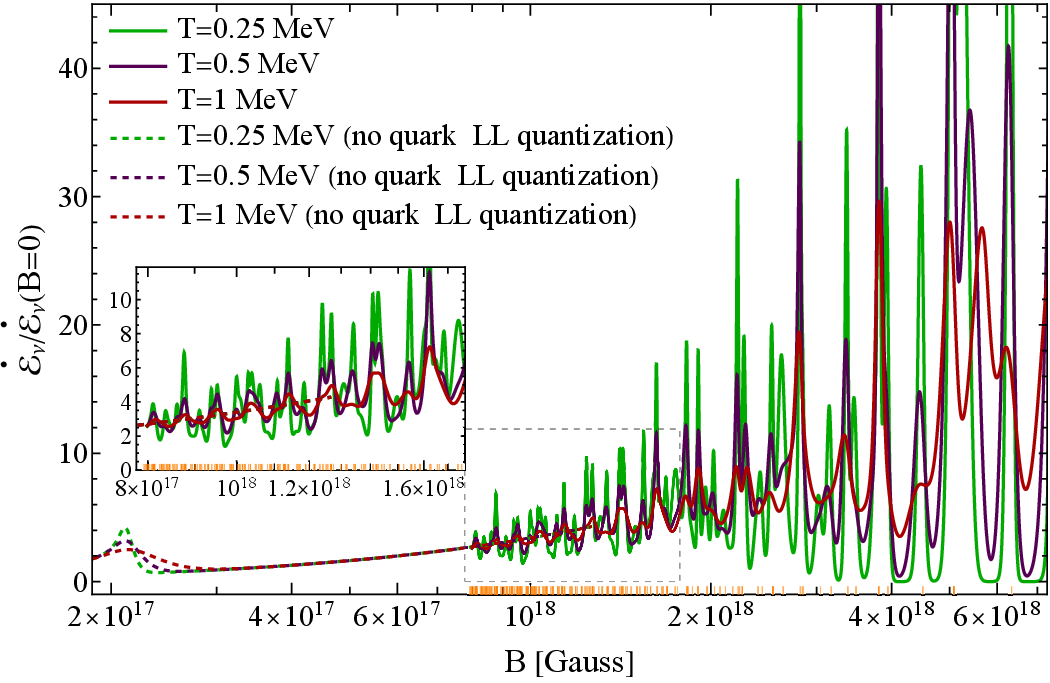}}
\\
\subfigure[]{\includegraphics[width=0.85\textwidth]{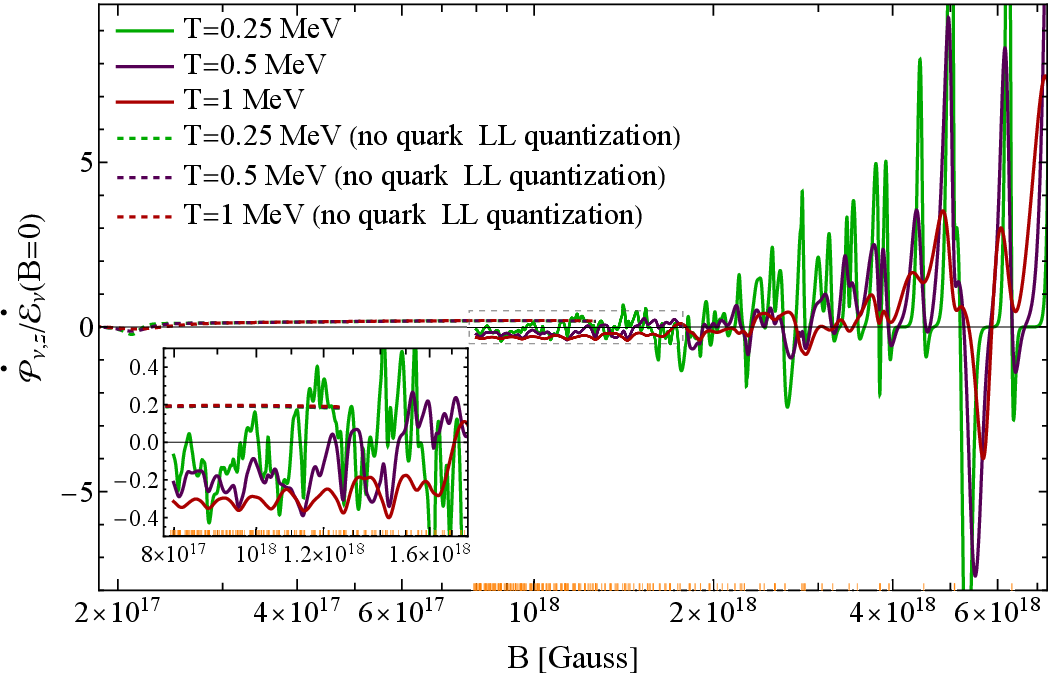}}
\caption{The neutrino energy emission rate (a) and longitudinal momentum emission rate (b) as functions of the magnetic field strength. Results are shown for three temperatures: $T=0.25~\mbox{MeV}$ (green), $T=0.5~\mbox{MeV}$ (purple), and $T=1~\mbox{MeV}$ (red). The solid lines show the results obtained in the present study with quark LL quantization included, whereas the dashed lines show the results of ref.~\cite{Shovkovy:2026aci} for the same model parameters but with quark LL quantization neglected. All rates are normalized by the zero-field energy emission rate $\dot{\cal E}_{\nu}(B=0)$ defined in Eq.~(\ref{E-dot-B0}). The insets provide enlarged views of the overlap regions.}
\label{fig.low-mag-Urca-rates}
\end{figure}

With decreasing magnetic field, we find that the neutrino energy emission rates not only approach the $T^6$ scaling but also become comparable in magnitude to those obtained in ref.~\cite{Shovkovy:2026aci}, which used the same model parameters but neglected quark LL quantization. The behavior of the rates obtained using the two approaches is illustrated in figure~\ref{fig.low-mag-Urca-rates}, where both are normalized by their zero-field limits. Here, we use the same improved expression for the zero-field rate as that obtained in ref.~\cite{Ghosh:2025sjn}, i.e.,
\begin{equation}
 \dot{\cal E}_{\nu}(B=0) \simeq C_T \frac{457\pi N_c}{2520}v_F (1-v_F^2)  G_F^2\cos^2\theta_C \mu_{u} \mu_{d} \mu_{e} T^6\left(1+\frac{\mu_{e}}{2\mu_{u}}\right).
 \label{E-dot-B0}
\end{equation}
Unlike the conventional Iwamoto rate, this expression includes an extra function $C_T$, which is of order $1$. Its approximate dependence on the temperature is given by 
 \begin{equation}
C_T \approx 1+ c_{1} \frac{T}{\mu_{e}}  +c_{2} \frac{T^2}{\mu_{e}^2} , 
 \label{CT-app-B0}
\end{equation}
where $c_{1}\approx 15.70$ and $c_{2}\approx 6.287$. Unlike our main result for the rate in Eq.~(\ref{der-rate-main}), the zero-field expression in Eq.~(\ref{E-dot-B0}) incorporates Fermi-liquid corrections for quarks. Following refs.~\cite{Ghosh:2025sjn,Shovkovy:2026aci}, we use the same value of the strong coupling constant, $\alpha_s=0.3$.

For simplicity, figure~\ref{fig.low-mag-Urca-rates} shows the results only for the three highest temperatures, $T=0.25~\mbox{MeV}$, $T=0.5~\mbox{MeV}$, and $T=1~\mbox{MeV}$. We found that the rates at lower temperatures, when normalized by $\dot{\cal E}_{\nu}(B=0)$, also converge to the same average values. Their fluctuations about these averages are considerably larger because the numerous Urca-resonant magnetic fields are better resolved at lower temperatures. The dashed lines in the weaker-field regime represent the results obtained without quark LL quantization \cite{Shovkovy:2026aci}, whereas the solid lines show the rates calculated in the present study with all quantization effects included. Considering how different the two approximations are, the corresponding energy emission rates are in remarkably good agreement. 

Overall, as shown in figure~\ref{fig.low-mag-Urca-rates}, the emission rate obtained with quark LL quantization exhibits substantial variations around its average value, with the oscillations becoming especially pronounced at lower temperatures. When averaged over these variations, the rate appears to increase approximately linearly with the magnetic field strength. For fields of order $10^{18}~\mbox{G}$ and above, however, the effects of quark LL quantization gradually become significant even at the highest temperature considered, $T=1~\mbox{MeV}$.

We should emphasize once again that quark Fermi-liquid corrections were included in refs.~\cite{Ghosh:2025sjn,Shovkovy:2026aci} but are neglected in the present study. Such corrections are not expected to play a significant role at very strong magnetic fields, particularly when quark LL quantization is essential. However, their importance may increase gradually as the magnetic field decreases. It is natural to expect that such corrections should slightly enhance the energy emission rates, but the effect is unlikely to be dramatic. 

\subsection{Asymmetry of momentum emission}
\label{sec:momentum-emission}

The numerical results for the longitudinal momentum emission rate shown in figure~\ref{fig.mag-Urca-rates}(b) reveal a pronounced anisotropy of neutrino emission. In particular, near the maxima of the energy emission peaks, neutrinos are emitted predominantly along the magnetic field, corresponding to $\dot{\cal P}_{\nu,z}>0$. The degree of asymmetry can be characterized by $\eta= \dot{\cal P}_{\nu,z}/\dot{\cal E}_{\nu}$. In the vicinity of the energy emission peaks near the largest Urca-resonant magnetic field values, $\eta$ reaches values of up to approximately $0.22$. Away from the centers of the energy emission peaks, and especially near their edges, the magnitude of the asymmetry can exceed $0.6$ and may also change sign. Owing to the strong dependence of the emission amplitude on the   kinematics of specific LL transitions, including the longitudinal momenta of the $u$- and $d$-quarks, the asymmetry varies substantially from one peak to another. A representative outlier is the second-largest energy emission peak near $B_{3,1}\approx 10.1\times10^{18}~\mbox{G}$, for which the asymmetry appears rather small, i.e., $\eta\lesssim0.05$ at the center of the energy emission peak. In general, the asymmetry becomes less pronounced as the temperature increases, consistent with the stronger thermal smearing of the underlying LL-resolved kinematics.

As the magnetic field decreases, an increasing number of LL transitions contribute simultaneously, each with its own characteristic kinematics and momentum asymmetry. Their superposition causes both the magnitude and sign of $\dot{\cal P}_{\nu,z}$ to vary strongly from one energy emission peak to another. The variability of the individual contributions from different LL transitions arises from the nontrivial interplay between quark LL quantization and the additional freedom provided by a small but nonzero neutrino longitudinal momentum, which helps reconcile the constraints imposed by energy and momentum conservation with the requirement that the quark states remain near their respective Fermi surfaces. Noting also that the left-handedness of weak interactions requires the longitudinal momentum of the $d$-quark, $p_z$, to be larger than that of the $u$-quark, $k_z$, one finds different degrees of flexibility for transitions with $p_z<0$ and $p_z>0$; see figure~\ref{fig.LLs-transitions}. This distinction becomes particularly important when the magnetic field deviates from the fine-tuned values obtained in the zero-neutrino-energy limit, and the effect becomes more pronounced with increasing temperature.

At sufficiently low temperatures, it is therefore no longer evident that a preferred emission direction exists as the magnetic field decreases into the regime with a high density of Urca-resonant field values; see figure~\ref{fig.mag-Urca-rates}(b). Somewhat surprisingly, however, the results show that the net longitudinal momentum gradually becomes negative for the highest temperature considered, $T=1~\mbox{MeV}$, as the field decreases. We find that a large fraction of the negative longitudinal momentum appears to originate from only a relatively small subset of the many LL transitions contributing to neutrino emission. Moreover, most of these transitions are not the same ones that give the largest contributions to the energy emission rate. We argue that the strongest emission asymmetry is produced by transitions for which the quantization constraints are substantially frustrated in the limit of vanishing neutrino energy, but can be satisfied for appropriate values of the neutrino energy and longitudinal momentum of a specific sign.

Overall, as seen in figure~\ref{fig.low-mag-Urca-rates}(b), the longitudinal momentum emission rates obtained with and without quark LL quantization do not agree well at intermediate magnetic fields. At higher temperatures, even their overall signs appear to differ. This contrasts with the energy emission rates, which show reasonably good agreement. The comparison suggests that the momentum emission asymmetry is more sensitive to the detailed quark LL structure than the total energy emission rate. The disagreement raises questions about the correct interpolation of the results for the momentum emission rates between the weak- and strong-field limits. 

At the highest temperature, $T=1~\mbox{MeV}$, the momentum emission also exhibits SdH-type oscillations similar to those observed in the energy emission. These oscillations result from $u$-quark quantization, which remains resolved, whereas the weaker $d$-quark LL quantization is largely washed out by thermal broadening.

As shown in refs.~\cite{Ghosh:2025sjn,Shovkovy:2026aci}, Fermi-liquid corrections in the absence of quark LL quantization can substantially enhance the net momentum emission along the magnetic-field direction. The same may also be true in the strong-field regime. It will therefore be of interest to perform a careful quantitative analysis of the combined effects of quark LL quantization and Fermi-liquid corrections on the momentum asymmetry in future studies. Naively, the resulting asymmetry, which is of order $10\%$ even under favorable assumptions, appears unlikely to be large enough to play a major role in compact-star phenomenology~\cite{Ghosh:2025sjn,Shovkovy:2026aci}. Nevertheless, a more complete treatment would be valuable for assessing the robustness of the predicted anisotropy and clarifying its microscopic origin.

\section{Discussion and Summary}
\label{sec:Summary}

In this study, we investigated direct Urca emission from strongly magnetized, unpaired quark matter by treating the Landau-level (LL) quantization of electrons and quarks on the same footing. Using the Kadanoff--Baym formalism, we derived the neutrino production rate and obtained the corresponding energy and longitudinal momentum emission rates. Our analysis extends earlier studies \cite{Ghosh:2025sjn,Shovkovy:2026aci}, in which the electron LL quantization was included but the quark spectrum was treated as quasicontinuous, to the regime of magnetic fields of order $10^{18}~\mbox{G}$ and higher, where only a limited number of quark LLs are occupied and their discrete structure becomes essential.

The main qualitative result is that quark LL quantization imposes a restrictive kinematic matching condition on the Urca processes. Since an appreciable rate requires the electron and both quark states to lie within an energy of order $T$ of their respective Fermi surfaces, this condition can be satisfied only near a discrete set of magnetic field strengths. For a transition characterized by the quark LL indices $(n_{d},n_{u})$, the corresponding Urca-resonant field is determined by Eq.~(\ref{resonant-B}) when the electrons occupy the LLL states [or Eq.~(\ref{resonant-B-nd-nu-ne}) in general], with the necessary condition $n_{d}>2n_{u}$. The term ``resonant'' is used here in a purely kinematic sense: the enhancement arises when the energy and momentum separations between the quantized initial and final states are compatible with the near-Fermi-surface Urca kinematics.

It should be emphasized that the corresponding LL quantization effects in dense quark matter are rather special, because all three charged particles participating in the direct Urca process, the $u$-quark, the $d$-quark, and the electron, occupy quantized Landau levels. An analogous situation does not occur in nuclear matter, where only the proton and the electron are Landau quantized, while the neutron does not undergo LL quantization. Since the neutron is not subject to Landau quantization, its energy and longitudinal momentum can be adjusted more freely, allowing the energies of all three particles to lie at their respective Fermi surfaces without conflicting with energy conservation. In other words, the kinematics in nuclear matter is not over-constrained. Consequently, there are no analogues of the Urca-resonant field values found in dense quark matter, where the simultaneous quantization of all three charged participants imposes much stronger restrictions on the allowed transitions.

Because of LL quantization effects in dense quark matter, the magnetic field dependence of the emission rate develops a sequence of pronounced peaks centered near the Urca-resonant field values. At low temperatures, the individual peaks are narrow and well resolved, whereas thermal broadening causes neighboring resonances to overlap as the temperature increases. The peak magnitudes vary considerably, because the transition amplitudes depend sensitively on the LL indices and longitudinal momenta of the participating quarks. In the strongly quantized regime, the characteristic energy emission rate scales approximately as $T^5$, rather than with the conventional $T^6$ dependence of the zero-field Urca emissivity. As the magnetic field decreases and the quark spectrum becomes increasingly quasicontinuous, the usual $T^6$ scaling is gradually recovered.

At moderately strong fields, we also identified quasiperiodic structures analogous to Shubnikov--de Haas oscillations. These oscillations arise when successive quark LLs cross the corresponding Fermi surfaces and modulate the density of available states. The clearest structures are associated with the $u$-quark LL thresholds, whose spacing near the Fermi surface is larger than that of the $d$-quark levels and is therefore less susceptible to thermal smearing. As the temperature decreases, such oscillations remain visible down to progressively weaker magnetic fields. Eventually, however, they are washed out when the LL spacing becomes smaller than the thermal broadening scale. In this weaker-field regime, our results approach those obtained when quark LL quantization is neglected.

At the opposite extreme, in the ultra-strong-field regime, LL quantization leads to a strong suppression of Urca emission. For magnetic fields above the largest Urca-resonant value, $ |eB|\simeq 6\mu_{e}\mu_{u} $, no electron and quark states near their respective Fermi surfaces can satisfy the required kinematic conditions. The emission rate is then exponentially suppressed. For the representative chemical potentials $\mu_{e}=50~\mbox{MeV}$ and $\mu_{u}=300~\mbox{MeV}$ used in our numerical analysis, this occurs at approximately $B\simeq 1.5\times10^{19}~\mbox{G}$. Thus, while selected magnetic field values can substantially enhance the Urca rate, sufficiently strong fields ultimately inhibit neutrino emission by eliminating the available near-Fermi-surface phase space.

It is natural to ask whether this conclusion changes when modified Urca processes involving spectator quarks are included. Although their calculation is considerably more involved and beyond the scope of this work, general kinematic arguments suggest that they are unlikely to eliminate the strong suppression caused by LL quantization, as the spectator quarks are themselves subject to LL quantization. All quarks must remain near their Fermi surfaces, which restricts their longitudinal momenta to discrete values $p^{(n)}_{F,z} \simeq \pm\sqrt{\mu_f^2-m_f^2-2n|e_fB|}$, up to thermal broadening. Consequently, spectator quarks can transfer only a discrete set of longitudinal momenta, rather than arbitrary momenta as in unmagnetized matter. While this may introduce additional kinematic matching conditions and resonant magnetic-field values, it does not restore the continuous longitudinal phase space available without LL quantization. This restriction is especially severe in the ultra-strong-field regime, where only the lowest few LLs are occupied. We therefore expect the qualitative suppression of neutrino emission to persist.

We also calculated the net longitudinal momentum carried by neutrinos and antineutrinos. Its dependence on the magnetic field exhibits the same resonance structure as the energy emission rate, while its sign and magnitude vary nonmonotonically between different LL transitions. This behavior reflects the anisotropic phase space induced by the magnetic field and the sensitivity of the longitudinal momentum asymmetry to the detailed kinematics of the participating states. A quantitative analysis of these LL-specific kinematics and their role in explaining the microscopic origin of the momentum anisotropy is an interesting topic for future studies.

In the present calculation, we neglected Fermi-liquid corrections to the quark dispersion relations. Such corrections are indispensable in the zero-field limit, where they relax the nearly collinear kinematics of noninteracting quark matter. In the strongly quantized regime considered here, however, the available phase space and the matching conditions are governed primarily by the discrete quark LL structure. We therefore expect Fermi-liquid effects mainly to shift the Urca-resonant field values and possibly split some resonances through interaction-induced spin-state splitting, without qualitatively changing the resonance-like pattern or the strong-field quenching. A quantitative treatment of these effects in LL-quantized interacting quark matter remains an important direction for future work.

The strong and nonmonotonic magnetic field dependence found here may have implications for the cooling and momentum evolution of compact stars containing quark matter. In particular, spatial variations of the magnetic field inside a stellar core could place different regions near or away from the Urca-resonant conditions, producing an inhomogeneous neutrino emissivity. Incorporating the resulting rates into dynamical cooling simulations, together with realistic equations of state, magnetic field profiles, interaction corrections, and possible color-superconducting phases, will be necessary to assess the corresponding astrophysical consequences.

It is interesting to speculate that the strongly nonmonotonic field dependence could produce intermittent cooling in compact stars containing strongly magnetized dense quark matter. Since the Urca-resonant fields scale parametrically with the square of the quark chemical potential, $B_{n_{d},n_{u}}\propto\mu_e \mu_u \propto\mu^2$, a substantial fraction of the quark core could approach a given resonant condition at approximately the same stage of its magnetic evolution. This  is plausible if the field strength increases with quark-matter density according to $B(r)\propto\mu^2(r)$, as expected under flux freezing and approximately isotropic compression.

If the initial field lies between the Urca-resonant values, or above the largest resonance, the neutrino emissivity may be strongly suppressed, leading to relatively slow cooling. As the field gradually decreases through resistive decay or other magnetohydrodynamic processes, it may enter the thermally broadened vicinity of an Urca resonance. The direct Urca channel would then become kinematically accessible, triggering a transient episode of enhanced neutrino emission and rapid cooling. Once the field moves away from the resonance, the emissivity would again be suppressed and slower cooling would resume. Passage through successive resonances could therefore generate rapid-cooling episodes separated by relatively quiescent intervals.

Although this heuristic scenario is intriguing, whether it can occur in a realistic compact star remains uncertain. Its realization requires a sufficiently large region of the quark core to enter an Urca-resonant regime. Spatial variations in composition and magnetic field \cite{Dexheimer:2016yqu,Chatterjee:2018prm}, magnetic heating, and processes in nuclear matter may weaken or obscure such cooling episodes. Moreover, the magnetic and thermal evolution must be treated self-consistently, since the field-decay rate depends on temperature, while the temperature controls the width and strength of the Urca resonances. Assessing the observability of this mechanism will therefore require coupled magneto-thermal simulations incorporating a realistic stellar structure and the Urca rates obtained here.

\acknowledgments

This research was funded in part by the U.S. National Science Foundation under Grant No.~PHY-2514933.

\appendix

\section{Derivation of the Quark Tensor}
\label{sec:Im-Pi}

In this appendix, we derive the expression for the imaginary (absorptive) part of the quark tensor given by Eq.~(\ref{Im-Pi}). In dense quark matter, the retarded self-energy of the gauge boson is determined by the simplest one-loop quark diagram, as illustrated in figure~\ref{fig.NuSelfEne}b. The corresponding expression reads
\begin{equation}
\bar{\Pi}_{R}^{\delta\sigma}(Q) = - i N_c  \sumint \frac{d^4K}{(2\pi)^4}\Tr\left[\gamma^\delta(1-\gamma_5) \bar{S}_{u}(K)\gamma^\sigma(1-\gamma_5) \bar{S}_{d}(P)\right],
\label{pimunu-app}
\end{equation}
where $P=K+Q$ and $N_c=3$ is the number of quark colors. Nonzero temperature effects are incorporated using the imaginary-time formalism, where the integration over the energy $k_{0}$ is replaced by a Matsubara sum, i.e.,
\begin{equation}
\sumint \frac{d^4 K}{(2\pi)^4} f(k_{0},\bm{k}) = T\sum_{k=-\infty}^{\infty}  i  \int \frac{d^3 \bm{k}}{(2\pi)^3} f(i\omega_k,\bm{k}).
\label{int-Matsubara}
\end{equation}
By definition, the fermionic Matsubara frequencies are $\omega_k=(2k+1)\pi T$. When calculating the sum, the external gauge-boson energy $q_{0}+i\epsilon$ is replaced with $i \Omega_m$, where $\Omega_m=2m\pi T$ is the bosonic Matsubara frequency. The dependence on $q_{0}$ is then restored at the end by performing the analytic continuation $i \Omega_m \to q_{0}+i\epsilon$.

In the presence of a background magnetic field, we must use the translation-invariant parts $\bar{S}_{f}(K)$ of the quark propagators in Eq.~\eqref{pimunu-app}, i.e.,
\begin{align}
    \bar{S}_{f} (k_{0},\bm k)
    &= i e^{-k_{\perp}^2\ell_{f}^2} \sum_{n=0}^\infty
    \sum_{\lambda_{f}=\pm 1}
    \frac{(-1)^n}
    {E_{f,n}
    \left[
        k_{0}+\mu_{f}+i\epsilon \, \sign(k_{0})-\lambda_{f} E_{f,n}
    \right]
    }
    \Big\{
        \left[
        E_{f,n} \gamma^0-\lambda_{f}  k_{f,z}\gamma^3+\lambda_{f} m_{f}
        \right]    \nonumber\\
    &\times 
        \left[
            {\cal P}_{+}L_{n}
            \left(2k_{\perp}^2\ell_{f}^2
            \right)
            -{\cal P}_{-}L_{n-1}
            \left(2k_{\perp}^2\ell_{f}^2
            \right)
        \right]
        +2\lambda_{f} (\bm k_{\perp}\cdot\bm\gamma_{\perp}) L_{n-1}^1
        \left(2 k_{\perp}^2\ell_{f}^2\right)
    \Big\},
\label{prop-quark}
\end{align}
with $\ell_{f} =1/\sqrt{|e_{f}B|}$, where $e_{f} = q_{f} |e| $, $q_{u}=2/3$, and $q_{d}=-1/3$. Using the spectral representation,
\begin{equation}
\bar{S}_{f} (k_{0},\bm k) =  i  \int_{-\infty}^{\infty} \frac{dk_{0}^\prime}{2\pi} \frac{ A_{f} (k_{0}^\prime + \mu_{f},\bm k) }{k_{0} - k_{0}^\prime},
  \label{prop-spectral-representation}
 \end{equation}
where
\begin{align}
    A_{f} (k_{0} + \mu_{f},\bm k) 
    &= 2\pi e^{-k_{\perp}^{2} \ell_{f}^2} 
    \sum_{\lambda_{f}=\pm 1} 
    \sum_{n_{f}=0}^\infty
    \frac{(-1)^{n_{f}}}{E_{f,n_{f}}}
    \delta
    \left(k_{0} + \mu_{f} - \lambda_{f} E_{f,n_{f}}\right)
    \Big\{
        \left[
            E_{f,n_{f}}\gamma^0-\lambda_{f}  k_{z}\gamma^3+\lambda_{f} m_{f}
        \right]    \nonumber \\ 
    & \times
        \left[
            {\cal P}_{f}^+ L_{n_{f}}
            \left(2k_{\perp}^{2} \ell_{f}^2
            \right)
            -{\cal P}_{f}^- L_{n_{f}-1}
            \left(2k_{\perp}^{2} \ell_{f}^2
            \right)
        \right]
        +  2\lambda_{f}
        (\bm k_{\perp} \cdot \bm \gamma_{\perp}) 
        L_{n_{f}-1}^1
        \left(2 k_{\perp}^{2} \ell_{f}^2
        \right)
    \Big\},
\label{quark-spectral-density-app}
\end{align}
we derive
{\small
\begin{equation}
\bar{\Pi}_{R}^{\delta\sigma}(Q) = - N_c T \sum_{k=-\infty}^{\infty} \int \frac{d^3 \bm{k}}{(2\pi)^3} \int_{-\infty}^{\infty} \frac{dk^\prime_{0}dp^\prime_{0}}{(2\pi)^2} \frac{ \Tr\left[\gamma^\delta(1-\gamma_5) A_{u} (k^\prime_{0}+\mu_{u},\bm{k}) \gamma^\sigma(1-\gamma_5) A_{d} (p^\prime_{0}+\mu_{d},\bm{p}) \right] }{(i\omega_k-k^\prime_{0})(i\omega_k+i \Omega_m-p^\prime_{0})},
\label{pimunu-app2}
\end{equation}}
with $\bm{p}=\bm{k}+\bm{q}$. Using the summation formula 
\begin{equation}
T \sum_{k=-\infty}^{\infty} \frac{1}{(i\omega_k-k^\prime_{0})(i\omega_k+i \Omega_m-p^\prime_{0})} 
=\frac{n_F(k^\prime_{0}) - n_F(p^\prime_{0}) }{ k^\prime_{0}-p^\prime_{0} +i \Omega_m  } ,
\end{equation}
we perform the Matsubara sum and obtain
\begin{align}
\bar{\Pi}_{R}^{\delta\sigma}(Q) &= - N_c \int \frac{d^3 \bm{k}}{(2\pi)^3} \int_{-\infty}^{\infty} \frac{dk^\prime_{0}dp^\prime_{0}}{(2\pi)^2} 
\frac{n_F(k^\prime_{0}) - n_F(p^\prime_{0}) }{ k^\prime_{0}-p^\prime_{0} +q_{0}+i\epsilon }
\Tr\Big[\gamma^\delta(1-\gamma_5) A_{u} (k^\prime_{0}+\mu_{u},\bm{k})    \nonumber \\ 
    & \times  \gamma^\sigma(1-\gamma_5) A_{d} (p^\prime_{0}+\mu_{d},\bm{p}) \Big],
\label{pimunu-app-sum}
\end{align}
where we also performed the analytic continuation $i \Omega_m \to q_{0}+i\epsilon$. Using the Sokhotski formula, we extract the imaginary (absorptive) part, thereby obtaining Eq.~(\ref{Im-Pi}):
\begin{align}
    \operatorname{Im}
    \left[\bar{\Pi}_R^{\delta\sigma}(Q)\right]
    &=N_c\int 
    \frac{d^3 \bm{k}}{(2\pi)^3} \int_{-\infty}^{\infty}\frac{dk_{0}}{4\pi}
    \left[n_F(k_{0})-n_F(p_{0})\right] 
    \Tr
    \Big[
        \gamma^\delta(1-\gamma_5) A_{u} (k_{0} + \mu_{u},\bm k) \nonumber \\ 
    & \times \gamma^\sigma(1-\gamma_5) A_{d} (p_{0} + \mu_{d},\bm p) 
    \Big],
\label{Im-Pi-App1}
\end{align}
with $p_{0}=k_{0}+q_{0}$.

\section{Tensor Contraction and Neutrino Production Rate}
\label{sec:L-Im-Pi}

In this appendix, we compute the tensor contraction, ${\cal L}_{n_{e},\lambda_{e}}^{\delta\sigma}(\bm{p}_{e},\bm{p}_\nu) \Im\left[ \bar{\Pi}^R_{\delta\sigma}(Q)\right]$, and then we integrate this expression over $\bm{p}_{e,\perp}$ to obtain an analytical formula for the neutrino-number production rate $\partial f_\nu(t, \bm{p}_\nu)/\partial t.$ For simplicity, from now on we neglect the antiparticle contributions coming from the terms with $\lambda_{d/u/e} = -1$. The corresponding formulas and derivations for these terms are easy to obtain, but their contribution is negligible since $T \ll \mu_{d/u/e}.$

\subsection{Calculation of lepton and quark tensor contraction}
\label{subsec:L-Im-Pi-1}

Performing the $k_{0}$ integral in Eq.~(\ref{Im-Pi-App1}), we obtain
\begin{align}
    \mbox{Im}\left[ \bar{\Pi}_R^{\delta\sigma}(Q)\right] 
    &= N_c 
    \sum_{n_{d},n_{u}=0}^{\infty}  
    \int \frac{d^3 \bm k}{8\pi^2}
    e^{-k_{\perp}^2\ell_{u}^2-p_{\perp}^2\ell_{d}^2}
    \frac{(-1)^{n_{u}+n_{d}}}
    {E_{u,n_{u}} E_{d,n_{d}}}   
    \delta\left[(E_{u,n_{u}}-\mu_{u})- (E_{d,n_{d}}-\mu_{d})+q_{0}\right]
    \nonumber \\ 
    &\times
    \left[n_F(E_{u,n_{u}}-\mu_{u}) - n_F(E_{d,n_{d}}-\mu_{d})\right]
    \Tr(T^{\delta\sigma}) ,
    \label{Im-Pi-App2}
\end{align}
where
\begin{align}
    T^{\delta\sigma}
    &=
    \gamma^\delta(1-\gamma_5) 
    \Big\{\left[E_{u,n_{u}} \gamma^0-k_{z}\gamma^3+m_{u}\right]
    \left[{\cal P}_{+}L_{n_{u}}\left(2 k_{\perp}^2\ell_{u}^2\right)
    -{\cal P}_{-}L_{n_{u}-1}\left(2 k_{\perp}^2\ell_{u}^2\right)\right]\nonumber \\ 
    &+2 (\bm k_{\perp}\cdot\bm\gamma_{\perp}) L_{n_{u}-1}^1\left(2 k_{\perp}^{2} \ell_{u}^2\right)\Big\}
    \gamma^\sigma(1-\gamma_5) 
    \Big\{\left[E_{d,n_{d}} \gamma^0-p_{z}\gamma^3+m_{d}\right]\nonumber \\ 
    &\times
    \left[{\cal P}_{-}L_{n_{d}}\left(2 p_{\perp}^2\ell_{d}^2\right)
    -{\cal P}_{+}L_{n_{d}-1}\left(2 p_{\perp}^2\ell_{d}^2\right)\right]
    +2 (\bm{p}_{\perp}\cdot\bm\gamma_{\perp}) L_{n_{d}-1}^1\left(2 p_{\perp}^{2} \ell_{d}^2\right)\Big\}.
\end{align}
Here, we use the simplified spin projector definitions ${\cal P}_{\pm}=(1\pm i \gamma^1\gamma^2)/2$ without an additional $s_{\perp}=\sign(e_{f} B)$, which is replaced with either $-1$ for $d$-quarks or $+1$ for $u$-quarks. Throughout this section, $\bm p$ should be considered shorthand for $\bm k + \bm q $, where $ \bm q = \bm{p}_{e} - \bm{p}_\nu$.

Recalling that the lepton tensor with $\lambda_{e} = 1$ is
\begin{align}
    {\cal L}^{\delta\sigma}_{n_{e}}(\bm{p}_{e},\bm{p}_\nu)
    &=
    \Tr
    \Big[
        \gamma^\delta (1-\gamma^5)
        \Big\{
            \left(E_{e,n_{e}}\gamma^0 -p_{e,z}\gamma^3+m_{e}\right) 
            \left[
                {\cal P}_{+} L_{n_{e}}\left(2 p_{e,\perp}^{2} \ell^2\right)
                -{\cal P}_{-}L_{n_{e}-1}\left(2 p_{e,\perp}^2\ell^2\right)
            \right] 
    \nonumber \\ 
    &{}+2       
            (\bm{p}_{e,\perp} \cdot\bm\gamma_{\perp}) L_{n_{e}-1}^1 \left(2 p_{e,\perp}^2\ell^2\right)
        \Big\}
        \gamma^\sigma(1-\gamma^5)(\gamma_{0} E_\nu-\bm\gamma\cdot \bm{p}_\nu)
    \Big],
\end{align}
one can compute the contraction of the two trace terms:
\begin{align}
    {\cal L} ^{\delta\sigma} _{n_{e}}(\bm{p}_{e},\bm{p}_\nu)
    \Tr(T_{\delta\sigma})
    &=
    64
    \Big[
        (E_{e,n_{e}}+p_{e,z})
        (E_{u,n_{u}}-k_{z})
        L_{n_{e}}(2p_{e,\perp}^2\ell^2)
        L_{n_{u}}(2k_{\perp}^2\ell_{u}^2)
        \nonumber \\ 
        &\hspace{30pt}+
        (E_{e,n_{e}}-p_{e,z})
        (E_{u,n_{u}}+k_{z})
        L_{n_{e}-1}(2p_{e,\perp}^2\ell^2)
        L_{n_{u}-1}(2k_{\perp}^2\ell_{u}^2)
        \nonumber \\ 
        &\hspace{30pt}-
        8(\bm{p}_{e,\perp}\cdot\bm k_{\perp})
        L_{n_{e}-1}^1(2p_{e,\perp}^2\ell^2)
        L_{n_{u}-1}^1(2k_{\perp}^2\ell_{u}^2)
    \Big]
    \nonumber \\ 
    & \times
    \Big[
        (E_\nu-p_{\nu,z})
        (E_{d,n_{d}}+p_{z})
        L_{n_{d}}(2p_{\perp}^2\ell_{d}^2)
        -(E_\nu+p_{\nu,z})
        (E_{d,n_{d}}-p_{z})
        L_{n_{d}-1}(2p_{\perp}^2\ell_{d}^2)
        \nonumber \\ 
        &\hspace{30pt}+
        4(\bm{p}_{\nu,\perp}\cdot\bm{p}_{\perp})
        L_{n_{d}-1}^1(2p_{\perp}^2\ell_{d}^2)
    \Big].
\end{align}
Expanding the product of the previous expression produces nine terms, each of which contains a product of three Laguerre polynomials, $L_{n_{e}-\alpha'}^\alpha(2p_{e,\perp}^2\ell^2) L_{n_{u}-\alpha'}^\alpha(2k_{\perp}^2\ell_{u}^2) L_{n_{d}-\beta'}^{\beta}(2p_{\perp}^2\ell_{d}^2)$, with various choices of $\alpha,\alpha',\beta,\beta' = 0, 1$. In fact, the product can be expressed concisely as a sum over these indices:
\begin{align}
    {\cal L} ^{\delta\sigma} _{n_{e},\lambda_{e}}(\bm{p}_{e},\bm{p}_\nu)
    \Tr(T_{\delta\sigma})
    &= 64
    \sum_{\alpha,\beta=0}^1
    \sum_{\alpha'=\alpha}^1
    \sum_{\beta'=\beta}^1
    C_{u}^{\alpha,\alpha'} C_{d}^{\beta,\beta'}
    (\bm{p}_{e,\perp}\cdot\bm k_{\perp})^\alpha
    (\bm{p}_{\nu,\perp}
    \cdot\bm{p}_{\perp})^\beta
    \nonumber \\ 
    &\times
    L_{n_{e}-\alpha'}^\alpha(2p_{e,\perp}^2\ell^2)
    L_{n_{u}-\alpha'}^\alpha(2k_{\perp}^2\ell_{u}^2)
    L_{n_{d}-\beta'}^\beta(2p_{\perp}^2\ell_{d}^2),
    \label{L-Tr-T}
\end{align}
where
\begin{align}
    C_{u}^{0,\alpha'}
    &= 
    (E_{e,n_{e}} + (-1)^{\alpha'} p_{e,z})
    (E_{u,n_{u}} - (-1)^{\alpha'} k_{z}), \label{C-u-0j}
    \\
    C_{d}^{0,\beta'}
    &= 
    ((-1)^{\beta'}E_\nu - p_{\nu,z})
    (E_{d,n_{d}} + (-1)^{\beta'} p_{z}) ,
    \label{C-d-0j}
    \\
    C_{u}^{1,1}
    &= -8 ,
    \label{C-u-11}
    \\
    C_{d}^{1,1}
    &= 4.
    \label{C-d-11}
\end{align}

Combining Eqs.~(\ref{Im-Pi-App2}) and (\ref{L-Tr-T}) gives
{\small
\begin{align}
    {\cal L}^{\delta\sigma}_{n_{e},\lambda_{e}}(\bm{p}_{e},\bm{p}_\nu)
    \mbox{Im}\left[ \bar{\Pi}_R^{\delta\sigma}(Q)\right]
    &= 64N_c 
    \sum_{n_{d},n_{u}=0}^{\infty}  
    \int \frac{dk_{z}}2
    \frac{(-1)^{n_{u}+n_{d}}}
    {E_{u,n_{u}} E_{d,n_{d}}}   
    \delta\left[(E_{u,n_{u}}-\mu_{u})- (E_{d,n_{d}}-\mu_{d})+q_{0}\right]
    \nonumber \\ 
    &\times
    \left[n_F(E_{u,n_{u}}-\mu_{u}) - n_F(E_{d,n_{d}}-\mu_{d})\right]
    \sum_{\alpha,\beta=0}^1
    \sum_{\alpha'=\alpha}^1
    \sum_{\beta'=\beta}^1
    C_{u}^{\alpha,\alpha'} C_{d}^{\beta,\beta'}
    \nonumber \\ 
    &\times
    \int \frac{d^{2} \bm k_{\perp}}{(2\pi)^2}
    e^{-k_{\perp}^2\ell_{u}^2-p_{\perp}^2\ell_{d}^2}
    (\bm{p}_{e,\perp}\cdot\bm k_{\perp})^\alpha
    (\bm{p}_{\nu,\perp}
    \cdot\bm{p}_{\perp})^\beta
    \nonumber \\ 
    &\times
    L_{n_{e}-\alpha'}^\alpha(2p_{e,\perp}^2\ell^2)
    L_{n_{u}-\alpha'}^\alpha(2k_{\perp}^2\ell_{u}^2)
    L_{n_{d}-\beta'}^\beta(2p_{\perp}^2\ell_{d}^2),
    \label{L-Im-Pi-app1}
\end{align}}
and we see there are four types of integrals to perform over $\bm k_{\perp}$ corresponding to the four possible assignments of $\alpha,\beta = 0, 1$ (whereas the overall form of the integrals does not depend on $\alpha'$ or $\beta'$). In appendix~\ref{subsec:Integrals-k-perp}, we show that these integrals have the following solutions:
\begin{align}
    &\mathcal{I}_{n_{u}-\alpha',n_{d}-\beta'}^{\alpha,\beta}
    =
    \int \frac{d^{2} \bm k_{\perp}}{(2\pi)^2}
    e^{-k_{\perp}^2\ell_{u}^2-p_{\perp}^2\ell_{d}^2}
    (\bm{p}_{e,\perp}\cdot\bm k_{\perp})^\alpha
    (\bm{p}_{\nu,\perp}
    \cdot\bm{p}_{\perp})^\beta
    L_{n_{u}-\alpha'}^\alpha(2k_{\perp}^2\ell_{u}^2)
    L_{n_{d}-\beta'}^\beta(2p_{\perp}^2\ell_{d}^2)
    \nonumber \\ 
    &\qquad =
    \frac{(-1)^{n_{u}-\alpha'+n_{d}-\beta'+\alpha}}{18\pi\ell^2}
    \frac{2^{\alpha}}{3^{\alpha+\beta}}
    e^{-q_{\perp}^2\ell^2}
    \sum_{i=0}^{n_{u}-\alpha'}
    \sum_{j=0}^{n_{d}-\beta'}
    (-1)^{i + j}
    \frac{2^{2i+j}}{3^{i+j}}
    \binom{n_{u}-\alpha'+\alpha}{i+\alpha}
    \binom{n_{d}-\beta'+\beta}{j+\beta}
    \binom{i+j}{i}
    \nonumber \\ 
    &\qquad\times
    \left[
    (\bm{p}_{e,\perp}\cdot\bm q_{\perp})^\alpha
    (\bm{p}_{\nu,\perp}\cdot\bm q_{\perp})^\beta
    L_{i+j}^{\alpha+\beta}(q_{\perp}^2\ell^2)
    - 
    \frac{\alpha\beta}{2\ell^2}
    (\bm{p}_{e,\perp}\cdot\bm{p}_{\nu,\perp})
    L_{i+j}^1(q_{\perp}^2\ell^2)
    \right],
    \label{I-integrals}
\end{align}
where $\bm p_\perp = \bm k_\perp + \bm q_\perp$, and therefore
\begin{align}
    {\cal L}^{\delta\sigma}_{n_{e},\lambda_{e}}&(\bm{p}_{e},\bm{p}_\nu)
    \mbox{Im}\left[ \bar{\Pi}_R^{\delta\sigma}(Q)\right]
    \nonumber \\ 
    &= \frac{16N_c}{9\pi\ell^2}
    \sum_{n_{d},n_{u}=0}^{\infty}  
    \int \frac{dk_{z}}{E_{u,n_{u}} E_{d,n_{d}}}  
    \delta\left[(E_{u,n_{u}}-\mu_{u})- (E_{d,n_{d}}-\mu_{d})+q_{0}\right]
    \nonumber \\ 
    &\times
    \left[n_F(E_{u,n_{u}}-\mu_{u}) - n_F(E_{d,n_{d}}-\mu_{d})\right]
    \sum_{\alpha,\beta=0}^1
    \sum_{\alpha'=\alpha}^1
    \sum_{\beta'=\beta}^1
     (-1)^{\alpha+\alpha'+\beta'}
    \frac{2^{\alpha}}{3^{\alpha+\beta}}
    C_{u}^{\alpha,\alpha'} C_{d}^{\beta,\beta'}
   \nonumber \\ 
    &\times
   \sum_{i=0}^{n_{u}-\alpha'}
    \sum_{j=0}^{n_{d}-\beta'}
    (-1)^{i+j}
    \frac{2^{2i+j}}{3^{i+j}}
     \binom{n_{u}-\alpha'+\alpha}{i+\alpha}
    \binom{n_{d}-\beta'+\beta}{j+\beta}
    \binom{i+j}{i}
    e^{-q_{\perp}^2\ell^2}
    L_{n_{e}-\alpha'}^\alpha(2p_{e,\perp}^2\ell^2)
   \nonumber \\ 
    &\times
    \left[
    (\bm{p}_{e,\perp}\cdot\bm q_{\perp})^\alpha
    (\bm{p}_{\nu,\perp}\cdot\bm q_{\perp})^\beta
    L_{i+j}^{\alpha+\beta}(q_{\perp}^2\ell^2)
    - 
    \frac{\alpha\beta}{2\ell^2}
    (\bm{p}_{e,\perp}\cdot\bm{p}_{\nu,\perp})
    L_{i+j}^1(q_{\perp}^2\ell^2)
    \right].
    \label{L-Im-Pi-Final}
\end{align}

\subsection{Calculation of neutrino-number production rate}
\label{subsec:calc-dfdt}

The neutrino-number production rate is given by Eq.~(\ref{rate-01}), which combined with Eq.~(\ref{L-Im-Pi-Final}) becomes
\begin{align}
    \frac{\partial f_\nu(t,\bm{p}_\nu)}{\partial t} 
    &= 
    -\frac{4N_cG_F^2\cos^2\theta_C}{9\pi^2\ell^2}
    \hspace{-4pt}
    \sum_{n_{e},n_{d},n_{u}=0}^\infty (-1)^{n_{e}} 
    \int \frac{dp_{e,z} dk_{z}}
        {E_\nu E_{e,n_{e}} E_{u,n_{u}}
        E_{d,n_{d}}
        }
    \nonumber \\ 
    &\times
   \delta\left[(E_{u,n_{u}}-\mu_{u})- (E_{d,n_{d}}-\mu_{d})+q_{0}\right]
     n_F(E_{e,n_{e}}-\mu_{e})  n_B(E_\nu+\mu_{e}-E_{e,n_{e}})
    \nonumber \\ 
    &\times
   \left[n_F(E_{u,n_{u}}-\mu_{u}) - n_F(E_{d,n_{d}}-\mu_{d})\right]
    \sum_{\alpha,\beta=0}^1
    \sum_{\alpha'=\alpha}^1
    \sum_{\beta'=\beta}^1
   (-1)^{\alpha+\alpha'+\beta'}
    \frac{2^{\alpha}}{3^{\alpha+\beta}}
    C_{u}^{\alpha,\alpha'} C_{d}^{\beta,\beta'}
    \nonumber \\ 
    &\times
     \sum_{i=0}^{n_{u}-\alpha'}
    \sum_{j=0}^{n_{d}-\beta'}
    (-1)^{i+j}
    \frac{2^{2i+j}}{3^{i+j}}
    \binom{n_{u}-\alpha'+\alpha}{i+\alpha}
    \binom{n_{d}-\beta'+\beta}{j+\beta}
    \binom{i+j}{i}
    \mathcal{J}_{n_{e}-\alpha',i+j}^{\alpha,\beta} ,
    \label{dfdt-app1}
\end{align}
where $C_{u}^{\alpha,\alpha'}$ and $C_{d}^{\beta,\beta'}$ are given by Eqs.~(\ref{C-u-0j})--(\ref{C-d-11}) and
\begin{align}
    \mathcal{J}_{n_{e}-\alpha',i+j}^{\alpha,\beta}
    &=
    \int
    \frac{d^2\bm{p}_{e,\perp}}{(2\pi)^2}
    e^{-(p_{e,\perp}^2+q_{\perp}^2)\ell^2}
    L_{n_{e}-\alpha'}^\alpha(2p_{e,\perp}^2\ell^2)
    \Big[
        (\bm{p}_{e,\perp}\cdot\bm q_{\perp})^\alpha
        (\bm{p}_{\nu,\perp}\cdot\bm q_{\perp})^\beta
        L_{i+j}^{\alpha+\beta}(q_{\perp}^2\ell^2) \nonumber \\ 
        & - \frac{\alpha\beta}{2\ell^2}
        (\bm{p}_{e,\perp}\cdot\bm{p}_{\nu,\perp})
        L_{i+j}^1(q_{\perp}^2\ell^2)
    \Big]
= \frac{(-1)^{n_{e}-\alpha'+\beta}}{2^{3+i+j+\alpha+\beta} \pi \ell^{2+2\alpha}}
    p_{\nu,\perp}^{2\beta}
    e^{-\frac{1}{2} p_{\nu,\perp}^2\ell^2}
    \nonumber \\ 
    &\times    \sum_{k=0}^{n_{e}-\alpha'}
    (-1)^{k}
    \binom{n_{e} - \alpha' + \alpha}{k + \alpha}
    \frac{(i+j+k+\alpha)!}{(i+j)!k!}
    L_{i+j+k+\alpha}^\beta
    (\tfrac12 p_{\nu,\perp}^2\ell^2),
    \label{master-J-formula}
\end{align}
where $\bm q_{\perp} = \bm p_{e,\perp} - \bm p_{\nu,\perp}$. The analytical solutions to the integrals defined above are derived in appendix~\ref{subsec:Integrals-p-e-perp}. Inserting these solutions into Eq.~(\ref{dfdt-app1}), we find
\begin{align}
    \frac{\partial f_\nu(t,\bm{p}_\nu)}{\partial t} 
    &=
    -\frac{N_cG_F^2\cos^2\theta_C}{18\pi^3\ell^4}
    \hspace{-4pt}
    \sum_{n_{e},n_{d},n_{u}=0}^\infty
    \int \frac{dp_{e,z} dk_{z}}
        {E_\nu E_{e,n_{e}} E_{u,n_{u}}
        E_{d,n_{d}}
        }
    \delta\left[(E_{u,n_{u}}-\mu_{u})- (E_{d,n_{d}}-\mu_{d})+q_{0}\right]
    \nonumber \\ 
    &\times
    n_F(E_{e,n_{e}}-\mu_{e})  n_B(E_\nu+\mu_{e}-E_{e,n_{e}})
    \left[n_F(E_{u,n_{u}}-\mu_{u}) - n_F(E_{d,n_{d}}-\mu_{d})\right]
    \nonumber \\ 
    &\times
    \sum_{\alpha,\beta=0}^1
    \sum_{\alpha'=\alpha}^1
    \sum_{\beta'=\beta}^1
    C_{u}^{\alpha,\alpha'} C_{d}^{\beta,\beta'}
    \frac{(-1)^{\alpha+\beta+\beta'}}{2^{\beta}3^{\alpha+\beta}}
    \frac{p_{\nu,\perp}^{2\beta}}{\ell^{2\alpha}}
    \mathcal{K}_{n_{e}-\alpha',n_{u}-\alpha',n_{d}-\beta'}^{\alpha,\beta}(\tfrac12 p_{\nu,\perp}^2\ell^2)  ,
    \label{dfdt-app2}
\end{align}
where $\mathcal{K}_{n_e,n_u,n_d}^{\alpha,\beta}(\xi)$ are the functions defined in Eq.~(\ref{K-functions}). 
These functions also admit the following (coordinate-space) integral representation:
\begin{align}
\mathcal{K}_{n_e,n_u,n_d}^{\alpha,\beta}(\xi)
&= \frac{1}{\xi^{\beta/2}} 
\int_0^\infty  d t \, e^{-t} t^{\alpha+\beta/2}
L_{n_e}^{\alpha}(t)
L_{n_u}^{\alpha}(2t/3)
L_{n_d}^{\beta}(t/3)
J_\beta(2\sqrt{\xi t}) ,
\end{align}
with $t=u_\perp^2/(2\ell^2)$, as well as the following alternative explicit expression:  
\begin{align}
\mathcal{K}^{\alpha,\beta}_{n_{e},n_{u},n_{d}} (\xi)
&=  (-1)^\beta 
\frac{3^{(\alpha+\beta)/2}}{2^{\alpha/2}\xi^{\beta/2}}
\sqrt{\frac{(n_e+\alpha)!}{n_e!}
\frac{(n_u+\alpha)!}{n_u!}
\frac{(n_d+\beta)!}{n_d!} }
\nonumber \\ 
    &\times 
    \sum_{m=0}^{n_u+n_d}
\mathcal{A}_m^{(n_u,n_d)}
\mathcal{A}_{m+\alpha+\beta}^{(n_u+\alpha,n_d+\beta)}
\varphi_{n_e,m}(\xi) \varphi_{n_e+\alpha,m+\alpha+\beta}(\xi),
\end{align}
where
\begin{align}
\mathcal{A}_m^{(n_u,n_d)}
&=(-1)^{n_u}\sqrt{\frac{2^{n_d-m}}{3^{n_u+n_d}} \frac{m!(n_u+n_d-m)!}{n_u!n_d!}}
\sum_{r=\max(0,m-n_d)}^{\min(n_u,m)}
(-2)^{r}\binom{n_u}{r}\binom{n_d}{m-r} \nonumber\\
&=(-1)^{n_u}\sqrt{\frac{2^{n_d-m}}{3^{n_u+n_d}} 
\binom{n_u+n_d}{n_u}
\binom{n_u+n_d}{m}}
{_{2}F_{1}}\left(-m , -n_u ; -n_u-n_d ; 3 \right), 
\end{align}
and 
\begin{align}
\varphi_{n_e,m}(\xi)
&= \begin{cases}
\sqrt{\frac{m!}{n_{e}!}}\, e^{-\xi/2} \,\xi^{(n_{e}-m)/2}
L_m^{n_{e}-m}(\xi),&n_{e}\geq m,\\[3mm]
(-1)^{m-n_{e}}\sqrt{\frac{n_{e}!}{m!}}\, e^{-\xi/2} \,\xi^{(m-n_{e})/2}
L_{n_{e}}^{m-n_{e}}(\xi),&n_{e}<m.
\end{cases}
\end{align}

Defining $\bar{q}_{0} =  q_{0} + \mu_{e} = E_{e,n_{e}}-E_\nu$, the argument of the $\delta$ function can be written as $E_{u,n_{u}} - E_{d,n_{d}} + \bar{q}_{0}$. The solutions of $E_{u,n_{u}} - E_{d,n_{d}} + \bar{q}_{0} = 0$ in $k_{z}$ and the corresponding values of $E_{u,n_{u}}$, $p_{z}$, and $E_{d,n_{d}}$ are given by
\begin{align}
    k_{z}^{(\pm)}
    &= -\frac{q_{z}}2
    \left(
        1-\frac{q_{+}q_{-}}{\bar{q}_{0}^2-q_{z}^2}
    \right)
    \pm\frac{1}{2}\frac{\bar{q}_{0}}{\bar{q}_{0}^2-q_{z}^2}
        \sqrt{
        (q_{+}^{2} - \bar{q}_{0}^{2} + q_{z}^2)
        (q_{-}^{2} - \bar{q}_{0}^{2} + q_{z}^2)
        },
    \label{kz-pm}
    \\
    E_{u,n_{u}}^{(\pm)}
    &= -\frac{\bar{q}_{0}}2
    \left(
        1-\frac{q_{+}q_{-}}{\bar{q}_{0}^2-q_{z}^2}
    \right)
    \pm
    \frac{1}{2}\frac{q_{z}}{\bar{q}_{0}^2-q_{z}^2}
        \sqrt{
        (q_{+}^{2} - \bar{q}_{0}^{2} + q_{z}^2)
        (q_{-}^{2} - \bar{q}_{0}^{2} + q_{z}^2)
        },
    \label{Eu-pm}
    \\
    p_{z}^{(\pm)}
    &= \frac{q_{z}}2
    \left(
        1+\frac{q_{+}q_{-}}{\bar{q}_{0}^2-q_{z}^2}
    \right)
    \pm
    \frac{1}{2}\frac{\bar{q}_{0}}{\bar{q}_{0}^2-q_{z}^2}
        \sqrt{
        (q_{+}^{2} - \bar{q}_{0}^{2} + q_{z}^2)
        (q_{-}^{2} - \bar{q}_{0}^{2} + q_{z}^2)
        },
    \label{pz-pm}
    \\
    E_{d,n_{d}}^{(\pm)}
    &= \frac{\bar{q}_{0}}2
    \left(
        1+\frac{q_{+}q_{-}}{\bar{q}_{0}^2-q_{z}^2}
    \right)
    \pm
    \frac{1}{2}\frac{q_{z}}{\bar{q}_{0}^2-q_{z}^2}
        \sqrt{
        (q_{+}^{2} - \bar{q}_{0}^{2} + q_{z}^2)
        (q_{-}^{2} - \bar{q}_{0}^{2} + q_{z}^2)
        },
    \label{Ed-pm}
\end{align}
where $q_{z} = p_{e,z}-p_{\nu,z}$ and 
\begin{align}
    q_\pm &= \sqrt{2|e_{d}B|n_{d}+m_{d}^2}
    \pm \sqrt{2|e_{u}B|n_{u}+m_{u}^2}.
\end{align}
Each solution $k_{z}^{(s)}$ exists if and only if three conditions are satisfied: (i) The argument of the square root in Eq.~(\ref{kz-pm}) is positive, (ii) $E_{u,n_{u}}^{(s)} > 0$, and (iii) $E_{d,n_{d}}^{(s)} > 0.$ Physically, the latter conditions result from restricting to quarks rather than anti-quarks. We therefore have the identity
\begin{align}
    \delta[E_{u,n_{u}} - E_{d,n_{d}} + \bar{q}_{0}]
    &= 
    \sum_{s=\pm}
    \frac{2E_{u,n_{u}}E_{d,n_{d}}
    \delta(k_{z} - k_{z}^{(s)})
    }
    {\sqrt{
    (q_{+}^2-\bar{q}_{0}^2+q_{z}^2)
    (q_{-}^2-\bar{q}_{0}^2+q_{z}^2)}}
    \Theta_{n_{u},n_{d}}^{(s)}(\bar{q}_{0},q_{z}) ,
    \label{app:energy-delta}
\end{align}
where we have introduced the following shorthand notation for the product of step functions:
\begin{equation}
\Theta_{n_{u},n_{d}}^{(s)}(\bar{q}_{0},q_{z}) = \theta\left[
        \left(q_{+}^2-\bar{q}_{0}^2+q_{z}^2\right)
        \left(q_{-}^2-\bar{q}_{0}^2+q_{z}^2\right)\right]
        \theta\left[E_{u,n_{u}}^{(s)}\right]
        \theta\left[E_{d,n_{d}}^{(s)}\right] .
\label{app:Theta-function}
\end{equation}
Utilizing energy conservation, we also obtain the following identity:
\begin{align}
    n_F(E_{e,n_{e}}-\mu_{e})
    &n_B(E_\nu+\mu_{e}-E_{e,n_{e}})  
    \left[n_F(E_{u,n_{u}}-\mu_{u}) - 
    n_F(E_{d,n_{d}}-\mu_{d})\right]
    \nonumber \\ 
    &\qquad=
    -n_F(E_{e,n_{e}}-\mu_{e})
    n_F(E_\nu-\mu_{u})  
    n_F(E_{d,n_{d}}-\mu_{d})
    \nonumber \\ 
    &\qquad=
    -\frac{e^{-\frac{E_\nu}{2T}}}
    {8\cosh\left(\frac{E_{e,n_{e}}-\mu_{e}}{2T}\right)
    \cosh\left(\frac{E_{u,n_{u}}-\mu_{u}}{2T}\right)
    \cosh\left(\frac{E_{d,n_{d}}-\mu_{d}}{2T}\right)}.
\end{align}
Then, using the result in Eq.~(\ref{app:energy-delta}), we can evaluate the $k_{z}$ integral in Eq.~(\ref{dfdt-app2}) and obtain
\begin{align}
    \frac{\partial f_\nu(t,\bm{p}_\nu)}{\partial t} 
    &= 
    \frac{N_cG_F^2\cos^2\theta_C}{72\pi^3\ell^4}
    \hspace{-4pt}
    \sum_{n_{e},n_{d},n_{u}=0}^\infty
    \sum_{s=\pm}
    \int dp_{e,z}
    \frac{
        e^{-\frac{E_\nu}{2T}}\,
        \Theta_{n_{u},n_{d}}^{(s)}(\bar{q}_{0},q_{z})
        }
    {\cosh\left(\frac{E_{e,n_{e}}-\mu_{e}}{2T}\right)
    \cosh\left(
        \frac{E_{u,n_{u}}^{(s)}-\mu_{u}}{2T}\right)
    \cosh\left(
        \frac{E_{d,n_{d}}^{(s)}-\mu_{d}}{2T}\right)}
    \nonumber \\ 
    &\times
    \sum_{\alpha,\beta=0}^1
    \sum_{\alpha'=\alpha}^1
    \sum_{\beta'=\beta}^1
    \frac{D_{u}^{\alpha,\alpha'} D_{d}^{\beta,\beta'}}
    {E_\nu E_{e,n_{e}}
    \sqrt{
    (q_{+}^2-\bar{q}_{0}^2+q_{z}^2)
    (q_{-}^2-\bar{q}_{0}^2+q_{z}^2)}}
    \mathcal{K}_{n_{e}-\alpha',n_{u}-\alpha',n_{d}-\beta'}^{\alpha,\beta}(\xi) ,
    \label{dfdt-app3}
\end{align}
where we have absorbed several coefficients into the definitions of $C_{u}^{\alpha,\alpha'},C_{d}^{\beta,\beta'}$ and renamed them as the following $D_{u}^{\alpha,\alpha'},D_{d}^{\beta,\beta'}$ functions:
\begin{align}
    D_{u}^{0,\alpha'}
    &= 
    (E_{e,n_{e}} + (-1)^{\alpha'} p_{e,z})
    (E_{u,n_{u}}^{(s)} - (-1)^{\alpha'} k_{z}^{(s)}) ,
    \label{D-u-0j}
    \\
    D_{d}^{0,\beta'}
    &= 
    (E_\nu - (-1)^{\beta'} p_{\nu,z})
    (E_{d,n_{d}}^{(s)} + (-1)^{\beta'} p_{z}^{(s)}),
    \label{D-d-0j}
    \\
    D_{u}^{1,1}
    &= \frac8{3\ell^2}, \label{D-u-11}
    \\
    D_{d}^{1,1}
    &= \frac23 p_{\nu,\perp}^2.
    \label{D-d-11}
\end{align}
For $B > \mu_{e}^{2}/(2|e|) \approx 2.11 \times 10^{17}$ G, only the LLL of the electron is populated, and it suffices to consider just the $n_{e}=0$ term. In this case, the sum $\sum_{k=0}^{n_{e}-\alpha'}$ shows that the $\alpha'=1$ term (and hence also the $\alpha=1$ term) vanishes, leaving
\begin{align}
    \frac{\partial f_\nu(t,\bm{p}_\nu)}{\partial t}
    &\approx
    \frac{G_F^2\cos^2\theta_C}{24\pi^3\ell^4}
    \hspace{-4pt}
    \sum_{n_{d},n_{u}=0}^\infty
    \sum_{s=\pm}
    \int dp_{e,z}
    \frac{
        e^{-\frac{E_\nu}{2T}}\,
        \Theta_{n_{u},n_{d}}^{(s)}(\bar{q}_{0},q_{z})
        }
    {\cosh\left(\frac{E_{e,n_{e}}-\mu_{e}}{2T}\right)
    \cosh\left(
        \frac{E_{u,n_{u}}^{(s)}-\mu_{u}}{2T}\right)
    \cosh\left(
        \frac{E_{d,n_{d}}^{(s)}-\mu_{d}}{2T}\right)}
    \nonumber \\ 
    &\times
    \frac{
        (E_{e} + p_{e,z})
        (E_{u,n_{u}}^{(s)} - k_{z}^{(s)})
        e^{-\frac{1}{2} p_{\nu,\perp}^2\ell^2}
        }
    {E_\nu E_{e,n_{e}}
    \sqrt{
    (q_{+}^2-\bar{q}_{0}^2+q_{z}^2)
    (q_{-}^2-\bar{q}_{0}^2+q_{z}^2)}}
    \nonumber \\ 
    &\times
    \bigg\{
        (E_\nu-p_{\nu,z})
        (E_{d,n_{d}}^{(s)}+p_{z}^{(s)})
        \sum_{i=0}^{n_{u}}
        \sum_{j=0}^{n_{d}}
        \frac{
        (-1)^{i+j}2^{i}}{3^{i+j}
        }
        \binom{n_{u}}{i}
        \binom{i+j}{i}
        \binom{n_{d}}{j}
        L_{i+j}
        (\tfrac12 p_{\nu,\perp}^2\ell^2)
    \nonumber \\ 
    &+
        (E_\nu+p_{\nu,z})
        (E_{d,n_{d}}^{(s)}-p_{z}^{(s)})
        \sum_{i=0}^{n_{u}}
        \sum_{j=0}^{n_{d}-1}
        \frac{
        (-1)^{i+j}2^{i}}{3^{i+j}
        }
        \binom{n_{u}}{i}
        \binom{i+j}{i}
        \binom{n_{d}-1}{j}
        L_{i+j}
        (\tfrac12 p_{\nu,\perp}^2\ell^2)
    \nonumber \\ 
    &+
        \frac23 p_{\nu,\perp}^2
        \sum_{i=0}^{n_{u}}
        \sum_{j=0}^{n_{d}-1}
        \frac{
        (-1)^{i+j}2^{i}}{3^{i+j}
        }
        \binom{n_{u}}{i}
        \binom{i+j}{i}
        \binom{n_{d}}{j+1}
        L_{i+j+1}^1
        (\tfrac12 p_{\nu,\perp}^2\ell^2)
    \bigg\} .
    \label{dfdt-app4}
\end{align}
The expression for $\partial f_\nu/\partial t$ in Eq.~(\ref{dfdt-app4}) can be further simplified under the approximation $p_{\nu,\perp}^{2} \ell^2, p_{\nu,z}^2\ell^{2} \to 0$. This approximation is valid for our choice of parameters, as the neutrino energies are of the order of temperature and $T^{2} \ll |eB|$. Applying the identities $L_{n}(0) = 1$, $L_{n}^1(0)=n$, and 
\begin{align}
    \sum_{i=0}^{n_{u}}
    \sum_{j=0}^{n_{d}}
    \frac{(-1)^{i+j}2^{i}}{3^{i+j}}
    \binom{n_{u}}{i}
    \binom{n_{d}}{j}
    \frac{(i+j)!}{i!j!}
    &=
    \frac{2^{n_{d}}}{3^{n_{u}+n_{d}}}
    \binom{n_{u}+n_{d}}{n_{u}} ,
    \label{sum-identity-1}
    \\
    \sum_{i=0}^{n_{u}}
    \sum_{j=0}^{n_{d}-1}
    \frac{(-1)^{i+j}2^{i}}{3^{i+j}}
    \binom{n_{u}}{i}
    \binom{n_{d}}{j+1}
    \frac{(i+j+1)!}{i!j!}
    &=
    \frac{2^{n_{d}-1}}{3^{n_{u}+n_{d}-1}}
    \binom{n_{u}+n_{d}-1}{n_{u}}
    (n_{d}-2n_{u}),
    \label{sum-identity-2}
\end{align}
we obtain 
\begin{align}
    &\frac{\partial f_\nu(t,\bm{p}_\nu)}{\partial t}
    \approx 
    \frac{N_cG_F^2\cos^2\theta_C}{72\pi^3\ell^4 E_\nu}
    \hspace{-6pt}
    \sum_{n_{d},n_{u}=0}^\infty
    \frac{2^{n_{d}}}{3^{n_{u}+n_{d}}}
    \binom{n_{u}+n_{d}-1}{n_{u}} \hspace{-2pt}
    \nonumber \\ 
    &\times
    \sum_{s=\pm} \hspace{-1pt}
    \int \hspace{-4pt} dp_{e,z}
    \frac{   e^{-\frac{E_\nu}{2T}} \, 
        \Theta_{n_{u},n_{d}}^{(s)}(\bar{q}_{0},q_{z})  }
    {\cosh\left(\frac{E_{e,n_{e}}-\mu_{e}}{2T}\right)
    \cosh\left(
        \frac{E_{u,n_{u}}^{(s)}-\mu_{u}}{2T}\right)
    \cosh\left(
        \frac{E_{d,n_{d}}^{(s)}-\mu_{d}}{2T}\right)}
    \frac{  (1 + p_{e,z}/E_{e,n_{e}})
        (E_{u,n_{u}}^{(s)} - k_{z}^{(s)})  }
    {    \sqrt{ (q_{+}^2-\bar{q}_{0}^2+q_{z}^2)
    (q_{-}^2-\bar{q}_{0}^2+q_{z}^2)}}
    \nonumber \\ 
    &\times
    \bigg\{
        \frac{n_{u}+n_{d}}{n_{d}}
        (E_\nu-p_{\nu,z})
        (E_{d,n_{d}}^{(s)}+p_{z}^{(s)})
        +
        \frac32
        (E_\nu+p_{\nu,z})
        (E_{d,n_{d}}^{(s)}-p_{z}^{(s)})
        +
        (n_{d}-2n_{u})
        p_{\nu,\perp}^2
    \bigg\}.
    \label{dfdt-app5}
\end{align}
Eqs.~(\ref{sum-identity-2}) and (\ref{dfdt-app5}) use the convention that $\binom{n_{u}+n_{d}-1}{n_u} = 0$ when $n_d = 0$. However, Eq.~(\ref{dfdt-app5}) also contains a term with the product $\binom{n_{u}+n_{d}-1}{n_u}\frac{n_{u}+n_{d}}{n_{d}}$, which is formally $0\times \infty$ when $n_{d} = 0$. This product should be interpreted as $\binom{n_{u}+n_{d}}{n_{u}}$, which equals $1$ when $n_{d}=0$.

The $\theta$ functions can be simplified due to various assumptions that hold in our region of interest. In particular, we assume $|p_{\nu,z}|, p_{\nu,\perp} \lesssim T$ and $|p_{e,z}| \approx \mu_{e} \gg T.$ We also consider only the case where $n_{e}=0$, hence $E_{e} \approx |p_{e,z}| \approx \mu_{e}.$ Since $m_{e} \ll \mu_{e}$, the term $(1+p_{e,z}/E_{e})$ in Eq.~(\ref{dfdt-app5}) strongly suppresses the contribution with $p_{e,z} \approx -\mu_{e}$, so we neglect this contribution and assume $p_{e,z} \approx \mu_{e}$. We then have $\bar{q}_{0} \approx q_{z} \approx \mu_{e} > 0$, hence the term $\theta\big[E_{d,n_{d}}^{(s)}\big] = \theta\big[E_{u,n_{u}}^{(s)} + \bar{q}_{0}\big]$ in Eq.~(\ref{app:Theta-function})  is redundant. We also have $|\bar{q}_{0}^2-q_{z}^2|\approx 2\mu_{e}(E_\nu - p_{\nu,z})\sim \mu_{e} T \ll |eB| \sim q_{+}^2$. Note that $q_{-}^{2} \sim |eB|$ except in the special case $n_{d} = 2n_{u}$. The contribution from the latter transitions ($n_{d} = 2n_{u}$) turns out to be relatively small, so we treat this case separately in appendix~\ref{sec:Anom-transitions}. For the rest of this section, we consider only the case where $n_{d} \neq 2n_{u}$, which accounts for the dominant contribution to $\partial f_\nu/\partial t$.


In this case we have $|\bar{q}_{0}^2-q_{z}^2|\ll q_\pm^2$, hence the $(q_{+}^2-\bar{q}_{0}^2+q_{z}^2)(q_{-}^2-\bar{q}_{0}^2+q_{z}^2) > 0$ and we need only consider the second $\theta$ function in Eq.~(\ref{app:Theta-function}). Taking $|\bar{q}_{0}^2-q_{z}^2|\to0$ in Eqs.~(\ref{kz-pm})--(\ref{Ed-pm}) gives
\begin{align}
    k_{z}^{(\pm)}
    &\approx 
    -\frac{q_{z}}2
    \pm \frac{|q_{+}q_{-}|}{2[\bar{q}_{0}\mp \sign(q_{+}q_{-})q_{z}]}
    \mp
    \frac{\bar{q}_{0}(q_{+}^2+q_{-}^2)}{4|q_{+}q_{-}|} ,
    \label{kz-pm-approx}
    \\
    E_{u,n_{u}}^{(\pm)}
    &\approx
    -\frac{\bar{q}_{0}}2
    + \frac{q_{+}q_{-}}{2[\bar{q}_{0}\mp \sign(q_{+}q_{-})q_{z}]}
    \mp
    \frac{q_{z}(q_{+}^2+q_{-}^2)}{4|q_{+}q_{-}|} ,
    \label{Eu-pm-approx}
    \\
    p_{z}^{(\pm)}
    &\approx 
    \frac{q_{z}}2
    \pm \frac{|q_{+}q_{-}|}{2[\bar{q}_{0}\mp \sign(q_{+}q_{-})q_{z}]}
    \mp
    \frac{\bar{q}_{0}(q_{+}^2+q_{-}^2)}{4|q_{+}q_{-}|} ,
    \label{pz-pm-approx}
    \\
    E_{d,n_{d}}^{(\pm)}
    &\approx
    \frac{\bar{q}_{0}}2
    + \frac{q_{+}q_{-}}{2[\bar{q}_{0}\mp \sign(q_{+}q_{-})q_{z}]}
    \mp
    \frac{q_{z}(q_{+}^2+q_{-}^2)}{4|q_{+}q_{-}|}.
    \label{Ed-pm-approx}
\end{align}
Although both solutions $k_{z}^{(\pm)}$ may be dynamically allowed, we see from Eq.~(\ref{Eu-pm-approx}) that one choice of $\mp \sign(q_{+}q_{-})$ produces asymptotically large energies that are far from the Fermi surface. Imposing the near-Fermi surface condition $E_{u}^{(\pm)} \approx \mu_{u}$, the only surviving solution $k_{z}^*$ is given by $k_{z}^{(s)}$ with $s = -\sign(q_{+}q_{-})$. Taking only these solutions, we define:
\begin{align}
    k_{z}^*
    &=
    -\frac{q_{z}}2
    \left(
        1-\frac{q_{+}q_{-}}{\bar{q}_{0}^2-q_{z}^2}
    \right)
    -\frac{1}{2}\frac{\bar{q}_{0}}{\bar{q}_{0}^2-q_{z}^2}
        \sqrt{
        (q_{+}^{2} - \bar{q}_{0}^{2} + q_{z}^2)
        (q_{-}^{2} - \bar{q}_{0}^{2} + q_{z}^2)
        }\nonumber \\ 
    & \approx 
    -\frac{q_{+}q_{-}}{4\bar{q}_{0}}
    +\frac{\bar{q}_{0}(q_{+}-q_{-})^2}{4q_{+}q_{-}} ,
    \label{kz-star-app1}
    \\
    E_{u,n_{u}}^*
    &=
    -\frac{\bar{q}_{0}}2
    \left(
        1-\frac{q_{+}q_{-}}{\bar{q}_{0}^2-q_{z}^2}
    \right)
    -\frac{1}{2}\frac{q_{z}}{\bar{q}_{0}^2-q_{z}^2}
        \sqrt{
        (q_{+}^{2} - \bar{q}_{0}^{2} + q_{z}^2)
        (q_{-}^{2} - \bar{q}_{0}^{2} + q_{z}^2)
        }\nonumber \\ 
    & \approx
    +\frac{q_{+}q_{-}}{4\bar{q}_{0}}
    +\frac{\bar{q}_{0}(q_{+}-q_{-})^2}{4q_{+}q_{-}} ,
    \label{Eu-star-app1}
    \\
    p_{z}^*
    &=
    \frac{q_{z}}2
    \left(
        1+\frac{q_{+}q_{-}}{\bar{q}_{0}^2-q_{z}^2}
    \right)
    -\frac{1}{2}\frac{\bar{q}_{0}}{\bar{q}_{0}^2-q_{z}^2}
        \sqrt{
        (q_{+}^{2} - \bar{q}_{0}^{2} + q_{z}^2)
        (q_{-}^{2} - \bar{q}_{0}^{2} + q_{z}^2)
        }\nonumber \\ 
    & \approx 
    -\frac{q_{+}q_{-}}{4\bar{q}_{0}}
    +\frac{\bar{q}_{0}(q_{+}+q_{-})^2}{4q_{+}q_{-}} ,
    \label{pz-star-app1}
    \\
    E_{d,n_{d}}^*
    &=
    \frac{\bar{q}_{0}}2
    \left(
        1+\frac{q_{+}q_{-}}{\bar{q}_{0}^2-q_{z}^2}
    \right)
    -\frac{1}{2}\frac{q_{z}}{\bar{q}_{0}^2-q_{z}^2}
        \sqrt{
        (q_{+}^{2} - \bar{q}_{0}^{2} + q_{z}^2)
        (q_{-}^{2} - \bar{q}_{0}^{2} + q_{z}^2)
        }\nonumber \\ 
    &\approx
    +\frac{q_{+}q_{-}}{4\bar{q}_{0}}
    +\frac{\bar{q}_{0}(q_{+}+q_{-})^2}{4q_{+}q_{-}}.
    \label{Ed-star-app1}
\end{align}
From the approximate expression in Eq.~(\ref{Eu-star-app1}) we see that $\sign(E_{u}^*) = \sign(q_{+}q_{-}) = \sign(n_{d}-2n_{u})$, up to subleading corrections of order $\left(m_{d}^2-m_{u}^2\right)/|eB|$. Therefore, the term $\theta\big[E_{u,n_{u}}^{(s)}\big]$ in Eq.~(\ref{app:Theta-function}) selects the transitions with $n_{d} > 2n_{u}$. Taking these simplifications into account and taking $m_{e} \to 0$, we find that Eq.~(\ref{dfdt-app5}) simplifies to
\begin{align}
    &\frac{\partial f_\nu(t,\bm{p}_\nu)}{\partial t}
    \approx 
    \frac{N_cG_F^2\cos^2\theta_C}{36\pi^3\ell^4 E_\nu }
    \hspace{-3pt}
    \sum_{n_{u}=0}^\infty
    \sum_{n_{d}=2n_{u}+1}^\infty
    \frac{2^{n_{d}}}{3^{n_{u}+n_{d}}}
    \binom{n_{u}+n_{d}-1}{n_{u}}    \nonumber \\ 
    &\times
    \int_{0}^\infty \hspace{-4pt} dp_{e,z}
    \frac{e^{-\frac{E_\nu}{2T}}}
    {\cosh\left(\frac{E_{e,n_{e}}-\mu_{e}}{2T}\right)
    \cosh\left(
        \frac{E_{u,n_{u}}^*-\mu_{u}}{2T}\right)
    \cosh\left(
        \frac{E_{d,n_{d}}^*-\mu_{d}}{2T}\right)}
            \frac{E_{u,n_{u}}^* - k_{z}^*}
    { \sqrt{
    (q_{+}^2-\bar{q}_{0}^2+q_{z}^2)
    (q_{-}^2-\bar{q}_{0}^2+q_{z}^2)}}
    \nonumber \\ 
    &\times
    \bigg[
        \frac{n_{u}+n_{d}}{n_{d}}
        (E_\nu-p_{\nu,z})
        (E_{d,n_{d}}^*+p_{z}^*)
        +
        \frac32
        (E_\nu+p_{\nu,z})
        (E_{d,n_{d}}^*-p_{z}^*)
        +
        (n_{d}-2n_{u})
        p_{\nu,\perp}^2
    \bigg].
    \label{dfdt-app6}
\end{align}
Finally, one can further simplify the expression by keeping only the highest-order terms in the integrand outside the Fermi distribution functions: Taking $p_{e,z} = \mu_{e}$ and considering that $|\bar{q}_{0}^2-q_{z}^2|\ll q_\pm^2$, $p_{\nu,\perp}^2\sim T^{2} \ll \mu_{e} T$, and $m_{u/d}^2\ll |eB|,$ we find
\begin{align}
    \frac{\partial f_\nu(t,\bm{p}_\nu)}{\partial t}
    &\approx
    \frac{N_cG_F^2\cos^2\theta_C}{36\pi^3\ell^4}
    \frac{e^{-\frac{E_\nu}{2T}}}{E_\nu}
    \hspace{-3pt}
    \sum_{n_{u}=0}^\infty
    \sum_{n_{d}=2n_{u}+1}^\infty
    \frac{2^{n_{d}}}{3^{n_{u}+n_{d}}}
    \binom{n_{u}+n_{d}-1}{n_{u}}
    \bigg[
        \frac{n_{u}+n_{d}}{n_{d}-2n_{u}}
        (E_\nu-p_{\nu,z})    \nonumber \\ 
    & +
        \frac{n_{d}-2n_{u}}{4\mu_{e}^2\ell^2}
        (E_\nu+p_{\nu,z})
    \bigg]
    \int_{0}^\infty \frac{dp_{e,z}}
    {\cosh\left(\frac{E_{e,n_{e}}-\mu_{e}}{2T}\right)
    \cosh\left(
        \frac{E_{u,n_{u}}^*-\mu_{u}}{2T}\right)
    \cosh\left(
        \frac{E_{d,n_{d}}^*-\mu_{d}}{2T}\right)}.
    \label{dfdt-app7}
\end{align}

\section{Landau-level transitions with $n_{d} = 2n_{u}$}
\label{sec:Anom-transitions}

In this appendix, we consider the contribution to $\partial f_\nu/\partial t$ from Urca processes in which the LLs of the participating $u$- and $d$-quarks satisfy $n_{d} = 2n_{u}.$ We show that the neutrino emission rate arising from such a transition is heavily suppressed except near a two-dimensional sub-manifold of the neutrino momentum phase space, which moves far away from the origin as $|eB|$ increases. Thus, these transitions typically account for at most a few percent of the total emission rate, and their contribution becomes vanishingly small for magnetic fields $B \gtrsim 1.2\times 10^{19}$ G.

\subsection{Special case: $n_{d} = 2n_{u} > 0$}

When $n_{d} = 2n_{u} > 0$, we have $q_{-} \approx (m_{d}^2-m_{u}^2)/\left(2\sqrt{\frac{4}{3}|eB|n_{u}}\right)$. On the other hand, $|\bar{q}_{0}^2-q_{z}^2|\sim \mu_{e} T$, hence $q_{-}^{2} \ll |\bar{q}_{0}^2-q_{z}^2| \ll q_{+}^2$ for typical parameters in the region of interest. Taking this into account, the solutions given by Eqs.~(\ref{kz-pm})--(\ref{Ed-pm}) can be approximated as
\begin{align}
    \tilde{k}_{z}^{*}
    &= -\frac{q_{z}}2
    + \frac{\bar{q}_{0} q_{+}}{2\sqrt{q_{z}^2-\bar{q}_{0}^2}} ,
    \label{kz-star-star}
    \\
    \tilde{E}_{u,n_{u}}^{*}
    &= -\frac{\bar{q}_{0}}2
    + \frac{q_{z} q_{+}}{2\sqrt{q_{z}^2-\bar{q}_{0}^2}},
    \label{Eu-star-star}
    \\
    \tilde{p}_{z}^{*}
    &= \frac{q_{z}}2
    +  \frac{\bar{q}_{0} q_{+}}{2\sqrt{q_{z}^2-\bar{q}_{0}^2}},
    \label{pz-star-star}
    \\
    \tilde{E}_{d,n_{d}}^{*}
    &= \frac{\bar{q}_{0}}2
    +  \frac{q_{z} q_{+}}{2\sqrt{q_{z}^2-\bar{q}_{0}^2}},
    \label{Ed-star-star}
\end{align}
where we have discarded the solutions for which $E_{u,n_{u}}<0$. Considering that $\bar{q}_{0} \ll q_{+}$ for our typical parameter choices, we see from Eq.~(\ref{Eu-star-star}) that $\tilde{E}_{u,n_{u}}^{*} > 0$, so the second $\theta$ function in Eq.~(\ref{app:Theta-function}) is satisfied. The condition that the electron and quarks participating in the Urca process lie at their respective Fermi surfaces is enforced by the equation $\tilde{E}_{u,n_{u}}^{*} = \mu_{u}$ with $p_{e,z} = \mu_{e}$, which gives a relation between $p_{\nu,z}$ and $p_{\nu,\perp}$ for each choice of $|eB|$. Explicitly, from Eq.~(\ref{Eu-star-star}) we obtain
\begin{align}
   -\frac{\mu_{e}-p_{\nu,z}}2
    +\frac{q_{+}}{2}
    \frac{\mu_{e}-p_{\nu,z}}{\sqrt{(\mu_{e}-p_{\nu,z})^2
    -\left(\mu_{e}-\sqrt{p_{\nu,z}^2+p_{\nu,\perp}^2}\right)^2}}
     =\mu_{u},
    \label{anom-Fermi-condition}
\end{align}
where, following the same approximation as in the main text, we have neglected the electron mass as $m_{e} \ll \mu_{e}$.

\begin{figure}
\centering
\includegraphics[width=0.5\textwidth]{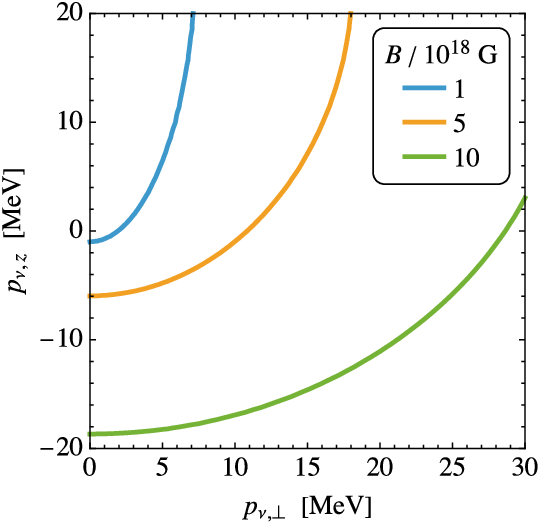}
\caption{Curves indicating values of $p_{\nu,z}$ and $p_{\nu,\perp}$ such that the electron and quarks lie at their respective Fermi surfaces. These curves are the solutions to Eq.~(\ref{anom-Fermi-condition}) for three magnetic field values, fixed chemical potentials $\mu_{e} = 50\mbox{ MeV},\mu_{u}=300\mbox{ MeV}$, and a fixed transition $(n_{d},n_{u}) = (2,1)$. No solutions exist for $B \gtrsim 1.16 \times 10^{19}$ G.}
\label{fig.anom_transitions}
\end{figure}

Solutions to Eq.~(\ref{anom-Fermi-condition}) are plotted in figure~\ref{fig.anom_transitions} for three sample magnetic field values. Assuming $|p_{\nu,z}|,p_{\nu,\perp}\ll\mu_{e}$, one can also expand Eq.~(\ref{anom-Fermi-condition}) in the small-$|p_{\nu,z}|,p_{\nu,\perp}$ limit to obtain
\begin{align}
    p_{\nu,z}
    \sim
    -\frac{\mu_{e}}4
    \left(
        \frac{q_{+}}{2\mu_{u}+\mu_{e}}
    \right)^2
    +
    \frac{p_{\nu,\perp}^2}{\mu_{e}}
    \left(
        \frac{2\mu_{u}+\mu_{e}}{q_{+}}  \right)^2,
    \label{approx-anom-Fermi-condition}
\end{align}
although the condition $|p_{\nu,z}|,p_{\nu,\perp}\ll\mu_{e}$ can only be justified \emph{a posteriori} by the solutions to Eq.~(\ref{anom-Fermi-condition}), and this begins to fail at only a few times $10^{18}$ G. Still, Eq.~(\ref{approx-anom-Fermi-condition}) and figure~\ref{fig.anom_transitions} capture the key qualitative features that (i) the neutrino production due to the $n_{d}=2n_{u}$ transitions is restricted to lie near a curve through $p_{\nu,z}$-$p_{\nu,\perp}$ phase space (equivalently, a surface through $\bm{p}_\nu$), and (ii) this curve moves away from the origin with increasing $|eB|n_{u}$. In fact, the solutions to Eq.~(\ref{anom-Fermi-condition}) contain a second branch that typically occurs at much larger values of $p_{\nu,z},p_{\nu,\perp}$, whose associated contribution is thermally suppressed. Moreover, these two branches merge near values $|\bm{p}_\nu|\sim40$ MeV and then vanish for $|eB|n_{u} \gtrsim 1.16 \times 10^{19}$ G, and hence the associated contribution to the emission rate is almost completely suppressed for sufficiently large magnetic fields or $n_{u}$.

We now let $\partial f_\nu^{(n_{d},n_{u})}/\partial t$ denote the contribution to the neutrino production rate arising only from Urca processes involving the transition $(n_{d},n_{u})$, such that
\begin{align}
    \frac{\partial f_\nu(t,\bm{p}_\nu)}
    {\partial t}
    =
    \sum_{n_{d},n_{u}=0}^\infty
    \frac{\partial f_\nu^{(n_{d},n_{u})}(t,\bm{p}_\nu)}{\partial t}.
\end{align}
Substituting the solutions in Eqs.~(\ref{kz-star-star})--(\ref{Ed-star-star}) into Eq.~(\ref{dfdt-app5}) and then applying the same approximations that led to Eq.~(\ref{dfdt-app7}), we find
\begin{align}
    &\frac{\partial f_\nu^{(2n_{u},n_{u})}(t,\bm{p}_\nu)}{\partial t}
    \approx 
    \frac{N_cG_F^2\cos^2\theta_C}{48\pi^3\ell^4 E_\nu}
    \frac{2^{2n_{u}}}{3^{3n_{u}}}
    \binom{3n_{u}-1}{n_{u}} \nonumber \\ 
&\qquad\qquad\times    \int_{0}^{\infty} dp_{e,z}
    \frac{e^{-\frac{E_\nu}{2T}} \sqrt{E_\nu-p_{\nu,z}} \left(
        \sqrt{ \frac{2|eB|n_{u}}{3\mu_{e}}}
        +\sqrt{E_\nu-p_{\nu,z}}
    \right)  }
    {\cosh\left(\frac{E_{e,n_{e}}-\mu_{e}}{2T}\right)
    \cosh\left(
        \frac{\tilde{E}_{u,n_{u}}^{*}-\mu_{u}}{2T}\right)
    \cosh\left(
        \frac{\tilde{E}_{d,n_{d}}^{*}-\mu_{d}}{2T}\right)} .
    \label{dfdt-nd=2nu}
\end{align}
In the numerator of the integrand, we have kept both the leading and subleading terms, which may formally go beyond the precision of the calculation but considerably improves the agreement with the numerical estimates.

The corresponding numerical results for the neutrino energy emission rate $\dot{\mathcal E}/{\mathcal E}_*$ at $10^{18}$ G are given in table~\ref{tab:anom-E-dot}. In particular, the total contribution to the emission rate coming from these transitions accounts for about $3\%$ and $0.6\%$ of the total rates at temperatures of 1 MeV and 0.5 MeV, respectively. As the magnetic field increases, the curves shown in figure~\ref{fig.anom_transitions} move farther from the origin, toward values of neutrino momentum that are thermally suppressed. On the other hand, the dominant term in Eq.~(\ref{dfdt-nd=2nu}) varies with $|eB|^{5/2}$, which grows faster than $\dot{\mathcal E}_* \sim |eB|$, so there are competing effects. We find that at $T = 1$ MeV, the contribution to $\dot{\mathcal E}/\dot{\mathcal E}_*$ coming from $n_{d}=2n_{u}$ transitions is maximized near $B = 4 \times 10^{18}$ G, with $(\dot{\mathcal E}/{\mathcal E}_*)_{n_{d}=2n_{u}} \approx 0.033$. While this contribution is somewhat sizable compared to a nearby minimum of $(\dot{\mathcal E}/{\mathcal E}_*)_{n_{d}>2n_{u}} \approx 0.2$ coming from transitions with $n_{d} > 2n_{u}$, we nonetheless find that the typical offset due to the $n_{d} = 2n_{u}$ transitions is only on the order of a few percent, and even less when $T < 1$ MeV. Therefore, we omit the $n_{d}=2n_{u}$ contribution from the main results presented in this paper.

\begin{table}
\centering 
\setlength{\tabcolsep}{0.5pt} 
{\small
\begin{tabular}{c *{11}{S[table-format=2.3]}} 
\toprule & 
\multicolumn{11}{c}{Emission due to transitions with $n_{d} = 2n_{u}$ at $B = 10^{18}$ G} \\
\cmidrule(lr){2-12} 
{$n_{d}$} & 2 & 4 & 6 & 8 & 10 & 12 & 14 & 16 & 18 & 20 & \\ 
{$n_{u}$} & 1 & 2 & 3 & 4 & 5 & 6 & 7 & 8 & 9 & 10 & \text{Total} \\ 
\cmidrule(lr){2-12} 
{$(\dot{\mathcal E}_{n_{d},n_{u}}/\dot{\mathcal E}_*)_{T = 1\text{ MeV}} / 10^{-3}$} & 1.47 & 3.58 & 4.64 & 4.24 & 2.92  & 1.54 & 0.61 & 0.18 & 0.04 &  0.00 & 19.21  \\
{$(\dot{\mathcal E}_{n_{d},n_{u}}/\dot{\mathcal E}_*)_{T = 0.5\text{ MeV}} / 10^{-3}$} & 0.56 & 0.70 & 0.37 & 0.11 & 0.02 & 0.00 & 0.00 & 0.00 & 0.00 & 0.00 & 1.76 \\ 
\bottomrule 
\end{tabular} }
\caption{Emission rates $\dot{\mathcal E}/\dot{\mathcal E}_*$ from transitions with $n_{d} = 2n_{u}$ at $B = 10^{18}$ G for two temperatures. Values are given for the first ten such transitions, with the sum given on the right. In total, these emission rates account for about $3\%$ and $0.6\%$ of the overall neutrino emission rate at $T = 1$ MeV and $0.5$ MeV, respectively.} 
\label{tab:anom-E-dot} 
\end{table}

\subsection{Special case: $n_{d} = n_{u} = 0$}

The case where $n_{d} = n_{u} = 0$ is more subtle than the previous cases, as the precise nature of the contribution to neutrino emission depends sensitively on the relative quark masses. Still, the main conclusion is that the contribution is suppressed in either case, and hence it can safely be neglected.

As in the preceding subsection, we use the approximation $m_{e} = 0$. We note that using  the more realistic finite value $m_{e} = 0.5$ MeV only slightly complicates the discussion but does not significantly affect the main conclusions. Let us rewrite the two possible $u$-quark energy solutions given by Eq.~(\ref{Eu-pm}) as
\begin{align}
    E_{u,n_{u}}^{(\pm)}
    &=
    -\frac{\bar{q}_{0}}2
    -\frac{\bar{q}_{0} q_{+}q_{-}
        \pm q_{z}\sqrt{
            (q_{+}^2+q_{z}^2-\bar{q}_{0}^2)
            (q_{-}^2+q_{z}^2-\bar{q}_{0}^2)
            }
    }{2(q_{z}^2-\bar{q}_{0}^2)}.
    \label{Eu-pm-var}
\end{align}
Considering again that $\bar{q}_{0} \sim q_{z} \sim \mu_{e} > 0$ and $q_{z}^{2} - \bar{q}_{0}^{2} = 2p_{e,z}(E_\nu - p_{\nu,z}) - p_{\nu,\perp}^{2} > 0$ for $|p_{\nu,z}|,p_{\nu,\perp}\ll p_{e,z}$, we see that the condition $E_{u,n_{u}} > 0$ can only be satisfied if the numerator of the second term in Eq.~(\ref{Eu-pm-var}) is negative. However, it is easy to see that the magnitude of the second term in this numerator is larger than that of the first. Thus, it follows that, regardless of the sign of $q_{+}q_{-},$ only the $s = -1$ solution has a chance to satisfy $E_{u,n_{u}}>0$. Restricting our attention to this solution, imposing the Fermi conditions $E_{u,n_{u}} = \mu_{u}$ and $p_{e,z} = \mu_{e}$, and using $q_\pm = m_{d}\pm m_{u}$, we have
\begin{align}
    \mu_{u}
    &=
    -\frac{\mu_{e} - E_\nu}2
    -\frac1{2[2\mu_{e}(E_\nu-p_{\nu,z})-p_{\nu,\perp}^2]}
    \bigg[
        (\mu_{e} - E_\nu)(m_{d}^2-m_{u}^2)
    \label{LLL-Fermi-condition}
\\ 
    &- (\mu_{e}-p_{\nu,z})\sqrt{
            [(m_{d}+m_{u})^2+2\mu_{e}(E_\nu-p_{\nu,z})-p_{\nu,\perp}^2]
            [(m_{d}-m_{u})^2+2\mu_{e}(E_\nu-p_{\nu,z})-p_{\nu,\perp}^2]
            }
    \bigg].
    \nonumber 
\end{align}

\begin{figure}
\centering
\subfigure[]{\includegraphics[width=0.45\textwidth]{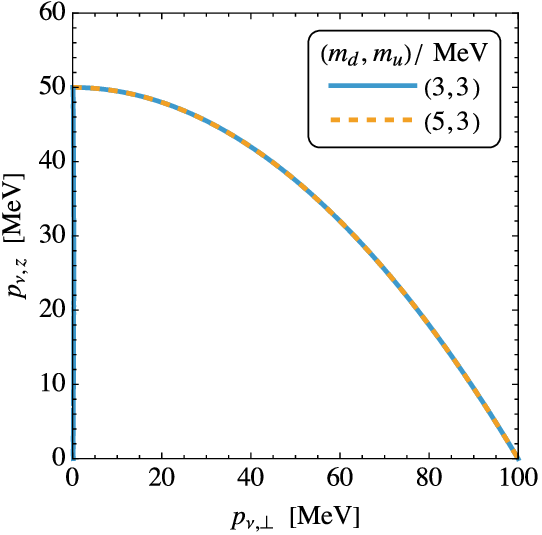}}
  \hspace{0.01\textwidth}
\subfigure[]{\includegraphics[width=0.45\textwidth]{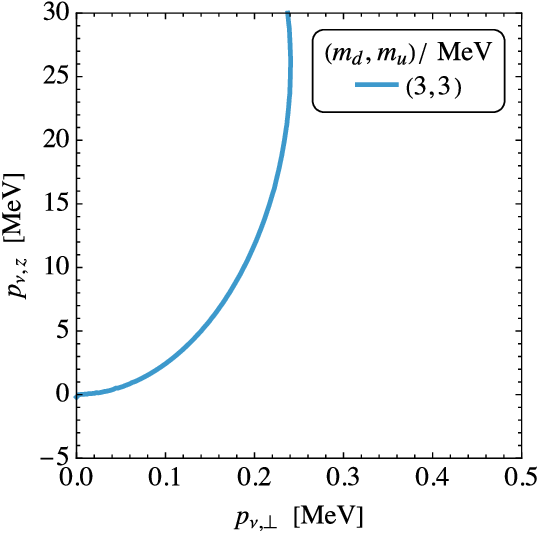}}
\caption{Curves indicating values of $p_{\nu,z}$ and $p_{\nu,\perp}$ such that the electron and quarks lie at their respective Fermi surfaces for transitions with $n_{d} = n_{u} = 0$. These curves are the solutions to Eq.~(\ref{LLL-Fermi-condition}) for sample values of quark masses.}
\label{fig.LLL-transitions}
\end{figure}

The nature of the solutions to Eq.~(\ref{LLL-Fermi-condition}) in $p_{\nu,\perp}$ and $p_{\nu,z}$ depends somewhat sensitively on the choices of quark masses. For the representative values $(m_{d},m_{u}) = (5~\text{MeV}, 3~\text{MeV})$ that we have used for the main results of this paper, the branch of solutions nearest to the origin is shown in figure~{\ref{fig.LLL-transitions}}(a) [dashed yellow line]. This curve is far from the origin, leading to a strong thermal suppression. Indeed, we estimated a contribution of only $\dot{\mathcal E}/\dot{\mathcal E}_* \approx 2 \times 10^{-23}$ coming from this transition at $B = 10^{18}$ G and $T = 1$ MeV. In the case where $(m_{d},m_{u}) = (3~\text{MeV}, 3~\text{MeV})$ for the same magnetic field and temperature, we find $\dot{\mathcal E}/\dot{\mathcal E}_* \approx 4 \times 10^{-9}$. The contribution is less strongly suppressed in this case, as the curve in figure~\ref{fig.LLL-transitions}(b) is much closer to the origin. Still, this value is negligible compared to other numerical results found in this work, so we are justified in neglecting the LLL contribution.

\section{Integral Formulas}
\label{sec:Integrals}

In this appendix, we present the main table integrals used in the calculation and outline the key steps in the derivation of the Urca emission rate in strongly magnetized dense quark matter.

\subsection{Integration over the transverse quark momentum in the $W$-boson self-energy}
\label{subsec:Integrals-k-perp}

When integrating over the $u$-quark momentum $\bm{k}_{\perp}$ in the $W$-boson self-energy in appendix~\ref{subsec:L-Im-Pi-1}, we encounter four distinct types of integrals, which were conveniently summarized by a master expression in Eq.~(\ref{I-integrals}). To keep this appendix self-contained, here we restate this master formula with slightly simplified and more explicit notation:
\begin{align}
    \mathcal{I}_{n,n'}^{\alpha,\beta}
    &=
    \int \frac{d^{2} \bm k_{\perp}}{(2\pi)^2}
    e^{-\frac32 \bm{k}_{\perp}^2\ell^2-3(\bm{k}_{\perp}+\bm{q}_{\perp})^2\ell^2}
    (\bm{p}_{e,\perp}\cdot\bm k_{\perp})^\alpha
    [\bm{p}_{\nu,\perp}
    \cdot(\bm{k}_{\perp}+\bm{q}_{\perp})]^\beta
    L_{n}^\alpha
    \left(3\bm{k}_{\perp}^2\ell^2\right)
    L_{n'}^\beta
    \left(
        6(\bm{k}_{\perp}
        +\bm{q}_{\perp})^2\ell^2
    \right)
    \nonumber \\ 
    &=
    \frac{(-1)^{n+n'+\alpha}}{18\pi\ell^2}
    \frac{2^\alpha}{3^{\alpha+\beta}}
    e^{-q_{\perp}^2\ell^2}
    \sum_{i=0}^{n}
    \sum_{j=0}^{n'}
    (-1)^{i + j}
    \frac{2^{2i+j}}{3^{i+j}}
    \binom{n+\alpha}{i+\alpha}
    \binom{n'+\beta}{j+\beta}
    \binom{i+j}{i}
    \nonumber \\ 
    &\times
    \left[
    (\bm{p}_{e,\perp}\cdot\bm q_{\perp})^\alpha
    (\bm{p}_{\nu,\perp}\cdot\bm q_{\perp})^\beta
    L_{i+j}^{\alpha+\beta}(q_{\perp}^2\ell^2)
    - 
    \frac{\alpha\beta}{2\ell^2}
    (\bm{p}_{e,\perp}\cdot\bm{p}_{\nu,\perp})
    L_{i+j}^1(q_{\perp}^2\ell^2)
    \right],
    \label{I-integrals-appD}
\end{align}
where $\alpha,\beta=0,1$. Here, we briefly outline the key steps in the derivation leading to this result.

Before proceeding to the four distinct types of integrals needed in the derivation of Eq.~(\ref{I-integrals-appD}), let us note the following auxiliary integral: 
\begin{align}
    \mathcal{N}^{\alpha,\beta}_{n,n',\mu}(\zeta)
    &=
    \int_{0}^{\infty}  dx\,  x^{\mu+1} e^{-\frac14 x^2}
    L^{\alpha}_{n}\left(\frac{x^2}{3}\right) 
    L^{\beta}_{n'}\left(\frac{x^2}{6}\right) J_{\mu}
    \left( x\sqrt{2\zeta} \right) 
    \nonumber \\ 
    &=
    2^{\mu+1}(2\zeta)^{\mu/2} e^{-2\zeta}\sum_{i=0}^{n} \sum_{j=0}^{n'}  (-1)^{i+j}
    \frac{2^{2i+j}}{3^{i+j}} 
    \binom{n+\alpha}{i+\alpha}
    \binom{n'+\beta}{j+\beta}
    \binom{i+j}{i}
    L_{i+j}^{\mu}(2\zeta),
    \label{eq:int-N-aux}
\end{align}
which is obtained by using the standard series representation for the generalized Laguerre polynomials, 
\begin{align}
L_{n}^{\alpha}(z) = \sum_{i=0}^{n} (-1)^{i} \binom{n+\alpha}{i+\alpha}\frac{z^i}{i!} ,
\label{Laguerre-series}
\end{align}
as well as the table integral 6.631~10 in ref.~\cite{Gradshteyn:1943cpj},
\begin{align}
\int_{0}^{\infty} dx \, e^{-x^2} x^{2n+\mu+1} J_{\mu}(2x\sqrt{z}) 
= \frac{n!}{2} z^{\mu/2} e^{-z} L_{n}^{\mu}(z).
\label{Gradshteyn-6.631.10}
\end{align}
The first type of integral over the transverse quark momentum appearing in the $W$-boson self-energy is given by
\begin{align}
    \mathcal{I}^{0,0}_{n,n'} 
    &= \int \frac{d^{2} \bm{k}_{\perp}}{(2\pi)^2}  
    e^{- \frac{3}{2}\bm{k}_{\perp}^2\ell^{2} -3(\bm{k}_{\perp} +\bm{q}_{\perp} )^2\ell^2} 
    L_{n} \left(3\bm{k}_{\perp}^{2} \ell^2\right) 
    L_{n'} \left(6(\bm{k}_{\perp} +\bm{q}_{\perp})^{2} \ell^2\right)
    \nonumber \\ 
    &= \frac{(-1)^{n+n'}}{72 \pi^2 \ell^4}  \int d^{2} \bm{u}_{\perp} 
    e^{- \frac{u_{\perp}^2}{4\ell^2}}
    L_{n}\left(\frac{u_{\perp}^2}{3\ell^2}\right)
    L_{n'}\left(\frac{u_{\perp}^2}{6\ell^2}\right)
    e^{ -i  \bm{q}_{\perp} \cdot \bm{u}_{\perp} } ,
\label{appD-I00a}
\end{align}
where the first representation is in momentum space and the second in coordinate space. The transformation from momentum to coordinate space can be carried out using the following Fourier transform:
\begin{align}
e^{- \frac{a}{2}p_{\perp}^2\ell^2}L_{n}\left(a p_{\perp}^2\ell^2\right)
&=\frac{(-1)^{n} }{2 a \pi \ell^2}\int d^{2} \bm{u}_{\perp} e^{-u_{\perp}^2/(2a\ell^2)} L_{n}\left(\frac{u_{\perp}^2}{a\ell^2}\right)
e^{ -i \bm{p}_{\perp}\cdot  \bm{u}_{\perp}}  . 
\label{FT-one} 
\end{align}
One can now use the identity
\begin{align}
    \int d^{2} \bm u_{\perp} 
    f(u_{\perp})
    e^{-i \bm q_{\perp} \cdot \bm u_{\perp} }
    &=
    2\pi
    \int_{0}^\infty u_{\perp} du_{\perp}
    f(u_{\perp})
    J_{0}(q_{\perp} u_{\perp}),
    \label{polar-coords-1}
\end{align}
to introduce polar coordinates and integrate over the angular coordinate in Eq.~(\ref{appD-I00a}). The final result is obtained using the auxiliary integral in Eq.~(\ref{eq:int-N-aux}), namely,
\begin{align}
    \mathcal{I}^{0,0}_{n,n'} 
    &=
    \frac{(-1)^{n+n'}}{36 \pi \ell^2} \mathcal{N}^{0,0}_{n,n',0}(\zeta) ,
    \label{appD-I00}
\end{align}
where $\zeta = \frac{1}{2} q_{\perp}^2\ell^2$. 

One can follow analogous steps to obtain the other three types of integrals, employing the additional Fourier transform,
\begin{align}
(\bm{p}_{\perp}\cdot\bm{b})  
e^{-\frac{a}{2} p_{\perp}^2\ell^2}L^{1}_{n}\left(ap_{\perp}^2\ell^2\right) 
&= i \frac{ (-1)^n }{2 a^{2} \pi \ell^4} \int d^{2} \bm{u}_{\perp}  (\bm{u}_{\perp}\cdot\bm{b})  e^{-u_{\perp}^2/(2a\ell^2)} L^{1}_{n}\left(\frac{u_{\perp}^2}{a\ell^2}\right) e^{ -i \bm{p}_{\perp}\cdot  \bm{u}_{\perp}},
\label{FT-two}
\end{align}
and the following analogues of Eq.~(\ref{polar-coords-1}),
\begin{align}
    \int d^{2} \bm u_{\perp} 
    (\bm u_{\perp}\cdot\bm{b})
    f(u_{\perp})
    e^{-i \bm q_{\perp} \cdot \bm u_{\perp} }
    &=
    2\pi
        (-i\bm{b}\cdot\bm{\hat q}_{\perp})
        \int_{0}^\infty u_{\perp}^{2} du_{\perp}
        f(u_{\perp})
        J_{1}(q_{\perp} u_{\perp}) ,
    \label{polar-coords-2}
    \\
    \int d^{2} \bm u_{\perp} 
    (\bm u_{\perp}\cdot\bm{b})
    (\bm u_{\perp}\cdot\bm{c})
    f(u_{\perp})
    e^{-i \bm q_{\perp} \cdot \bm u_{\perp} }
    &=
    2\pi
    \Big[
        (-i\bm{b}\cdot\bm{\hat q}_{\perp})
        (-i\bm{c}\cdot\bm{\hat q}_{\perp})
        \int_{0}^\infty u_{\perp}^3 du_{\perp}
        f(u_{\perp})
        J_{2}(q_{\perp} u_{\perp})
    \nonumber \\ 
    &+
    (\bm{b}\cdot \bm{c})
    \int_{0}^\infty u_{\perp}^{2} du_{\perp}
    \frac1{q_{\perp}}
    f(u_{\perp})
    J_{1}(q_{\perp} u_{\perp})
    \Big] ,
    \label{polar-coords-3}
\end{align}
where $\bm{b}$ and $\bm{c}$ denote arbitrary constant vectors in the plane transverse to the magnetic field. Using these formulas, we find:
\begin{align}
    \mathcal{I}^{0,1}_{n,n'} 
    &= \int \frac{d^{2} \bm{k}_{\perp}}{(2\pi)^2}  
    e^{- \frac{3}{2}\bm{k}_{\perp}^2\ell^{2} -3(\bm{k}_{\perp} +\bm{q}_{\perp} )^2\ell^2} 
    \left[\bm p_{\nu,\perp}\cdot (\bm{k}_{\perp} +\bm{q}_{\perp}) \right]
    L_{n} \left(3\bm{k}_{\perp}^{2} \ell^2\right) 
    L_{n'}^{1} \left(6(\bm{k}_{\perp} +\bm{q}_{\perp})^{2} \ell^2\right)\nonumber \\ 
    &= i \frac{(-1)^{n+n'} }{ 432 \pi^2 \ell^6}  \int d^{2} \bm{u}_{\perp} 
    e^{- u_{\perp}^2/(4\ell^2)}(\bm{u}_{\perp} \cdot\bm{p}_{\nu,\perp}) 
    L_{n}\left(\frac{u_{\perp}^2}{3\ell^2}\right)
    L_{n'}^{1}\left(\frac{u_{\perp}^2}{6\ell^2}\right)
    e^{-i  \bm{q}_{\perp} \cdot \bm{u}_{\perp} }
    \nonumber \\ 
    &= \frac{(-1)^{n+n'}}{216 \pi \ell^{2} \sqrt{2\zeta}}   (\bm{q}_{\perp} \cdot\bm{p}_{\nu,\perp}) 
    \mathcal{N}^{0,1}_{n,n',1}(\zeta) ,
   \label{appD-I01}
\end{align}
\begin{align}
    \mathcal{I}^{1,0}_{n,n'} 
    &= \int \frac{d^{2} \bm{k}_{\perp}}{(2\pi)^2}
    e^{- \frac{3}{2}\bm{k}_{\perp}^2\ell^{2} -3(\bm{k}_{\perp} +\bm{q}_{\perp} )^2\ell^2} 
    \left(\bm p_{e,\perp} \cdot \bm k_{\perp} \right)
    L_{n}^{1} \left(3\bm{k}_{\perp}^{2} \ell^2\right) 
    L_{n'} \left(6(\bm{k}_{\perp} +\bm{q}_{\perp})^{2} \ell^2\right)\nonumber \\ 
    &= -i \frac{(-1)^{n+n'} }{216 \pi^2 \ell^6}   \int d^{2} \bm{u}_{\perp}
    e^{-u_{\perp}^2/(4\ell^2)}
    (\bm{u}_{\perp} \cdot\bm{p}_{e,\perp}) 
    L_{n}^{1}\left(\frac{u_{\perp}^2}{3\ell^2}\right)
    L_{n'}\left(\frac{u_{\perp}^2}{6\ell^2}\right)
    e^{ -i  \bm{q}_{\perp} \cdot \bm{u}_{\perp} }
    \nonumber \\ 
    &=  -\frac{(-1)^{n+n'}}{108 \pi \ell^{2}  \sqrt{2\zeta} }  (\bm{q}_{\perp} \cdot\bm{p}_{e,\perp}) 
    \mathcal{N}^{1,0}_{n,n',1}(\zeta) ,
\end{align}
and
\begin{align}
    \mathcal{I}^{1,1}_{n,n'} 
    &= \int \frac{d^{2} \bm{k}_{\perp}}{(2\pi)^2}  
    e^{- \frac{3}{2}\bm{k}_{\perp}^2\ell^{2} -3(\bm{k}_{\perp} +\bm{q}_{\perp} )^2\ell^2} 
    \left(\bm p_{e,\perp} \cdot \bm k_{\perp} \right)
    \left[\bm p_{\nu,\perp}\cdot (\bm{k}_{\perp} +\bm{q}_{\perp}) \right]
    L_{n}^{1} \left(3\bm{k}_{\perp}^{2} \ell^2\right) 
    L_{n'}^{1}  \left(6(\bm{k}_{\perp} +\bm{q}_{\perp})^{2} \ell^2\right)\nonumber \\ 
    &= \frac{(-1)^{n+n'} }{1296 \pi^2  \ell^8 }  \int d^{2} \bm{u}_{\perp} 
    e^{-u_{\perp}^2/(4\ell^2)}
    (\bm{u}_{\perp} \cdot\bm{p}_{e,\perp}) 
    (\bm{u}_{\perp} \cdot\bm{p}_{\nu,\perp}) 
    L_{n}^{1}\left(\frac{u_{\perp}^2}{3\ell^2}\right)
    L_{n'}^{1}\left(\frac{u_{\perp}^2}{6\ell^2}\right)
    e^{ -i  \bm{q}_{\perp} \cdot \bm{u}_{\perp} }
    \nonumber \\ 
    &=  -\frac{(-1)^{n+n'}}{648 \pi \ell^2} \left[
    \frac{1}{2\zeta}
    (\bm{q}_{\perp} \cdot\bm{p}_{e,\perp}) 
    (\bm{q}_{\perp} \cdot\bm{p}_{\nu,\perp}) 
    \mathcal{N}^{1,1}_{n,n',2}(\zeta) 
    -  \frac{(\bm{p}_{\nu,\perp} \cdot\bm{p}_{e,\perp}) }{\ell^{2} \sqrt{2 \zeta} }
    \mathcal{N}^{1,1}_{n,n',1}(\zeta) \right].
\end{align}
As can be easily verified, upon using the series representation of the functions $\mathcal{N}^{\alpha,\beta}_{n,n',\mu}(\zeta)$ given in Eq.~(\ref{eq:int-N-aux}), all four formulas for $\mathcal{I}_{n,n'}^{\alpha,\beta}$ with $\alpha,\beta =0, 1$ can be cast into the master expression in Eq.~(\ref{I-integrals-appD}). Also, their derivations can be streamlined into a single unified derivation by using the following combined forms of the Fourier transforms in Eqs.~(\ref{FT-one}) and (\ref{FT-two}):
\begin{align}
    (\bm{p}_{\perp}\cdot\bm{b}) ^\alpha
    e^{-\frac a2 p_{\perp}^2\ell^2} 
    L^\alpha_{n}\left(ap_{\perp}^2\ell^2\right) 
    &= 
    \frac{(-1)^n}{2 a \pi \ell^2}
    \left(\frac{i }{a\ell^2}\right)^\alpha
    \int d^{2} \bm{u}_{\perp}  
    (\bm{u}_{\perp}\cdot\bm{b})^\alpha
    e^{-u_{\perp}^2/(2a\ell^2)}
    L^\alpha_{n}\left(\frac{u_{\perp}^2}{a\ell^2}\right)
    e^{-i  \bm{p}_{\perp}\cdot \bm{u}_{\perp}} ,
    \label{FT-general}
\end{align}
and the following integrations over the polar angle in Eqs.~(\ref{polar-coords-1}) and (\ref{polar-coords-2})--(\ref{polar-coords-3}):
{\small 
\begin{align}
    \int d^{2} \bm u_{\perp} 
    (\bm u_{\perp}\cdot\bm{b})^\alpha
    (\bm u_{\perp}\cdot\bm{c})^{\beta}
    f(u_{\perp})
    e^{-i  \bm q_{\perp} \cdot \bm u_{\perp} }
    &=
    2\pi
    \Big[
        (-i \bm{b}\cdot\bm{\hat q}_{\perp})^\alpha
        (-i \bm{c}\cdot\bm{\hat q}_{\perp})^{\beta}
        \int_{0}^\infty u_{\perp}^{\alpha+\beta+1} du_{\perp}
        f(u_{\perp})
        J_{\alpha+\beta}(q_{\perp} u_{\perp})
    \nonumber \\ 
    &+
    \alpha \beta \frac{(\bm{b}\cdot \bm{c})}{q_{\perp}}
    \int_{0}^\infty u_{\perp}^{2} du_{\perp}
    f(u_{\perp})
    J_{1}(q_{\perp} u_{\perp})
    \Big],
    \label{polar-coords-general}
\end{align}}
with $\alpha,\beta = 0, 1$. In particular, we find
\begin{align}
    \mathcal{I}_{n,n'}^{\alpha,\beta}
    =
    \frac{(-1)^{n+n'+\alpha}}{36 \pi \ell^2 3^\alpha 6^\beta}
    &\bigg[
        (\bm{q}_{\perp} \cdot \bm{p}_{e,\perp})^\alpha
        (\bm{q}_{\perp} \cdot \bm{p}_{\nu,\perp})^\beta
        (2\zeta)^{-(\alpha+\beta)/2}
        \mathcal{N}_{n,n',\alpha+\beta}^{\alpha,\beta}(\zeta) \nonumber \\ 
        & -\alpha\beta 
        \frac{(\bm{p}_{e,\perp} \cdot \bm{p}_{\nu,\perp})}{\ell^2}
        (2\zeta)^{-1/2}
        \mathcal{N}_{n,n',1}^{1,1}(\zeta)
    \bigg],
    \label{master-I-formula-with-N}
\end{align}
where $\zeta = \frac12 q_\perp^2\ell^2$. Substituting the expression for $\mathcal{N}_{n,n',\mu}^{\alpha,\beta}(\zeta)$ from Eq.~(\ref{eq:int-N-aux}) into Eq.~(\ref{master-I-formula-with-N}) gives Eq.~(\ref{I-integrals-appD}).

\subsection{Integration over the transverse electron momentum in the rate}
\label{subsec:Integrals-p-e-perp}

When integrating over the transverse electron momentum $\bm{p}_{e,\perp}$ in the final expression for the rate in appendix~\ref{subsec:calc-dfdt}, we encounter four additional types of integrals, which are described by the following expression:
\begin{align}
    \mathcal{J}_{n,n'}^{\alpha,\beta}
    &=
    \int\frac{d^2\bm p_{e,\perp}}{(2\pi)^2}
    e^{-\bm p_{e,\perp}^2\ell^{2} 
    -(\bm p_{e,\perp}-\bm p_{\nu,\perp})^{2} \ell^2}
    L_{n}^\alpha \left(2\bm p_{e,\perp}^2\ell^2\right)
    \bigg[
        [\bm{p}_{e,\perp}\cdot (\bm p_{e,\perp}-\bm p_{\nu,\perp})]^\alpha
        [\bm{p}_{\nu,\perp}\cdot (\bm p_{e,\perp}-\bm p_{\nu,\perp})]^{\beta}\nonumber \\ 
        &\times
        L_{n'}^{\alpha+\beta}\left((\bm p_{e,\perp}-\bm p_{\nu,\perp})^2\ell^2\right)
        - \frac{\alpha \beta}{2\ell^2}
        (\bm{p}_{e,\perp}\cdot\bm{p}_{\nu,\perp})
        L_{n'}^1\left((\bm p_{e,\perp}-\bm p_{\nu,\perp})^2\ell^2\right)
    \bigg]\nonumber \\ 
        & = \frac{(-1)^{n+\beta} \xi^{\beta} }{2^{3+n'+\alpha} \pi \ell^{2+2(\alpha+\beta)}}
    e^{-\xi}
    \sum_{k=0}^n (-1)^k
    \binom{n+\alpha}{k+\alpha}
    \frac{(n'+k+\alpha)!}{(n')!k!}
    L_{n'+k+\alpha}^\beta(\xi)   ,
    \label{J-def-appD}
\end{align}
for each possible choice of $\alpha,\beta = 0, 1$.  
Here, we outline the key steps in the calculating these integrals. 

As in the previous section, we start from an auxiliary integral,
\begin{align}
    \mathcal{M}_{n,n'}^{\alpha,\beta}(\xi)
    &=
    \int_{0}^\infty dx \, x^{\alpha+1}
    e^{-\frac14 x^2} 
    L_{n}^\alpha\left(\frac{x^2}2\right)
    J_\beta\left(x\sqrt{2\xi}\right)
    \int_{0}^\infty dy \, y^{\alpha+\beta+1}
    e^{-y^2} L_{n'}^{\alpha+\beta}(y^2)
    J_{\alpha+\beta}(xy)
    \nonumber \\ 
    &=
    \frac1{2^{n'+1}}
    \left( \frac{\xi}{2} \right)^{\beta/2} 
    e^{-\xi}
    \sum_{k=0}^n (-1)^k
    \binom{n+\alpha}{k+\alpha}
    \frac{(n'+k+\alpha)!}{(n')!k!}
    L_{n'+k+\alpha}^\beta(\xi) ,
    \label{int-M-aux}
\end{align}
which is obtained using Eqs.~(\ref{Laguerre-series}) and (\ref{Gradshteyn-6.631.10}), and table integral 7.421 6 from ref.~\cite{Gradshteyn:1943cpj},
\begin{align}
    \int_{0}^\infty dy \, y^{\mu+1} e^{-y^2}
    L_{n}^\mu(y^2) J_\mu(xy)
    &= \frac1{2\, n!}\left(\frac x2\right)^{2n+\mu}
    e^{-\frac14 x^2}.
    \label{int-Hankel-Laguerre}
\end{align}

Starting with the $\alpha=\beta = 0$ case, we have
\begin{align}
    \mathcal{J}_{n,n'}^{0,0}
    &=
    \int\frac{d^2\bm p_{e,\perp}}{(2\pi)^2}
    e^{-\bm p_{e,\perp}^2\ell^{2} 
    -(\bm p_{e,\perp}-\bm p_{\nu,\perp})^{2} \ell^2}
    L_{n}\left(2\bm p_{e,\perp}^2\ell^2\right)
    L_{n'}\left((\bm p_{e,\perp}-\bm p_{\nu,\perp})^2\ell^2\right)
    \nonumber \\ 
    &= 
    \frac{(-1)^n}{4\pi\ell^2}\! 
    \int\! \frac{d^2\bm p_{e,\perp}}{(2\pi)^2}
    e^{-(\bm p_{e,\perp}-\bm p_{\nu,\perp})^{2} \ell^2}\! 
        \int\!  d^2\bm u_{\perp}
        e^{-u_{\perp}^2/(4\ell^2)}
        L_{n}\left(\frac{u_{\perp}^2}{2\ell^2}\right)
        e^{-i  \bm p_{e,\perp}\cdot \bm u}
    L_{n'}\left((\bm p_{e,\perp}-\bm p_{\nu,\perp})^2\ell^2\right),
\end{align}
where we have applied the Fourier transform in Eq.~(\ref{FT-one}) once. Unlike in the previous section, we cannot straightforwardly apply the Fourier transform to the second factor due to the mismatch in pre-factors of the arguments of the exponential and Laguerre functions. Instead, we change the integration variable $\bm p_{e,\perp} \to \bm p_{e,\perp} + \bm p_{\nu,\perp}$ and then switch to polar coordinates, and integrate over the polar angle using Eq.~(\ref{polar-coords-1}). The result reads
{\small 
\begin{align}
    \mathcal{J}_{n,n'}^{0,0}
    &=
    \frac{(-1)^n}{4\pi\ell^2}
    \int_{0}^\infty u_{\perp}  du_{\perp}
    e^{-u_{\perp}^2/(4\ell^2)}
    L_{n}\left(\frac{u_{\perp}^2}{2\ell^2}\right)
    J_{0}(u_{\perp} p_{\nu,\perp})
    \int_{0}^\infty p_{e,\perp} dp_{e,\perp}
    e^{-p_{e,\perp}^{2} \ell^2}
    L_{n'}(p_{e,\perp}^2\ell^2)
    J_{0}(u_{\perp} p_{e,\perp})
    \nonumber \\ 
    &=
    \frac{(-1)^n}{4\pi\ell^2}
    \mathcal{M}_{n,n'}^{0,0}(\xi),
\end{align}}
where $\xi = \frac{1}{2} p_{\nu,\perp}^2\ell^2$. 

Following analogous steps but using the Fourier transform in Eqs.~(\ref{FT-two}) and the integral over the polar angle in Eq.~(\ref{polar-coords-2}), we find:
{\small 
\begin{align}
    \mathcal{J}_{n,n'}^{0,1}
    &=
    \int\frac{d^2\bm p_{e,\perp}}{(2\pi)^2}
    e^{-\bm p_{e,\perp}^2\ell^{2} 
    -(\bm p_{e,\perp}-\bm p_{\nu,\perp})^{2} \ell^2}
    [\bm p_{\nu,\perp}\cdot(\bm p_{e,\perp}-\bm p_{\nu,\perp})]
    L_{n}\left(2\bm p_{e,\perp}^2\ell^2\right)
    L_{n'}^1\left((\bm p_{e,\perp}-\bm p_{\nu,\perp})^2\ell^2\right)
    \nonumber \\ 
    &= 
    \frac{(-1)^n}{4\pi\ell^2} (-p_{\nu,\perp})
    \int_{0}^\infty u_{\perp}  du_{\perp}
    e^{-u_{\perp}^2/(4\ell^2)}
    L_{n}\left(\frac{u_{\perp}^2}{2\ell^2}\right)
    J_{1}(u_{\perp} p_{\nu,\perp})
    \int_{0}^\infty p_{e,\perp}^{2} dp_{e,\perp}
    e^{-p_{e,\perp}^{2} \ell^2}
    L_{n'}^1(p_{e,\perp}^2\ell^2)
    J_{1}(u_{\perp} p_{e,\perp})
    \nonumber \\ 
    &=
    -\frac{(-1)^n p_{\nu,\perp}}{4\pi\ell^3}
    \mathcal{M}_{n,n'}^{0,1}(\xi) ,
\end{align}}
{\small 
\begin{align}
    \mathcal{J}_{n,n'}^{1,0}
    &=
    \int\frac{d^2\bm p_{e,\perp}}{(2\pi)^2}
    e^{-\bm p_{e,\perp}^2\ell^{2} 
    -(\bm p_{e,\perp}-\bm p_{\nu,\perp})^{2} \ell^2}
    [\bm p_{e,\perp}\cdot(\bm p_{e,\perp}-\bm p_{\nu,\perp})]
    L_{n}^1\left(2\bm p_{e,\perp}^2\ell^2\right)
    L_{n'}\left((\bm p_{e,\perp}-\bm p_{\nu,\perp})^2\ell^2\right)
    \nonumber \\ 
    &= 
    \frac{(-1)^n}{4\pi\ell^2}
    \frac 1{2\ell^2}
    \int_{0}^\infty u_{\perp}^{2}  du_{\perp}
    e^{-u_{\perp}^2/(4\ell^2)}
    L_{n}^1\left(\frac{u_{\perp}^2}{2\ell^2}\right)
    J_{0}(u_{\perp} p_{\nu,\perp})
    \int_{0}^\infty p_{e,\perp}^{2} dp_{e,\perp}
    e^{-p_{e,\perp}^{2} \ell^2}
    L_{n'}^1(p_{e,\perp}^2\ell^2)
    J_{1}(u_{\perp} p_{e,\perp})
    \nonumber \\ 
    &=
    \frac{(-1)^n}{8\pi\ell^4}
    \mathcal{M}_{n,n'}^{1,0}(\xi).
\end{align}}

The case with $\alpha=\beta = 1$ is more subtle, because we have two additional terms to consider: one appears directly in the definition of $\mathcal{J}_{n,n'}^{1,1}$, namely the second term in the integrand of Eq.~(\ref{J-def-appD}), and the other results from transforming to polar coordinates, as the second term on the right-hand side of Eq.~(\ref{polar-coords-3}) has no analogue in the previous cases. It turns out that these two extra terms exactly cancel.

To see this, let
\begin{align}
    \mathcal{J}_{n,n'}^{1,1}
    &= \mathcal{A}_{n,n'}
    -\mathcal{B}_{n,n'},
    \label{app:Jnn-11}
\end{align}
where
\begin{align}
    \mathcal{A}_{n,n'}
    &=
    \int\frac{d^2\bm p_{e,\perp}}{(2\pi)^2}
    e^{-\bm p_{e,\perp}^2\ell^{2} 
    -(\bm p_{e,\perp}-\bm p_{\nu,\perp})^{2} \ell^2}
    [\bm{p}_{e,\perp}\cdot (\bm p_{e,\perp}-\bm p_{\nu,\perp})]
    [\bm{p}_{\nu,\perp}\cdot (\bm p_{e,\perp}-\bm p_{\nu,\perp})]
\nonumber \\ 
&\times  
    L_{n}^1\left(2\bm p_{e,\perp}^2\ell^2\right)
    L_{n'}^2\left((\bm p_{e,\perp}-\bm p_{\nu,\perp})^2\ell^2\right) ,
    \\
    \mathcal{B}_{n,n'}
    &=
    \frac1{2\ell^2}
    \int\frac{d^2\bm p_{e,\perp}}{(2\pi)^2}
    e^{-\bm p_{e,\perp}^2\ell^{2} 
    -(\bm p_{e,\perp}-\bm p_{\nu,\perp})^{2} \ell^2}
    (\bm{p}_{e,\perp}\cdot\bm{p}_{\nu,\perp})  
    L_{n}^1\left(2\bm p_{e,\perp}^2\ell^2\right)
    L_{n'}^1\left((\bm p_{e,\perp}-\bm p_{\nu,\perp})^2\ell^2\right).
\end{align}
Following the same steps as above, we find
{\small
\begin{align}
    \mathcal{A}_{n,n'}
    &= 
    \mathcal{A}_{n,n'}^{(1)} 
    + \mathcal{A}_{n,n'}^{(2)} , 
    \\
    \mathcal{A}_{n,n'}^{(1)} 
    &=
    -\frac{(-1)^n p_{\nu,\perp}}{8\pi\ell^4}
    \int_{0}^\infty u_{\perp}^{2} du_{\perp}
    e^{-u_{\perp}^2/(4\ell^2)}
    L_{n}^1\left(\frac{u_{\perp}^2}{2\ell^2}\right)
    J_{1}(u_{\perp} p_{\nu,\perp})
    \int_{0}^\infty p_{e,\perp}^3 dp_{e,\perp}
    e^{-p_{e,\perp}^2\ell^2}
    L_{n'}^2(p_{e,\perp}^2\ell^2)
    J_{2}(u_{\perp} p_{e,\perp})
    \nonumber \\ 
    &=
    -\frac{(-1)^n p_{\nu,\perp}}{8\pi\ell^5}
    \mathcal{M}_{n,n'}^{1,1}(\xi) ,
    \\
    \mathcal{A}_{n,n'}^{(2)} 
    &=
    \frac{(-1)^n p_{\nu,\perp}}{8\pi\ell^4}
    \int_{0}^\infty u_{\perp} du_{\perp}
    e^{-u_{\perp}^2/(4\ell^2)}
    L_{n}^1\left(\frac{u_{\perp}^2}{2\ell^2}\right)
    J_{1}(u_{\perp} p_{\nu,\perp})
    \int_{0}^\infty p_{e,\perp}^{2} dp_{e,\perp}
    e^{-p_{e,\perp}^2\ell^2}
    L_{n'}^2(p_{e,\perp}^2\ell^2)
    J_{1}(u_{\perp} p_{e,\perp}) ,
    \label{appD-A2nn}
\end{align}}
and
{\small 
\begin{align}
    \mathcal{B}_{n,n'}
    &=
    \frac{(-1)^n p_{\nu,\perp}}{16\pi\ell^6}
    \int_{0}^\infty u_{\perp}^{2} du_{\perp}
    e^{-u_{\perp}^2/(4\ell^2)}
    L_{n}^1\left(\frac{u_{\perp}^2}{2\ell^2}\right)
    J_{1}(u_{\perp} p_{\nu,\perp})
    \int_{0}^\infty p_{e,\perp} dp_{e,\perp}
    e^{-p_{e,\perp}^2\ell^2}
    L_{n'}^1(p_{e,\perp}^2\ell^2)
    J_{0}(u_{\perp} p_{e,\perp}).
        \label{appD-Bnn}
\end{align}}
Using the following relation for the generalized Laguerre polynomials:
$L_{n}^{\alpha +1 }\!\left(p_{e,\perp}^2\ell^2\right)=\sum_{m=0}^{n} L_{m}^{\alpha}\!\left(p_{e,\perp}^2\ell^2\right)$ together with the table integral in Eq.~(\ref{int-Hankel-Laguerre}), the integration over $p_{e,\perp}$ in Eqs.~(\ref{appD-A2nn}) and (\ref{appD-Bnn}) can be carried out straightforwardly using the following results:
\begin{align}
    \frac y2
    \int_{0}^\infty dx \, x e^{-x^2}
    L_{n}^1(x^2) J_{0}(xy)
    &=
    \int_{0}^\infty dx \, x^{2} e^{-x^2}
    L_{n}^2(x^2) J_{1}(xy)
    =\frac{y}{4}
e^{-\frac{y^{2}}{4}}
\sum_{k=0}^{n}
\frac{1}{k!} \left(\frac{y}{2}\right)^{2k}.
\end{align}
Then, it follows immediately that $\mathcal{A}_{n,n'}^{(2)} = \mathcal{B}_{n,n'}$, so that the corresponding contributions cancel in the expression for $\mathcal{J}_{n,n'}^{1,1}$ in Eq.~(\ref{app:Jnn-11}). We therefore have
\begin{align}
    \mathcal{J}_{n,n'}^{1,1}
    = \mathcal{A}_{n,n'}^{(1)}
    = -\frac{(-1)^n p_{\nu,\perp}}{8\pi\ell^5}
    \mathcal{M}_{n,n'}^{1,1}(\xi).
\end{align}
As in the previous section, using the series representation of $\mathcal{M}_{n,n'}^{\alpha,\beta}(\xi)$ in Eq.~(\ref{int-M-aux}), the four integrals $\mathcal{J}_{n,n'}^{\alpha,\beta}$ reproduce the master formula given in Eq.~(\ref{J-def-appD}). We also note that, apart from the cancellation of the extra terms in $\mathcal{J}_{n,n'}^{1,1}$, the formulas for $\mathcal{J}_{n,n'}^{\alpha,\beta}$ and their derivations can again be streamlined using Eqs.~(\ref{FT-general}) and (\ref{polar-coords-general}). Proceeding in this fashion, one can show
\begin{align}
    \mathcal{J}_{n,n'}^{\alpha,\beta}(\xi)
    &=
    \frac{(-1)^{n+\beta}}{2^{2+\alpha-\beta}\pi \ell^{2+2(\alpha+\beta)}}
    \left(\frac{\xi}{2}\right)^{\beta/2}
    \mathcal{M}_{n,n'}^{\alpha,\beta}(\xi)     .
    \label{master-J-formula-with-M}
\end{align}
Substituting the expression for $\mathcal{M}_{n,n'}^{\alpha,\beta}(\xi)$ from Eq.~(\ref{int-M-aux}) into Eq.~(\ref{master-J-formula-with-M}) gives Eq.~(\ref{J-def-appD}).

It is useful to note that the following relation holds:
\begin{align}
\mathcal{M}_{n,n'}^{1,1}(\xi)
    &=(n'+1)\left[
\mathcal{M}_{n,n'}^{0,1}(\xi)
+\sqrt{2\xi} \mathcal{M}_{n,n'+1}^{0,0}(\xi)
\right].
\end{align}
This relation also implies that
\begin{align}
\mathcal{J}_{n,n'}^{1,1}(\xi)
    &=\frac{n'+1}{2\ell^2}\left[
\mathcal{J}_{n,n'}^{0,1}(\xi)
-\frac{2\xi}{\ell^2 } \mathcal{J}_{n,n'+1}^{0,0}(\xi)
\right].
\end{align}
Let us mention in passing that there is an alternative derivation of $\mathcal{M}_{n,n'}^{\alpha,\beta}$, based on the generating function of the Laguerre polynomials, which yields the following expression:
\begin{align}
\mathcal{M}_{n,n'}^{\alpha,\beta}(\xi)
    &=\frac{(-1)^{n}(n'+1)^{\alpha}}{2^{n'+1}}
    \left(\frac{\xi}{2}\right)^{\beta/2}
    e^{-\xi}
    \sum_{k=0}^{n'+\alpha}(-1)^{k}
    \binom{n'+\alpha+\beta}{k+\beta}
    \frac{\xi^{k}}{k!}
    L_{n}^{n'-n+k+\beta}(\xi),
\end{align}
where $\alpha,\beta=0,1$. (The details of the generating-function method are omitted for brevity.) Although this expression appears quite different from that obtained in Eq.~(\ref{int-M-aux}), the two representations are equivalent. Using the alternative form of $\mathcal{M}_{n,n'}^{\alpha,\beta}$ and the relation in Eq.~(\ref{master-J-formula-with-M}), the result for  $\mathcal{J}_{n,n'}^{\alpha,\beta}(\xi)$ can be written as follows:
\begin{align}
\mathcal{J}_{n,n'}^{\alpha,\beta}(\xi)
    &=\frac{(-1)^{\beta}(n'+1)^{\alpha}\xi^{\beta}}
    {2^{n'+\alpha+3}\pi\ell^{2+2(\alpha+\beta)}}
    e^{-\xi}
    \sum_{k=0}^{n'+\alpha}(-1)^{k}
    \binom{n'+\alpha+\beta}{k+\beta}
    \frac{\xi^{k}}{k!}
    L_{n}^{n'-n+k+\beta}(\xi).
\end{align}


\providecommand{\href}[2]{#2}\begingroup\raggedright\endgroup

\end{document}